\documentclass{article}

\usepackage{arxiv}

\usepackage[utf8]{inputenc} 
\usepackage[T1]{fontenc}    
\usepackage{hyperref}       
\usepackage{url}            
\usepackage{booktabs}       
\usepackage{amsfonts}       
\usepackage{nicefrac}       
\usepackage{microtype}      
\usepackage{lipsum}		
\usepackage{graphicx}
\usepackage{natbib}
\usepackage{doi}
\usepackage{amsmath}
\usepackage{bm}
\usepackage{subcaption} 
\usepackage[most]{tcolorbox}

\title{A Misleading Gallery of Fluid Motion by Generative Artificial Intelligence}

\date{} 					

\author{ 
{Ali Kashefi}\thanks{Corresponding author.} \\
	Department of Civil \& Environmental Engineering\\ 
 Stanford University, 
	Stanford, CA 94305, USA \\
	\texttt{kashefi@stanford.edu} \\
 }

\renewcommand{\headeright}{}
\renewcommand{\undertitle}{}
\renewcommand{\shorttitle}{A Misleading Gallery of Fluid Motion by Generative AI}

\hypersetup{
pdftitle={A template for the arxiv style},
pdfsubject={q-bio.NC, q-bio.QM},
pdfauthor={David S.~Hippocampus, Elias D.~Striatum},
pdfkeywords={First keyword, Second keyword, More},
}

\begin{document}
\maketitle

\begin{abstract}
In this technical report, we extensively investigate the accuracy of outputs from well-known generative artificial intelligence (AI) applications in response to prompts describing common fluid motion phenomena familiar to the fluid mechanics community. We examine a range of applications, including Midjourney, Dall·E, Runway ML, Microsoft Designer, Gemini, Meta AI, and Leonardo AI, introduced by prominent companies such as Google, OpenAI, Meta, and Microsoft. Our text prompts for generating images or videos include examples such as ``Von Kármán vortex street'', ``flow past an airfoil'', ``Kelvin-Helmholtz instability'', ``shock waves on a sharp-nosed supersonic body'', etc. We compare the images generated by these applications with real images from laboratory experiments and numerical software. Our findings indicate that these generative AI models are not adequately trained in fluid dynamics imagery, leading to potentially misleading outputs. Beyond text-to-image/video generation, we further explore the transition from image/video to text generation using these AI tools, aiming to investigate the accuracy of their descriptions of fluid motion phenomena. This report serves as a cautionary note for educators in academic institutions, highlighting the potential for these tools to mislead students. It also aims to inform researchers at these renowned companies, encouraging them to address this issue. We conjecture that a primary reason for this shortcoming is the limited access to copyright-protected fluid motion images from scientific journals.
\end{abstract}


\section{Introduction and motivation}
\label{Sect1}

Generative AI models \citep{gozalo2023chatgpt,feuerriegel2024generative,lim2023generative,cao2023comprehensive,epstein2023art,kashefi2023chatgpt,brynjolfsson2023generative,fui2023generative} are nowadays popular and are used in various applications with transforming text, images, video, and sounds based on diverse input prompts. A subset of generative models is Large Language Models (LLMs) \citep{xu2022systematic,hoffmann2022training,thirunavukarasu2023large,kasneci2023chatgpt,chang2023survey,chen2021evaluating,kojima2022large,yang2023context,touvron2023llama,roziere2023code,ccelen2024design}, which have seen substantial development from famous tech companies such as Google, Microsoft, Open AI, Midjourney, and Runway ML.

One specific task of these models is to generate images from text prompts. As researchers in the fluid mechanics and applied mathematics communities, we are interested in investigating the capability of these generative AI models to produce accurate images based on textual prompts describing fluid mechanics phenomena. For our investigation, we use common textual prompts that are almost universally known in the community for a fair assessment.

Image generation using textual prompts can be assessed on two levels. At the first level, the question is whether the generative AI models can generate at least a relevant image. A relevant image can be thought of as something that first comes to mind for a fluid mechanics expert. For instance, if the prompt is ``Von-Karmann Vortex Street'', \citep{tritton1959experiments,kashefi2018finite,kashefi2020coarse,kashefi2022coarseMHD} an image similar to the cover of the ``An Album of Fluid Motion'' book \citep{van1982album} might be expected. At a deeper level, the images should not only be relevant but also accurately describe the characteristic features of fluid motion. For example, for the prompt ``Flow over a backward-facing step at Reynolds number of 400'', \citep{kim1985application,armaly1983experimental,le1997direct,kashefi2018finite,kashefi2020coarseB} it is crucial to assess whether the reattachment length in the generated image is correct. This technical report focuses on the first level. We compare generative images with those found in open-source articles from the Journal of Fluid Mechanics, which are derived from lab experiments or numerical software.

The remainder of this technical report is organized as follows. Sect. \ref{Sect2prime} offers a brief review of the governing equations of fluid motion along with two special cases of compressible and incompressible flow. Sect. \ref{Sect2}, which is the main part of this report, focuses on text-to-image generation for fluid motion using well-known AI tools. In Sect. \ref{Sect3}, an example of text-to-video generation relevant to fluid motion is provided. Image-to-text and video-to-text generation, as applied to fluid dynamic motion, are investigated in Sect. \ref{Sect4} and Sect. \ref{Sect5}, respectively. Lastly, we provide a summary and concluding remarks of our technical report in Sect. \ref{Sect6}.


\section{Brief overview of fluid dynamic motion}
\label{Sect2prime}

In this subsection, we briefly review the governing equations for fluids at the most general level. The two main principles are the continuity equation and the Cauchy momentum equation \citep{euler1757principes,lukaszewicz2016navier,constantin1988navier,chorin1990mathematical,batchelor1967introduction} and can be written respectively as
\begin{equation}
    \frac{D\rho}{Dt} + \rho \left (\nabla \cdot \bm{\mathit{u}}\right) = 0,
    \label{Eq1}
\end{equation}

\begin{equation}
    \rho  \frac{D\bm{\mathit{u}}}{Dt}  = \rho \bm{\mathit{f}} + \nabla \cdot \bm{\mathit{T}},
    \label{Eq2}
\end{equation}
where $\rho$ represents the fluid density, and $\bm{\mathit{u}}$ represents the velocity vector. The total derivative with respect to time is denoted by $\frac{D}{Dt}$. The body force is represented by $\bm{\mathit{f}}$. $\bm{\mathit{T}}$ represents the most general form of the stress tensor. Under different assumptions, these two equations (Eqs. \ref{Eq1}--\ref{Eq2}) can be simplified. For instance, in the case of compressible flow \citep{feistauer2003mathematical,novotny2004introduction}, the stress tensor is expressed as
\begin{equation}
    \bm{\mathit{T}} = -p \bm{\mathit{I}} + \lambda \left(\nabla \cdot \bm{\mathit{u}}\right) \bm{\mathit{I}} + \mu \left(\nabla \bm{\mathit{u}}+ \nabla \bm{\mathit{u}}^T \right),
    \label{Eq3}
\end{equation}
where $p$ indicates the fluid pressure. Dynamic viscosity and the second viscosity are respectively shown by $\mu$ and $\lambda$. $\bm{\mathit{I}}$ is the identity tensor. As another simplification, in the case of incompressible flow \citep{majda2002vorticity,panton2024incompressible}, the velocity vector is divergence-free, which is mathematically represented as
\begin{equation}
    \nabla \cdot \bm{\mathit{u}} = 0,
    \label{Eq4}
\end{equation}
leading to further simplification of Eq. \ref{Eq3}. These equations (Eqs. \ref{Eq1}--\ref{Eq2}), which may be coupled with others, describe the behavior of fluid motion in an Eulerian system. Similarly, fluid motion can be described in a Lagrangian system. In a specified time-space domain, along with boundary and initial conditions, the equations can be solved numerically, and the results are illustrated using visualization software \citep{matsson2023introduction,squillacote2007paraview,ayachit2015paraview,yang2020applied,jasak2009openfoam,jasak2007openfoam,heidbach2020manual,pryor2009multiphysics,flaischlen2019synthetic,mullen2011mastering}. In this technical report, we consider images of fluid motions obtained by both experimental \citep{huang1997errors,westerweel1997effect,willert1991digital,van1982album,adrian1991particle,tropea2007springer,raffel2018particle} and numerical tools \citep{hirsch2007numerical,patankar2018numerical,osher2004level,anderson1995computational,reddy2010finite}.


\section{Text to image}
\label{Sect2}

In this subsection, we first briefly introduce different generative applications and then compare their outputs for eleven different text prompts related to fluid dynamics motion. Specifically, we consider Midjourney \footnote{\href{https://www.midjourney.com/home}{https://www.midjourney.com/home}} (Sect. \ref{Sect21}), ChatGPT-4 \footnote{\href{https://chat.openai.com}{https://chat.openai.com}} and Microsoft Designer \footnote{\href{https://designer.microsoft.com}{https://designer.microsoft.com}} with DALL·E 3 (Sect. \ref{Sect22}), Runway ML \footnote{\href{https://runwayml.com/}{https://runwayml.com/}} (Sect. \ref{Sect23}), Gemini \footnote{\href{https://gemini.google.com/app}{https://gemini.google.com/app}} (Sect. \ref{Sect24}), Meta AI \footnote{\href{https://ai.meta.com/meta-ai/}{https://ai.meta.com/meta-ai/}} (Sect. \ref{Sect25}), and Leonardo AI 
\footnote{\href{https://runwayml.com/}{https://leonardo.ai}} (Sect. \ref{Sect26}).

\subsection{Midjourney}
\label{Sect21}

Midjourney \citep{borji2022generated,hanna2023use,huang2023crossing,henderson2023foundation} was released in July 2022 by the research lab Midjourney. Its primary goal is to generate images from users' textual prompts. The platform offers different plans, with the basic one priced at \$10 per month. Access to this AI platform is provided through Discord. Additionally, users can view the prompts and images generated by others. The capabilities of Midjourney have been investigated by other researchers in areas such as graphic design \citep{mansour2023intelligent}, architecture \citep{jaruga2022artificial}, industrial design \citep{yin2023exploration}, etc.

\subsection{DALL·E (embedded in ChatGPT 4, ChatGPT 4o, and Microsoft Designer)}
\label{Sect22}

DALL·E was developed and released by OpenAI in 2021 \citep{kang2023beyond,leivada2023dall,borji2022generated,marcus2022very,betker2023improving,adams2023does,hwang2023brief,henderson2023foundation,ajmera2024validity}. The most recent version, DALL·E 3, is now integrated with ChatGPT 4, and recently ChatGPT 4o. As a result, users of ChatGPT 4 and ChatGPT 4o can generate images using textual prompts. Currently, the basic plan for ChatGPT 4 and ChatGPT 4o is priced at \$20 per month, with a limit of 40 messages every three hours.

It is noted that Microsoft Designer currently uses DALL·E 3 for generating images as well. Microsoft Designer \citep{hwang2023brief,henderson2023foundation,pearson2023rise} was launched by Microsoft and is currently still in its experimental phase. This AI tool generates photos based on user-generated textual prompts, and the platform is currently available for free to users.

\subsection{Runway ML}
\label{Sect23}

Runway ML \citep{hales2021artificial,othman2023ai,gupta2021agency} was launched in 2018 by Runway AI, Inc. This platform has a variety of applications, but we specifically explore its capabilities in text-to-image and text-to-video generation, particularly in the context of fluid motion. The current version of Runway operates based on the deep learning model of Stable Diffusion \citep{rombach2022high,esser2024scaling,meng2023distillatio,ho2022imagen}. Notably, researchers from Runway AI, Inc. contributed to the development of the Stable Diffusion algorithm \citep{borji2022generated}. It is worth noting that Stability AI is another notable company that contributed to the development of the Stable Diffusion algorithm \citep{esser2024scaling}. However, in this work, our examination is focused on Runway ML. Note that Runway ML offers different styles for generating images. In the current article, we use the \textit{None} style.

\subsection{Gemini Advanced}
\label{Sect24}

Gemini, a large language model developed by Google DeepMind, originally known as Bard, can generate images from textual descriptions and describe image content in text \citep{masalkhi2024google,mcintosh2023google,singh2023chat,aydin2023google,campesato2024google,carla2024exploring,buscemi2024chatgpt,lu2024gpt}. This report explores both capabilities of Gemini: text-to-image in this subsection and image-to-text in Sect. \ref{Sect43}. The basic version of Gemini is free, while its advanced version is free for two months and then costs \$20 per month. We use the advanced version (i.e., Gemini Advanced) for the current study.

\subsection{Meta AI (built on Meta LLaMA 3)}
\label{Sect25}

Meta AI, released by Meta, is a large language model built on the Meta LLaMA 3 architecture \citep{touvron2023llama,bhatt2024cyberseceval,li2024nuteprune,harper2024autogenesisagent}. The model is currently available for free, but only in specific countries. The model has the capability to generate images based on text prompts.

\subsection{Leonardo Ai}
\label{Sect26}

Leonardo Ai is another tool that operates using deep learning models \citep{ruiz2023empowering,wang2023investigating,moreno2023artificial}. For the purposes of this technical report, we examine its capabilities for text-to-image generation. Additionally, we utilize the paid version of the service (\$12 per month) for this exploration. Note that Leonardo Ai offers different models with different modes. In this study, we use the model of ``Leonardo Lightning XL'' with the ``Dynamics" mode.

\subsection{Comparison}
\label{Sect27}

\begin{figure}[t]
    \centering
    \begin{subfigure}[b]{0.16\textwidth}
        \centering
        \includegraphics[width=\textwidth]{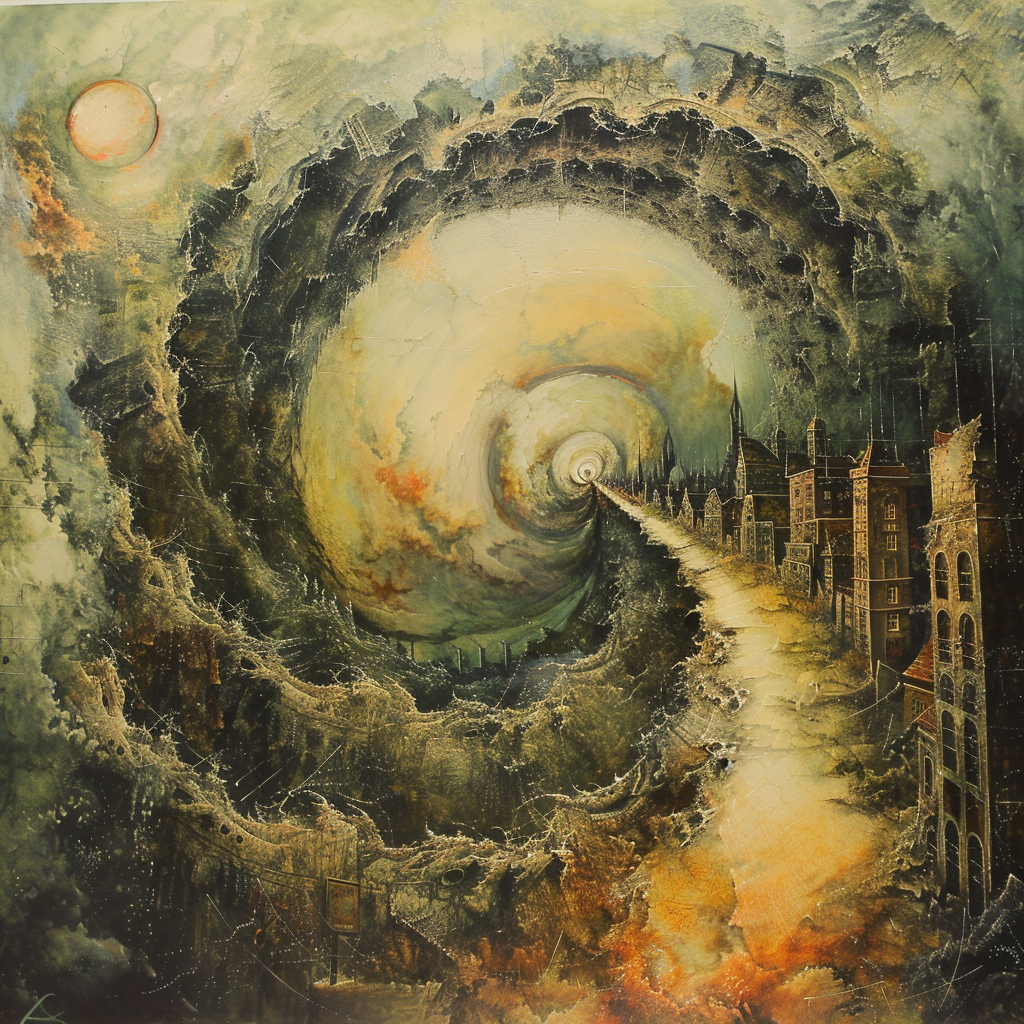}
        \caption{Midjourney}
    \end{subfigure}
    \begin{subfigure}[b]{0.16\textwidth}
        \centering
        \includegraphics[width=\textwidth]{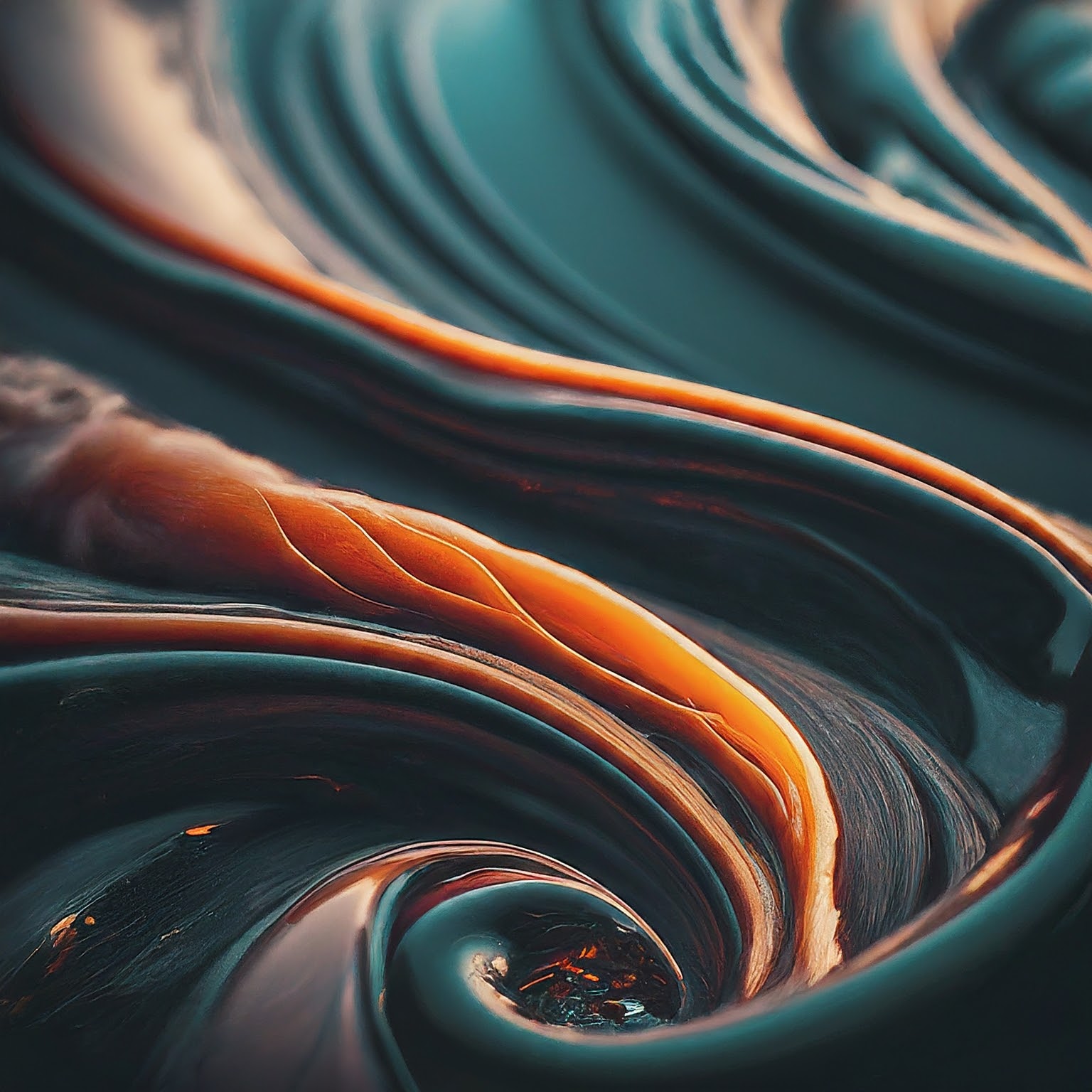}
        \caption{Gemini}
    \end{subfigure}
    \begin{subfigure}[b]{0.16\textwidth}
        \centering
        \includegraphics[width=\textwidth]{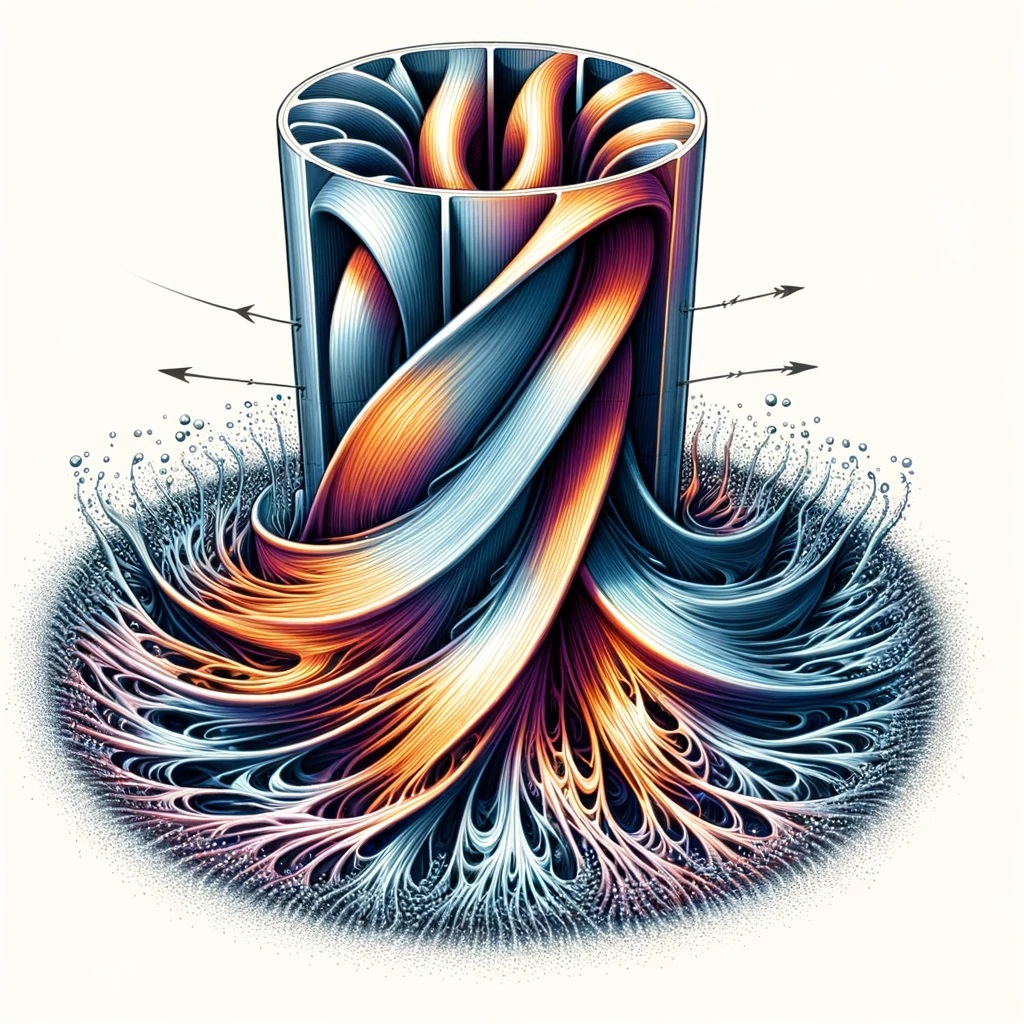}
        \caption{DALL·E 3}
    \end{subfigure}
    \begin{subfigure}[b]{0.16\textwidth}
        \centering
        \includegraphics[width=\textwidth]{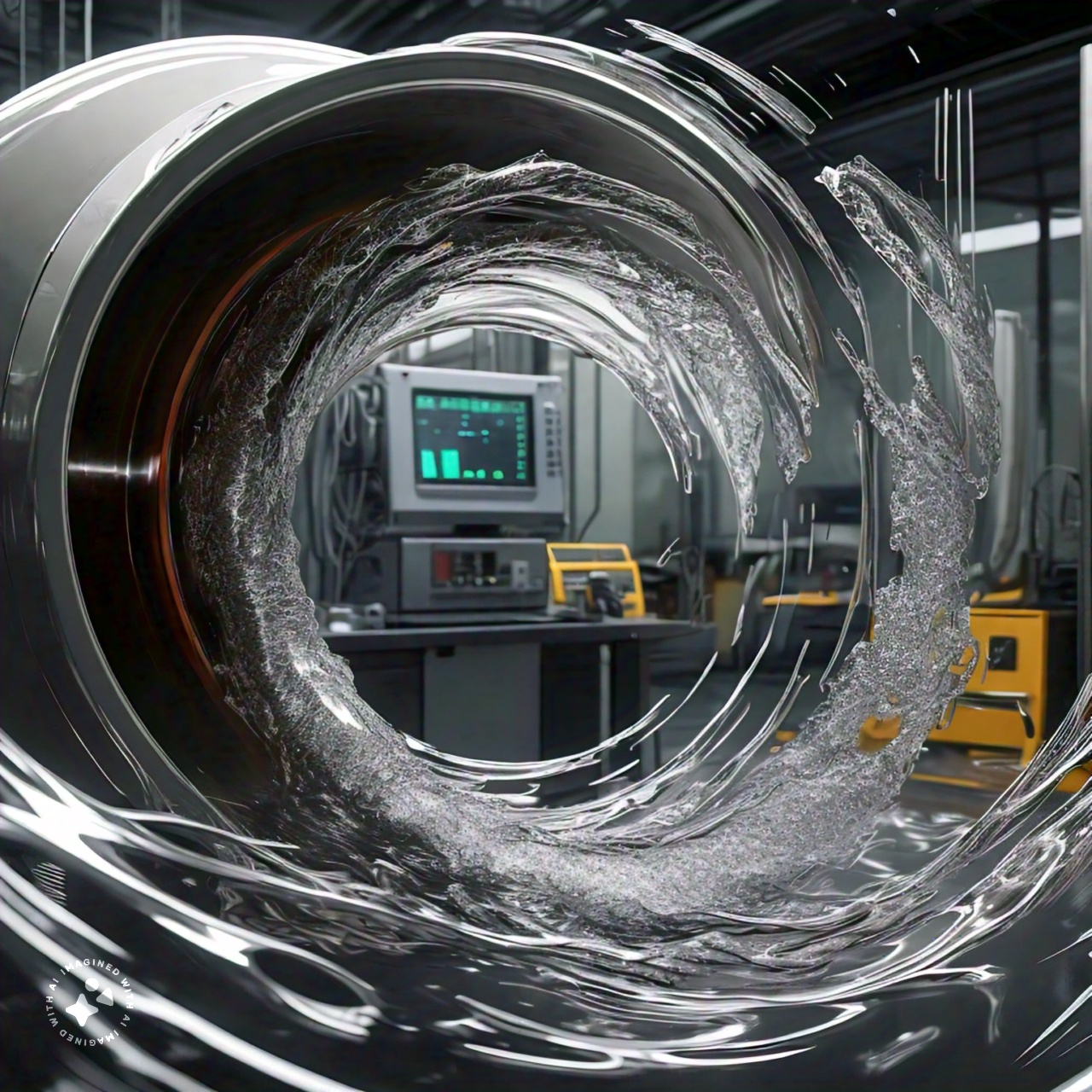}
        \caption{Meta AI}
    \end{subfigure}
    \begin{subfigure}[b]{0.16\textwidth}
        \centering
        \includegraphics[width=\textwidth]{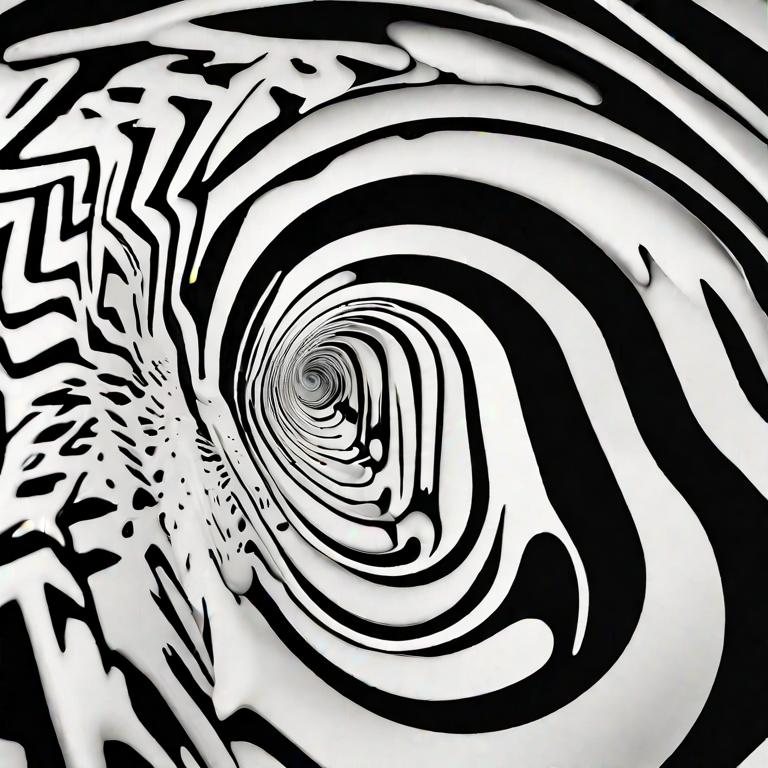}
        \caption{Runway}
    \end{subfigure}
    \begin{subfigure}[b]{0.16\textwidth}
        \centering
        \includegraphics[width=\textwidth]{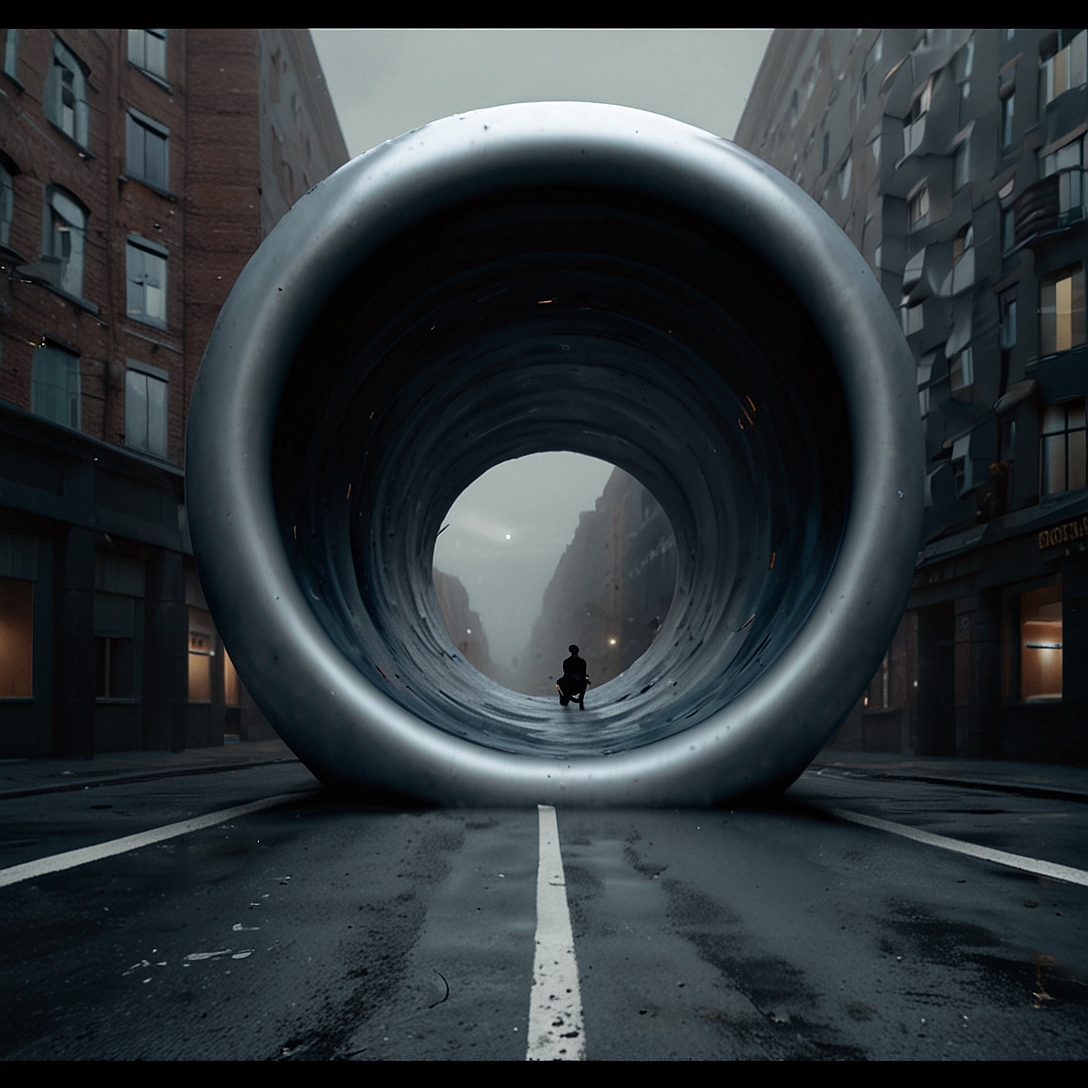}
        \caption{Leonardo Ai}
    \end{subfigure}

    \begin{subfigure}[b]{\textwidth}
        \centering
        \includegraphics[width=0.4\textwidth]{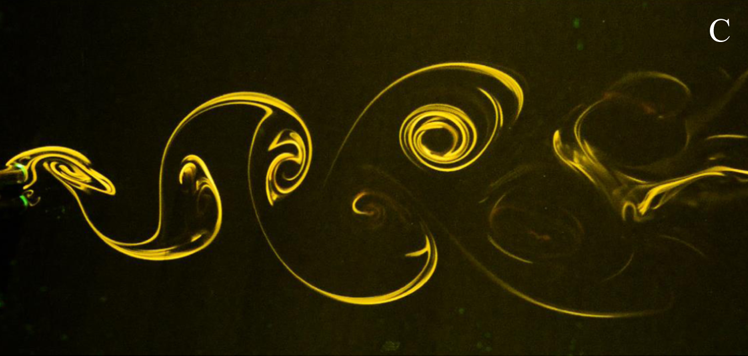}
        \caption{Von Kármán vortex street behind an oscillating circular cylinder in a lab experimental setup; this image is taken from part (c) of Fig. 25 in \cite{khan2023flow}, published as open access under a Creative Commons license in the Journal of Fluid Mechanics, allowing readers to distribute the content freely.}
    \end{subfigure}

    \caption{A comparison between the AI-generated images and the lap experimental result (i.e., ground truth) for the prompt ``Von Kármán vortex street''}
    \label{Fig7}
\end{figure}

In this subsection, we compare the performance of these six different generative models. As a general structure for this subsection, we present images generated by these generative models, along with a sub-figure in part (g) of Figs. \ref{Fig7}--\ref{Fig6} that shows an accurate image (as ground truth) associated with the prompt relevant to that fluid dynamics phenomenon.


\subsubsection{Von Kármán vortex street}
\label{Sect271}

\begin{figure}[t]
    \centering
    \begin{subfigure}[b]{0.16\textwidth}
        \centering
        \includegraphics[width=\textwidth]{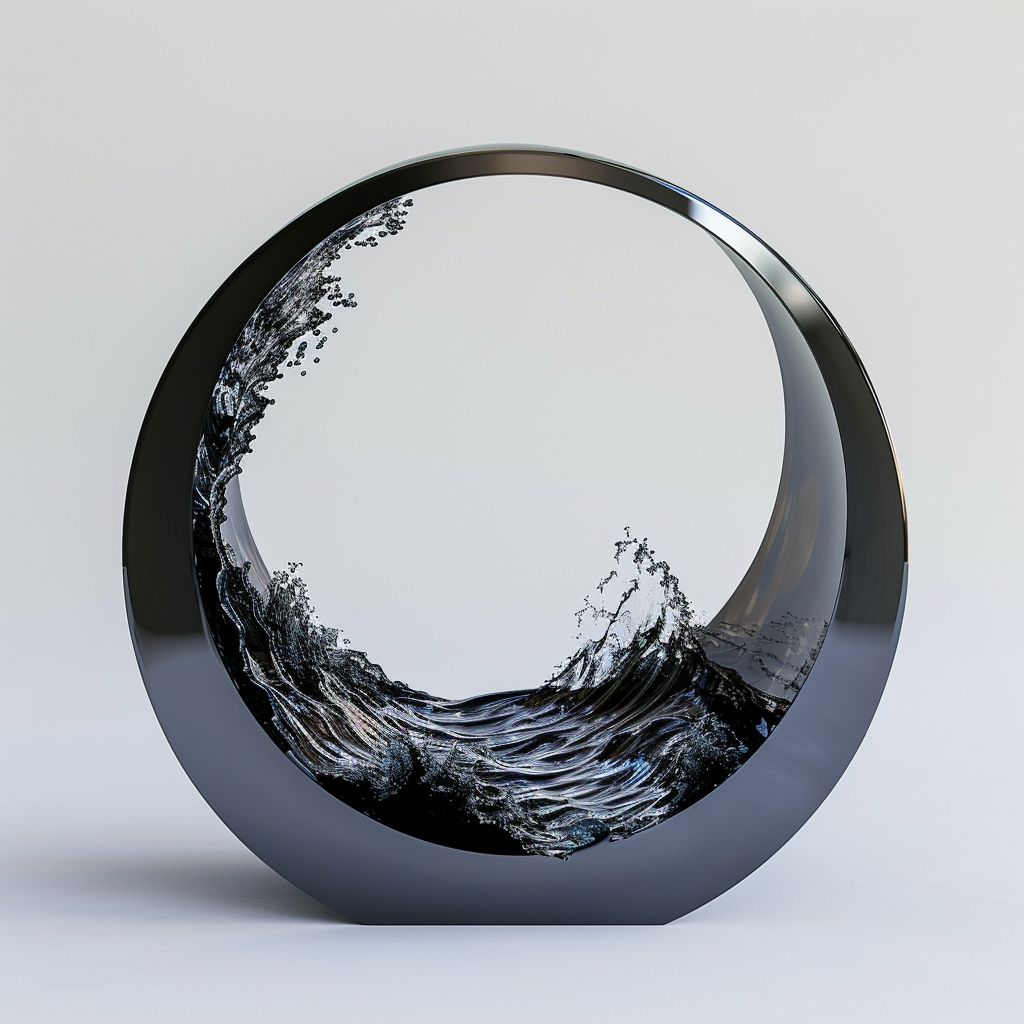}
        \caption{Midjourney}
    \end{subfigure}
    \begin{subfigure}[b]{0.16\textwidth}
        \centering
        \includegraphics[width=\textwidth]{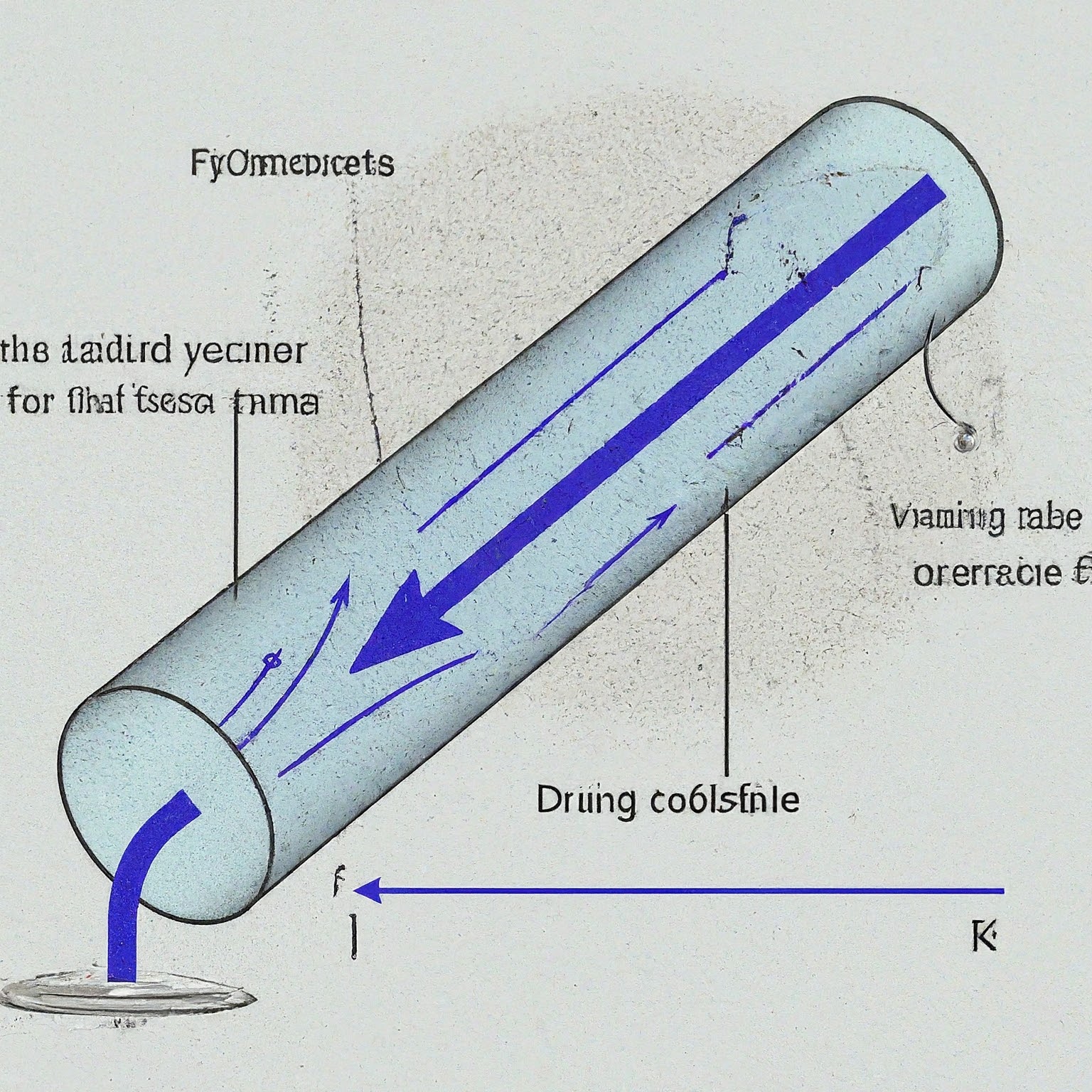}
        \caption{Gemini}
    \end{subfigure}
    \begin{subfigure}[b]{0.16\textwidth}
        \centering
        \includegraphics[width=\textwidth]{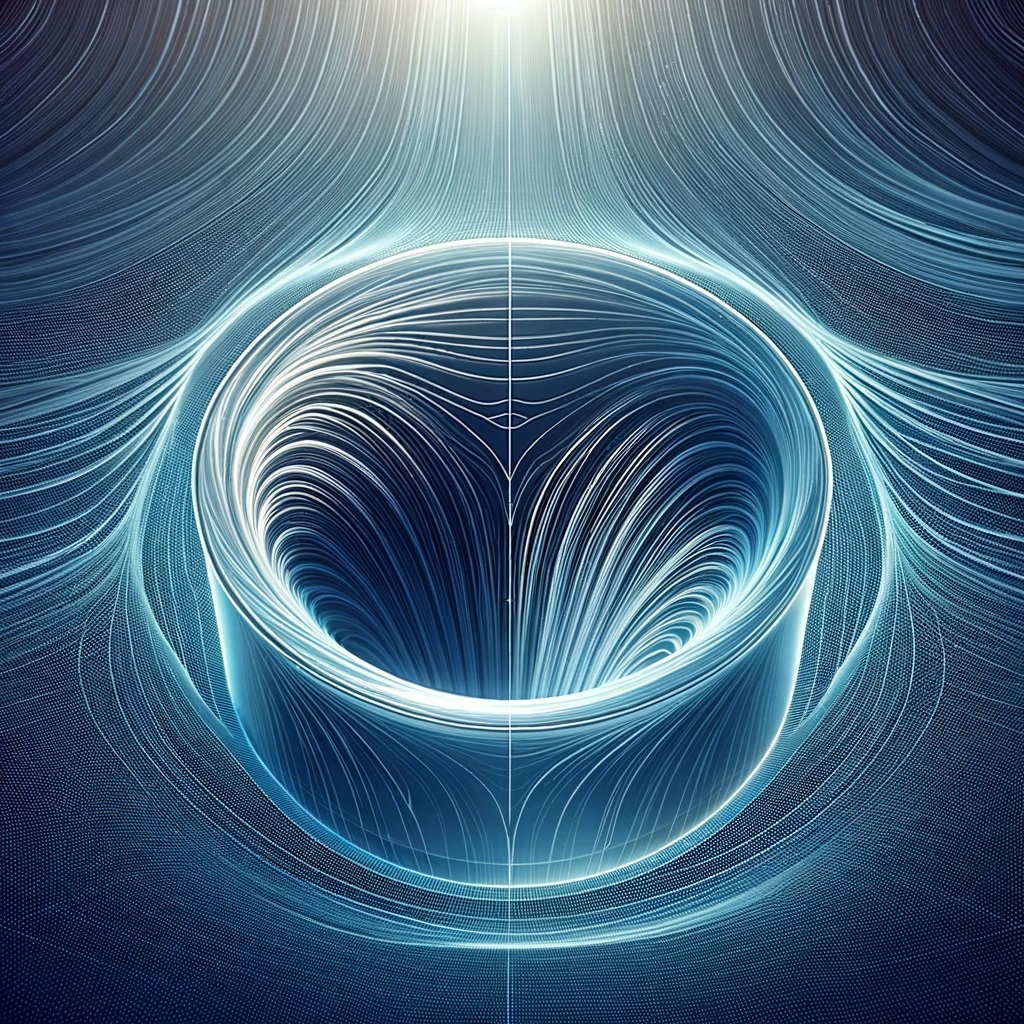}
        \caption{DALL·E 3}
    \end{subfigure}
    \begin{subfigure}[b]{0.16\textwidth}
        \centering
        \includegraphics[width=\textwidth]{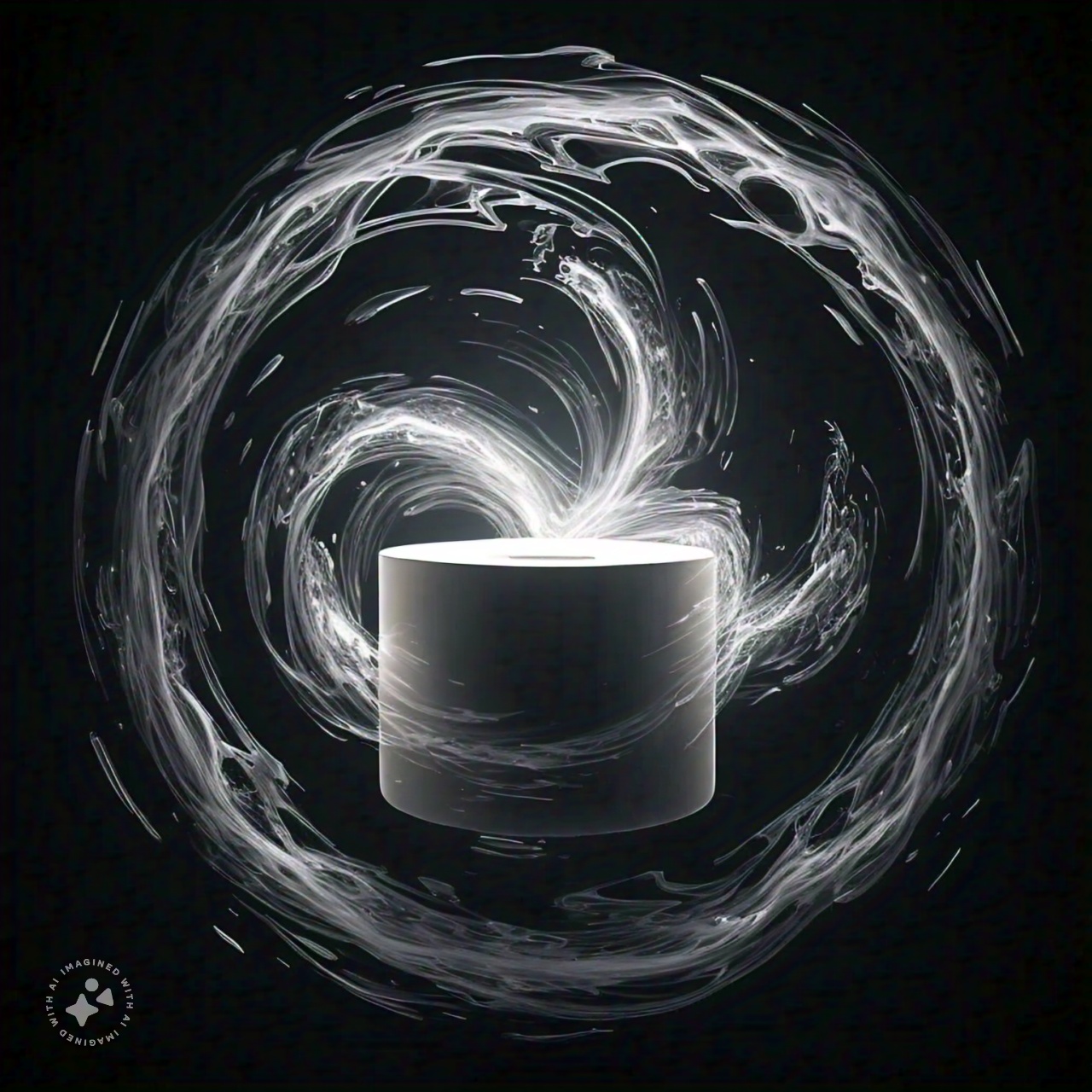}
        \caption{Meta AI}
    \end{subfigure}
    \begin{subfigure}[b]{0.16\textwidth}
        \centering
        \includegraphics[width=\textwidth]{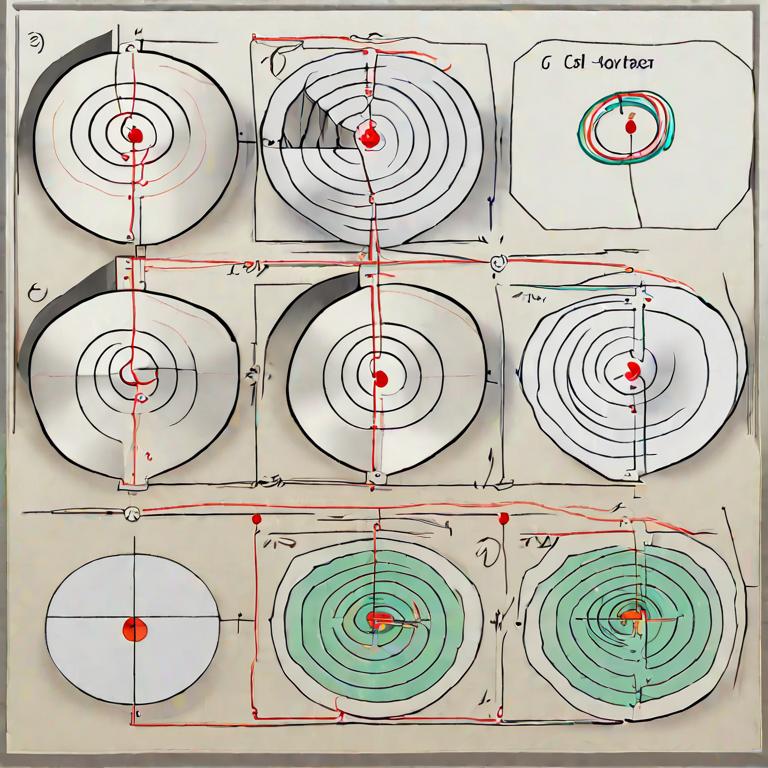}
        \caption{Runway}
    \end{subfigure}
    \begin{subfigure}[b]{0.16\textwidth}
        \centering
        \includegraphics[width=\textwidth]{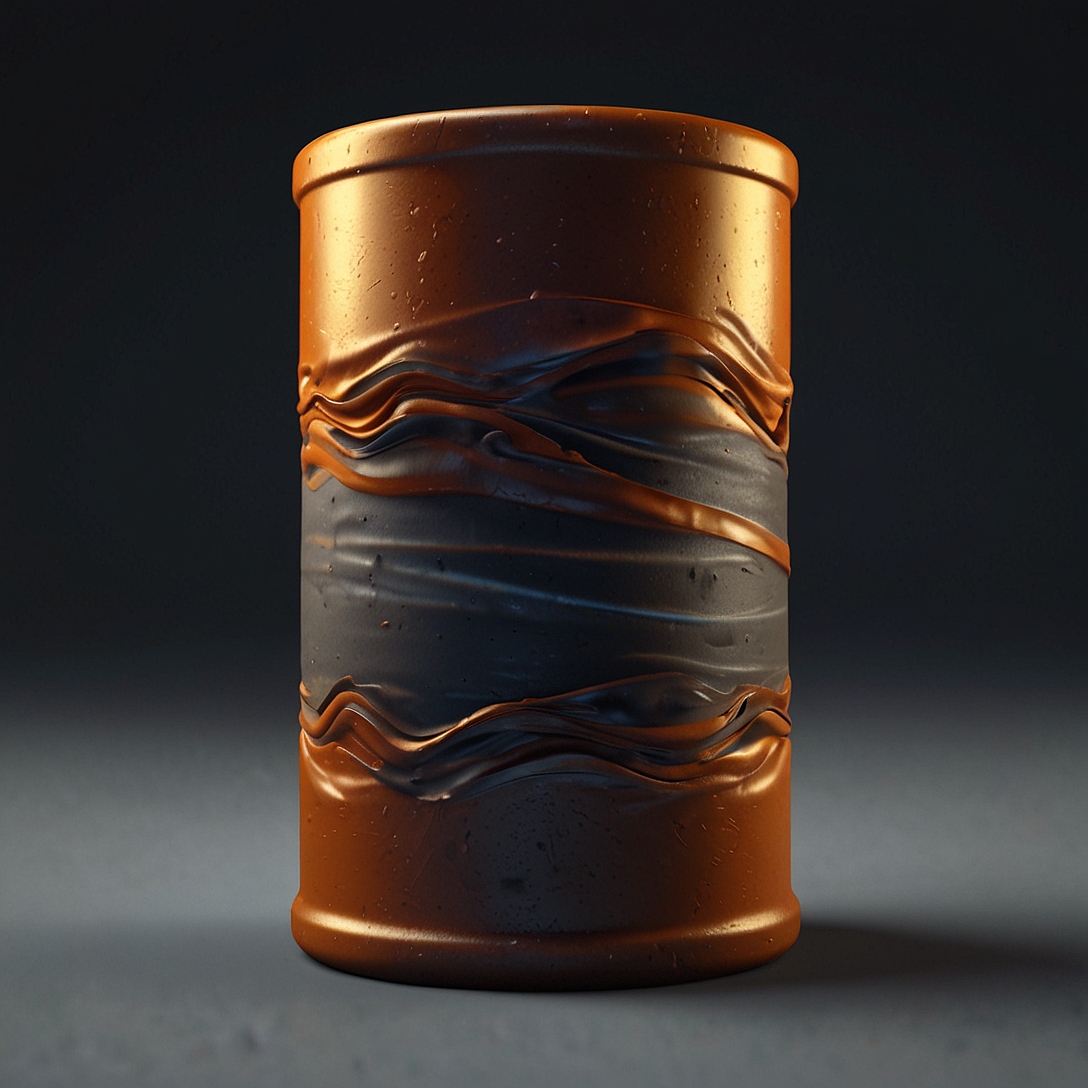}
        \caption{Leonardo Ai}
    \end{subfigure}

    \begin{subfigure}[b]{\textwidth}
        \centering
        \includegraphics[width=0.3\textwidth]{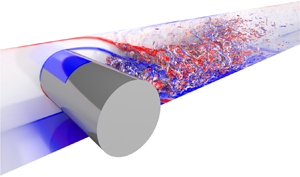}
        \caption{Flow past a three-dimensional circular cylinder at a Reynolds number of $Re = 3900$; this image, showing vorticity and produced by a numerical simulation, serves as the abstract photo in \cite{lu2023asymmetric}, published as open access under a Creative Commons license in the Journal of Fluid Mechanics, allowing readers to distribute the content freely.}
    \end{subfigure}

    \caption{A comparison between the AI-generated images and the numerical simulation result (i.e., ground truth) for the prompt ``Flow past a circular cylinder''}
    \label{Fig9}
\end{figure}

The first textual prompt we investigate in this subsection is ``Von Kármán vortex street''. Before analyzing the outputs from the generative models, we highlight a few aspects of this fluid dynamics problem, specifically by mentioning the Reynolds number ($Re$). An important dimensionless number in fluid mechanics is the Reynolds number (see e.g., \citep{rott1990note,johnson1999flow,purcell2014life,smits2011high,drela1989xfoil,kashefi2021PointNet,kashefi2020coarseRe}). Depending on the context of the problem, it can be written in different forms. For instance, for the problem of flow past a circular cylinder, which can be seen in part (g) of Fig. \ref{Fig7} and part (g) of Fig. \ref{Fig9}, it can be expressed as follows:

\begin{equation}
     Re = \frac{\rho d U_\infty}{\mu},
    \label{Eq5}
\end{equation}
where $\rho$ and $\mu$ are respectively the density and the dynamic viscosity of the fluid. The free stream velocity and the cylinder diameter are shown by $U_\infty$ and $d$, respectively. The Reynolds number ($Re$) is used to characterize other important fluid motion problems, such as flow over an airfoil (see part (g) of Fig. \ref{Fig10}), flow over an airplane wing (see part (g) of Fig. \ref{Fig6}), and Kelvin–Helmholtz instability (see part (g) of Fig. \ref{Fig8}), among others.

One of the dimensionless numbers relevant to the Von Kármán vortex street (see e.g., \citep{von2004aerodynamics}), which can be seen in part (g) of Fig. \ref{Fig7}, is the Strouhal number (see e.g., \citep{okajima1982strouhal}), which is formulated as follows:
\begin{equation}
St= \frac{f_s d}{u_\infty},
\label{Eq6}
\end{equation}
where $f_s$ represents the shedding frequency, $d$ is the cylinder diameter similar to what we define in Eq. \ref{Eq5}, and $u_\infty$ is the free stream velocity of the fluid.

Figure \ref{Fig7} depicts responses to the prompt ``Von Kármán vortex street''. In images generated by Midjourney and Leonardo Ai, a street is visible, indicating that these generative models are misled by the term ``street'' in the prompt. The concept of ``vortex'' is observed in images generated by Midjourney, Gemini Advanced, DALL·E 3, Meta AI, and Runway; however, none of these representations acceptably depict the vortex in a Von Kármán vortex street.

The image generated by Meta AI is interesting in that it depicts water circulation in a pipe within a laboratory environment. Although this image does not represent the Von Kármán vortex street, it is relevant to fluid mechanics. Interestingly, the image generated by DALL·E 3 could be considered relevant to the rotating disk problem or the Von Kármán swirling flow problem \citep{cochran1934flow,karman1921laminare}.

As a general observation, it seems that these generative models do not interpret ``Von Kármán vortex street'' as a cohesive expression. Instead, they treat it as a combination of the individual words ``vortex'' and ``street'', attempting to generate images that merge these two concepts.



\subsubsection{Flow past a circular cylinder}
\label{Sect272}

Figure \ref{Fig9} presents images generated in response to the prompt ``flow past a circular cylinder''. None of the generative models under investigation accurately captures the phenomenon. Among the generated images, the result from Midjourney is notable as it features fluid rotating in a hollow cylinder.

In the fluid mechanics community, the term ``flow'' typically implies fluid flow, such as water or air. Bearing this in mind, although some types of flow are visible in the images generated by DALL·E 3 and Meta AI, they do not depict actual fluid flow, perhaps due to the fact that the term ``fluid'' is not included in the prompt ``flow past a circular cylinder''. In all of these generated images, we can observe the presence of a cylinder as an object. Interestingly, each generative model depicts the cylinder from a unique perspective and angle.

The image generated by Gemini Advanced is interesting because, although it is not directly relevant to ``flow past a circular cylinder'', it depicts a schematic of flow inside an inclined cylinder. Additionally, this image includes some written letters. This type of figure can often be found in fluid mechanics textbooks (see, for example, Fig. P5.131 of \citep{munson2013fundamentals}). A similar observation applies to the image generated by Runway, which can be interpreted as showing the cross-section of a cylinder. The figure generated by Leonardo Ai is remarkable because it exhibits flow evolving on the outer surface of a cylinder, perhaps the closest representation to what we expected, yet it remains significantly distant from the actual phenomenon.


\begin{figure}[t]
    \centering
    \begin{subfigure}[b]{0.16\textwidth}
        \centering
        \includegraphics[width=\textwidth]{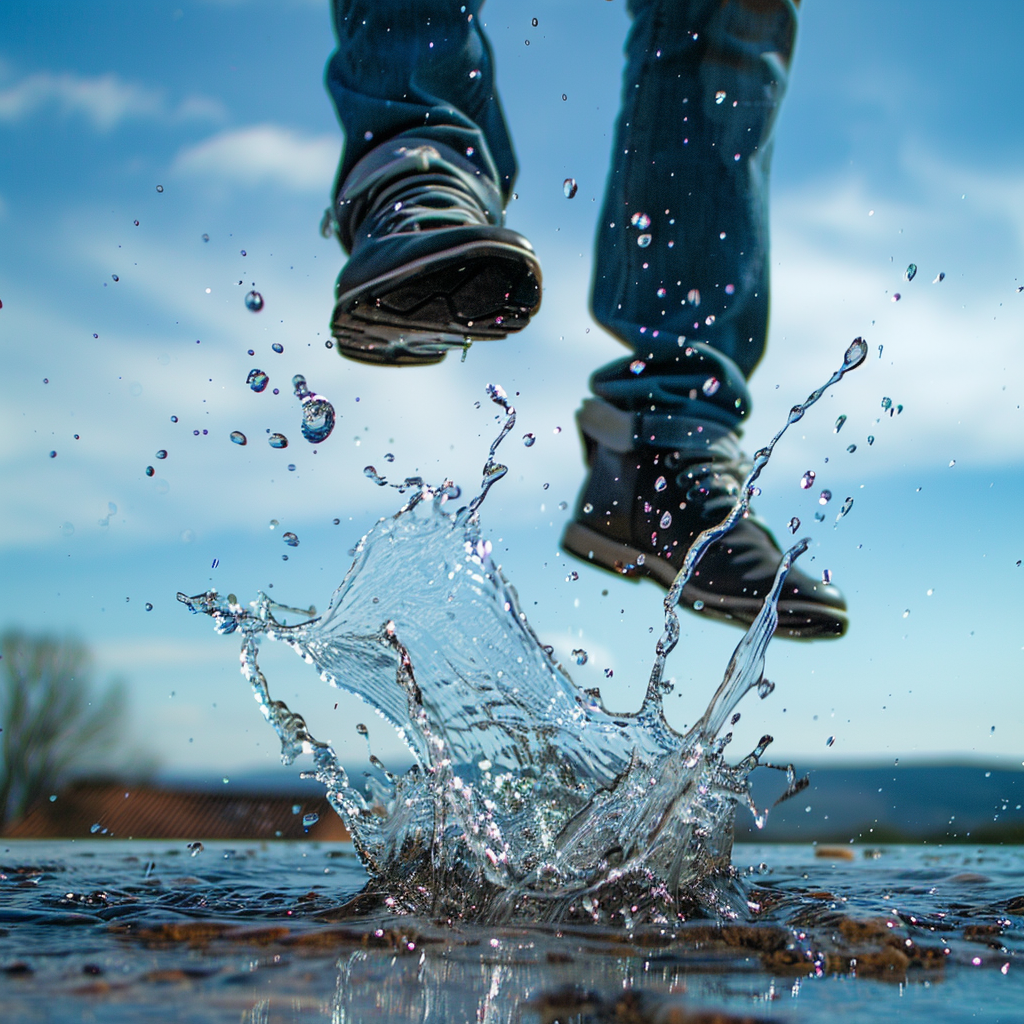}
        \caption{Midjourney}
    \end{subfigure}
    \begin{subfigure}[b]{0.16\textwidth}
        \centering
        \includegraphics[width=\textwidth]{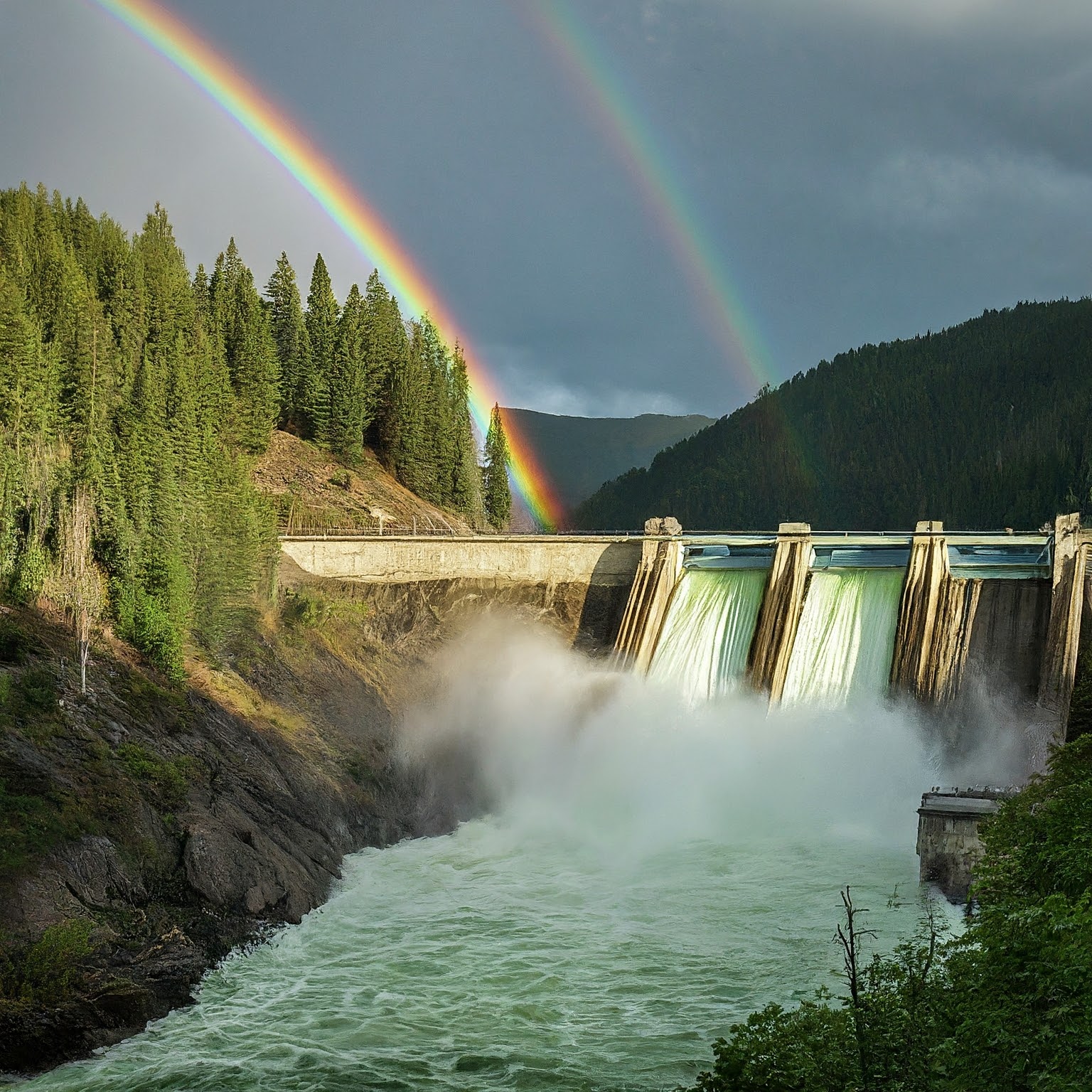}
        \caption{Gemini}
    \end{subfigure}
    \begin{subfigure}[b]{0.16\textwidth}
        \centering
        \includegraphics[width=\textwidth]{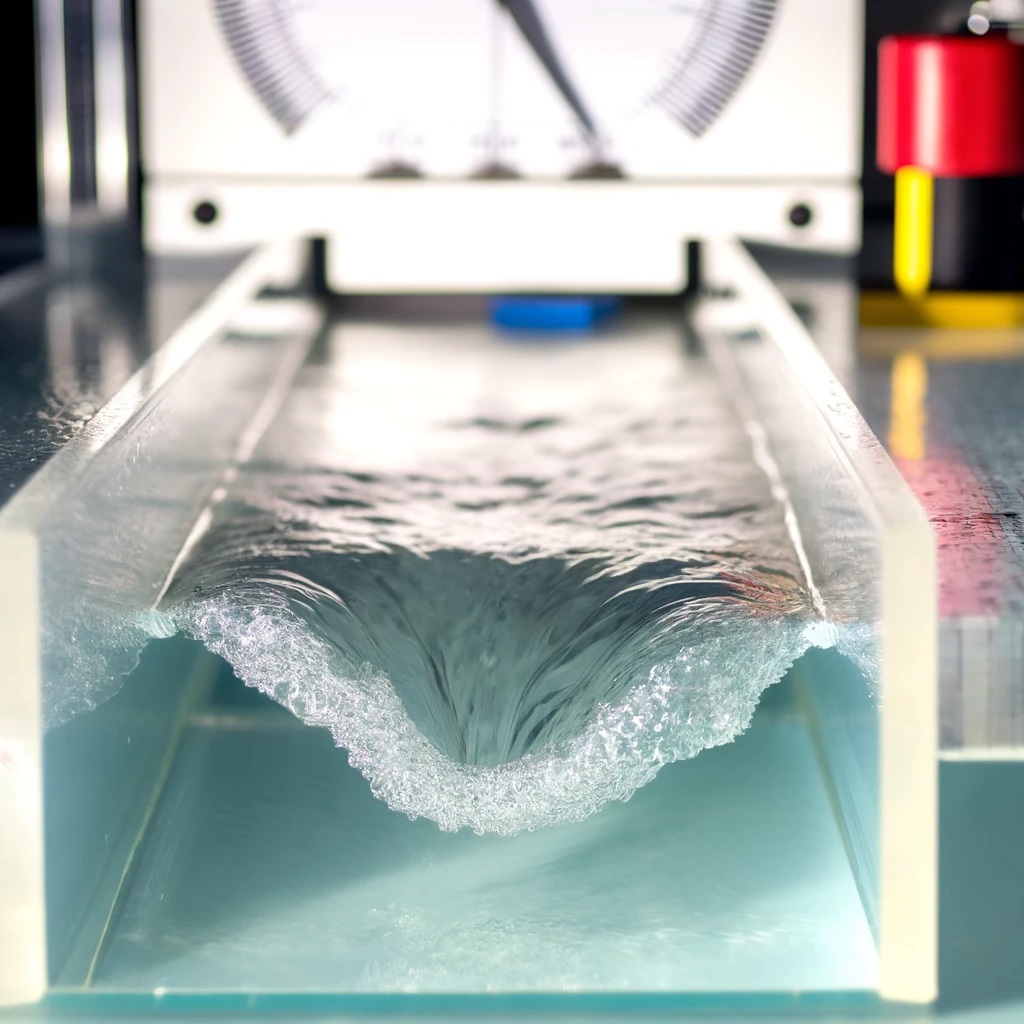}
        \caption{DALL·E 3}
    \end{subfigure}
    \begin{subfigure}[b]{0.16\textwidth}
        \centering
        \includegraphics[width=\textwidth]{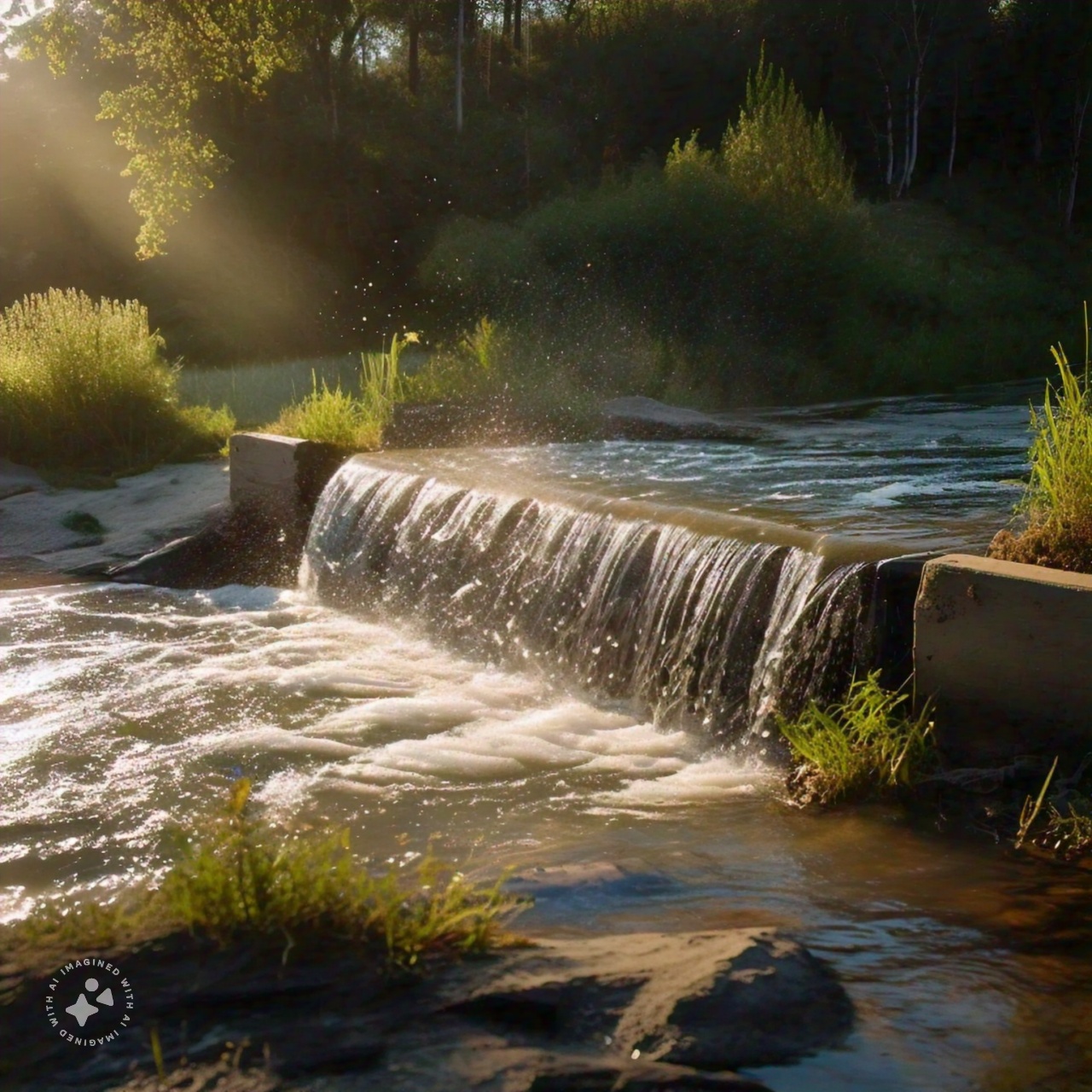}
        \caption{Meta AI}
    \end{subfigure}
    \begin{subfigure}[b]{0.16\textwidth}
        \centering
        \includegraphics[width=\textwidth]{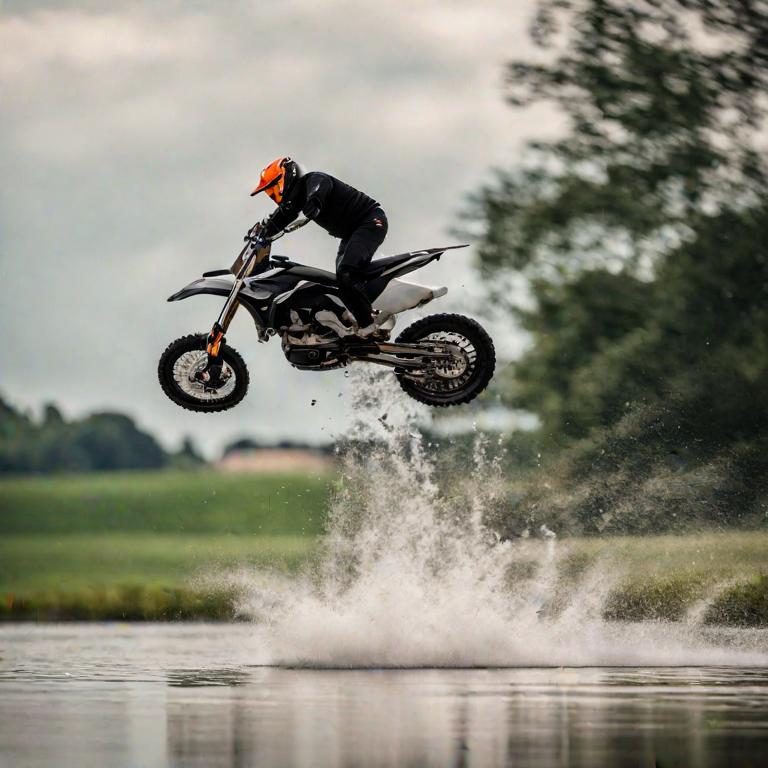}
        \caption{Runway}
    \end{subfigure}
    \begin{subfigure}[b]{0.16\textwidth}
        \centering
        \includegraphics[width=\textwidth]{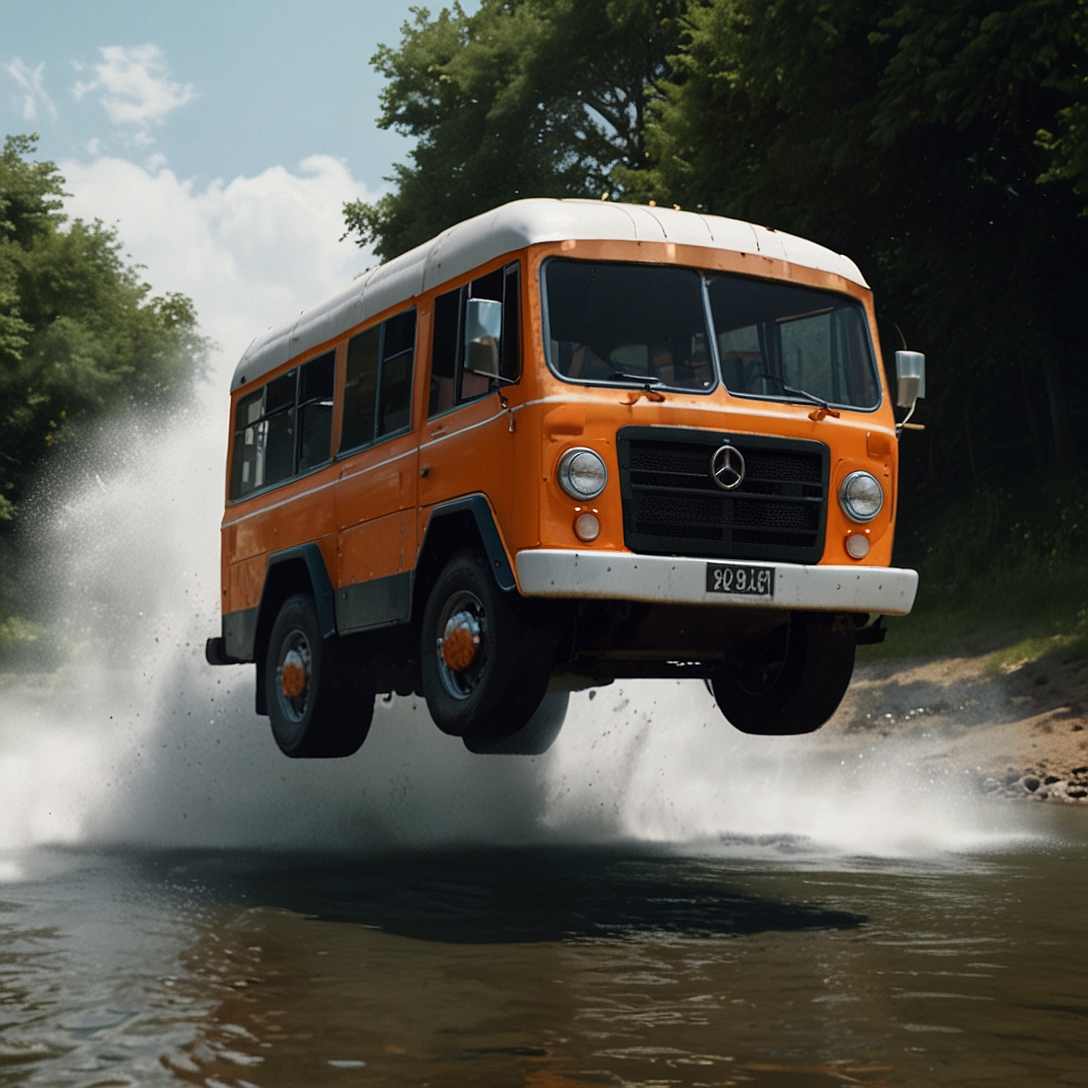}
        \caption{Leonardo Ai}
    \end{subfigure}

    \begin{subfigure}[b]{\textwidth}
        \centering
        \includegraphics[width=0.4\textwidth]{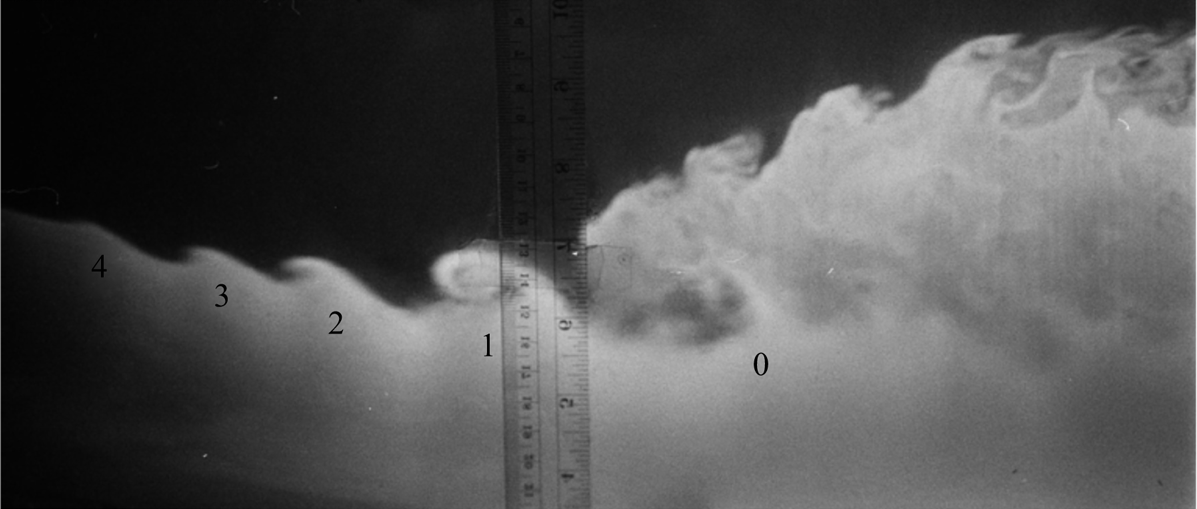}
        \caption{Hydraulic jump in a lab experimental setup; a series of Kelvin–Helmholtz billows is also observable. This image is taken from part (b) of Fig. 9 in \cite{lawrence2022stationary}, published as open access under a Creative Commons license in the Journal of Fluid Mechanics, allowing readers to distribute the content freely.}
    \end{subfigure}

    \caption{A comparison between the AI-generated images and the lap experimental result (i.e., ground truth) for the prompt ``Hydraulic jump''}
    \label{Fig11}
\end{figure}


\subsubsection{Hydraulic jump}
\label{Sect273}

The next prompt considered in this technical report is ``hydraulic jump'', with the corresponding images displayed in Fig. \ref{Fig11}. The occurrence of the hydraulic jump, as observed in part (g) of Fig. \ref{Fig11} and part (g) of Fig. \ref{Fig3}, can be identified using the dimensionless Froude number (see e.g., \citep{hornung1995flow}), defined as:

\begin{equation}
Fr= \frac{U_u^2}{gh_u},
\label{Eq7}
\end{equation}
where $U_u$ and $h_u$ respectively represent the upstream velocity and height of the fluid. For $Fr>1$, a hydraulic jump is expected.

As can be seen from Fig. \ref{Fig11}, the images generated by Midjourney, Runway, and Leonardo Ai are inaccurate. Similar to some results discussed for previous prompts, these generative models interpret ``hydraulic jump'' as two separate words, ``hydraulic'' and ``jump''. Consequently, they depict unrelated objects such as motorcycles, minibus, or even humans jumping over a river.

On the other hand, the results by Gemini Advanced and Meta AI could be considered moderately relevant. These two generative models produce images that are partially accurate, as they depict water exiting a dam and a step in a river, two scenarios in nature where a hydraulic jump can occur, based on our knowledge. The image generated by DALL·E 3, however, does not seem informative and is difficult to judge.


\begin{figure}[t]
    \centering
    \begin{subfigure}[b]{0.16\textwidth}
        \centering
        \includegraphics[width=\textwidth]{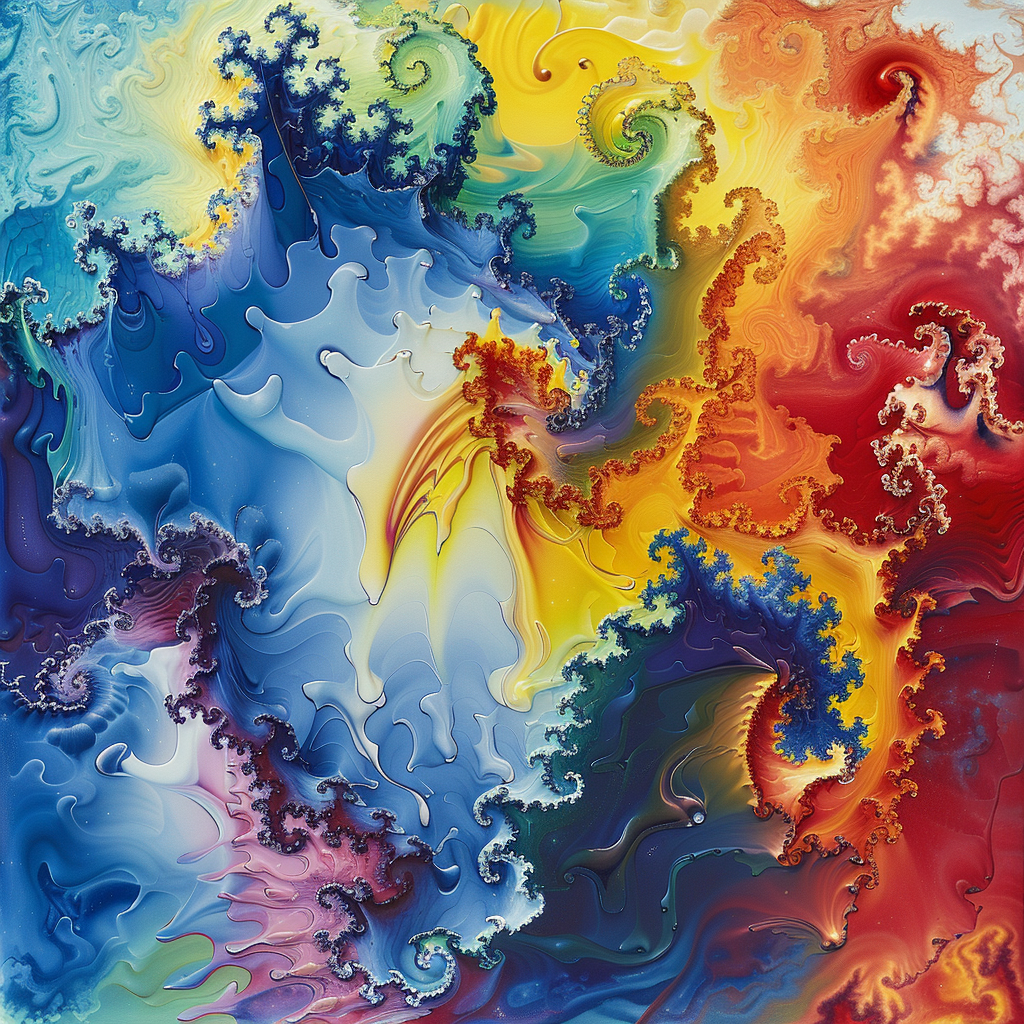}
        \caption{Midjourney}
    \end{subfigure}
    \begin{subfigure}[b]{0.16\textwidth}
        \centering
        \includegraphics[width=\textwidth]{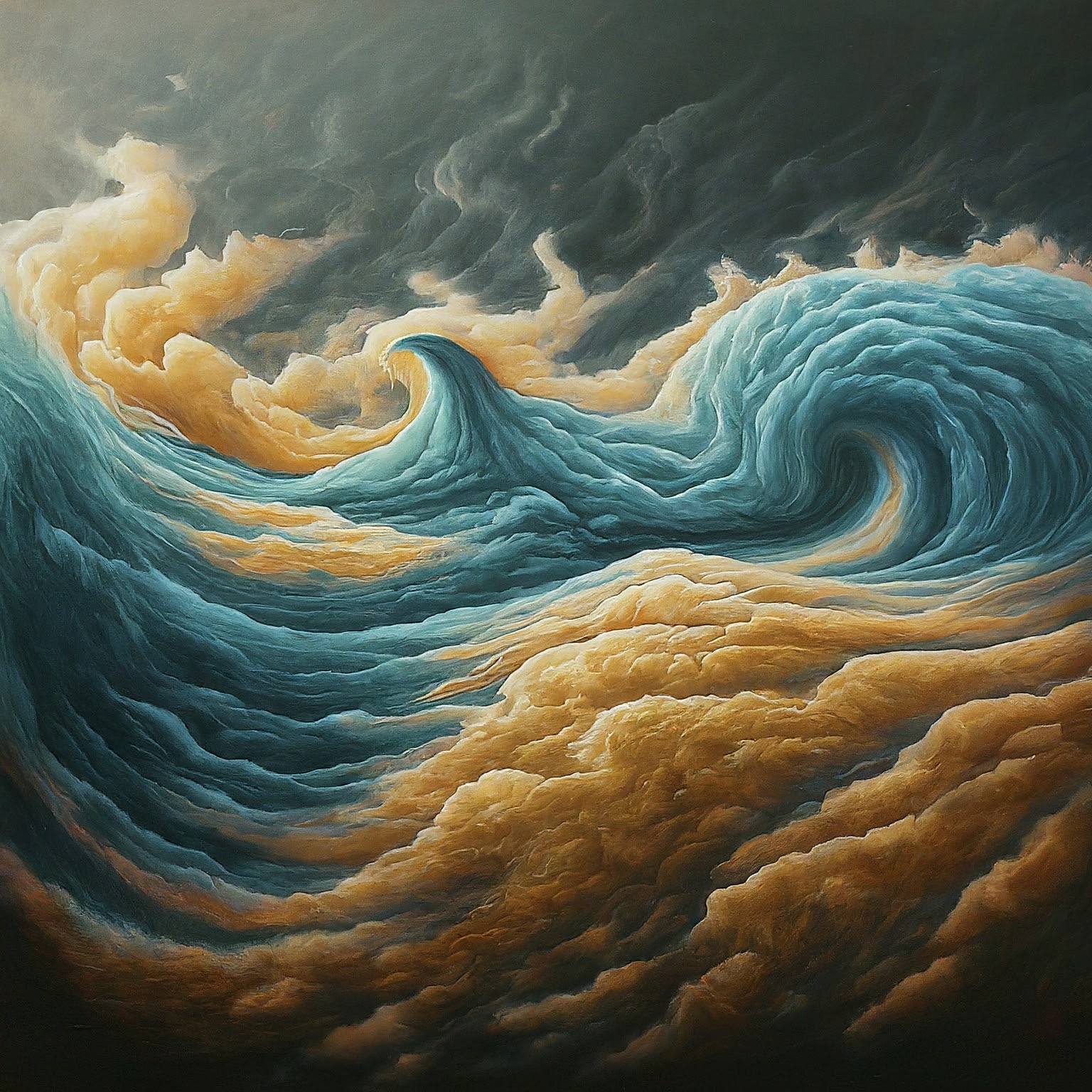}
        \caption{Gemini}
    \end{subfigure}
    \begin{subfigure}[b]{0.16\textwidth}
        \centering
        \includegraphics[width=\textwidth]{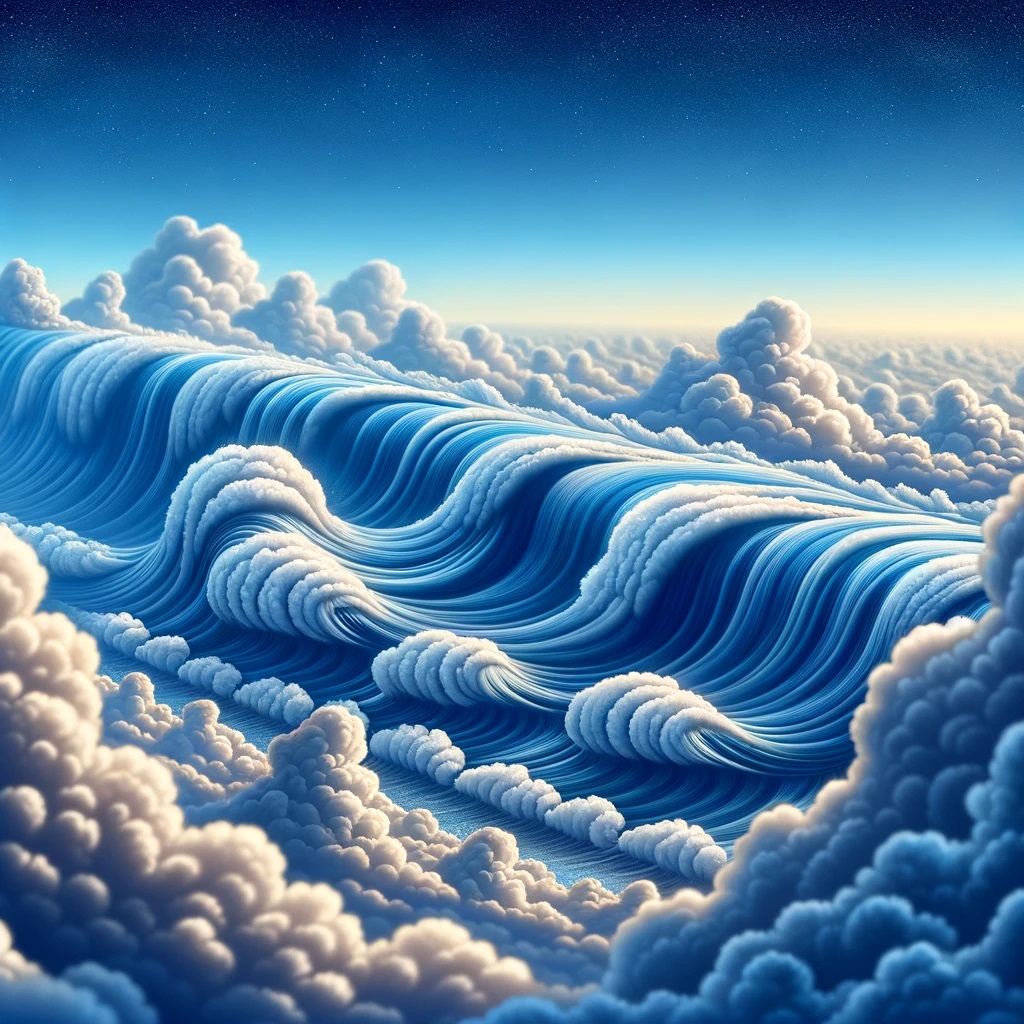}
        \caption{DALL·E 3}
    \end{subfigure}
    \begin{subfigure}[b]{0.16\textwidth}
        \centering
        \includegraphics[width=\textwidth]{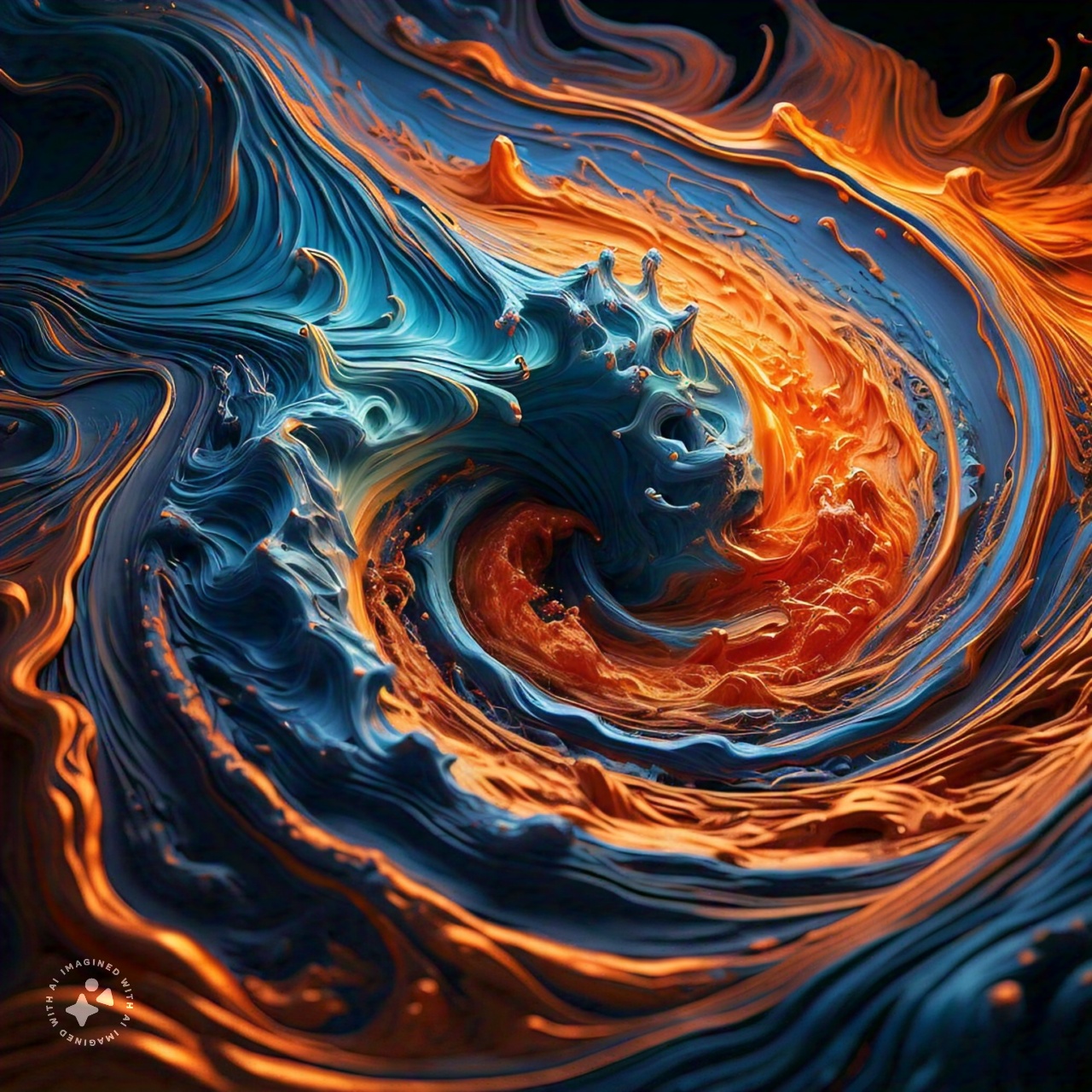}
        \caption{Meta AI}
    \end{subfigure}
    \begin{subfigure}[b]{0.16\textwidth}
        \centering
        \includegraphics[width=\textwidth]{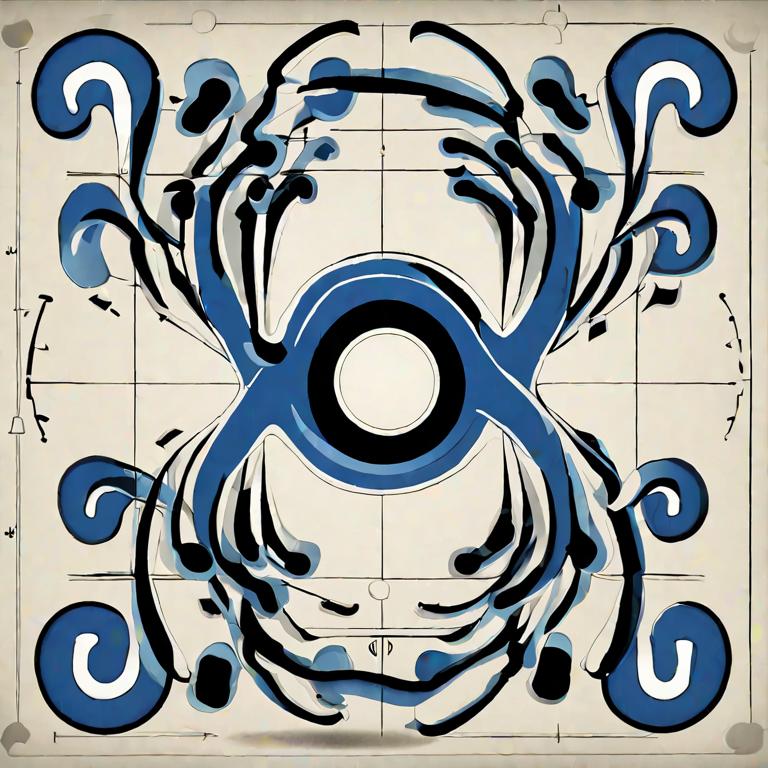}
        \caption{Runway}
    \end{subfigure}
    \begin{subfigure}[b]{0.16\textwidth}
        \centering
        \includegraphics[width=\textwidth]{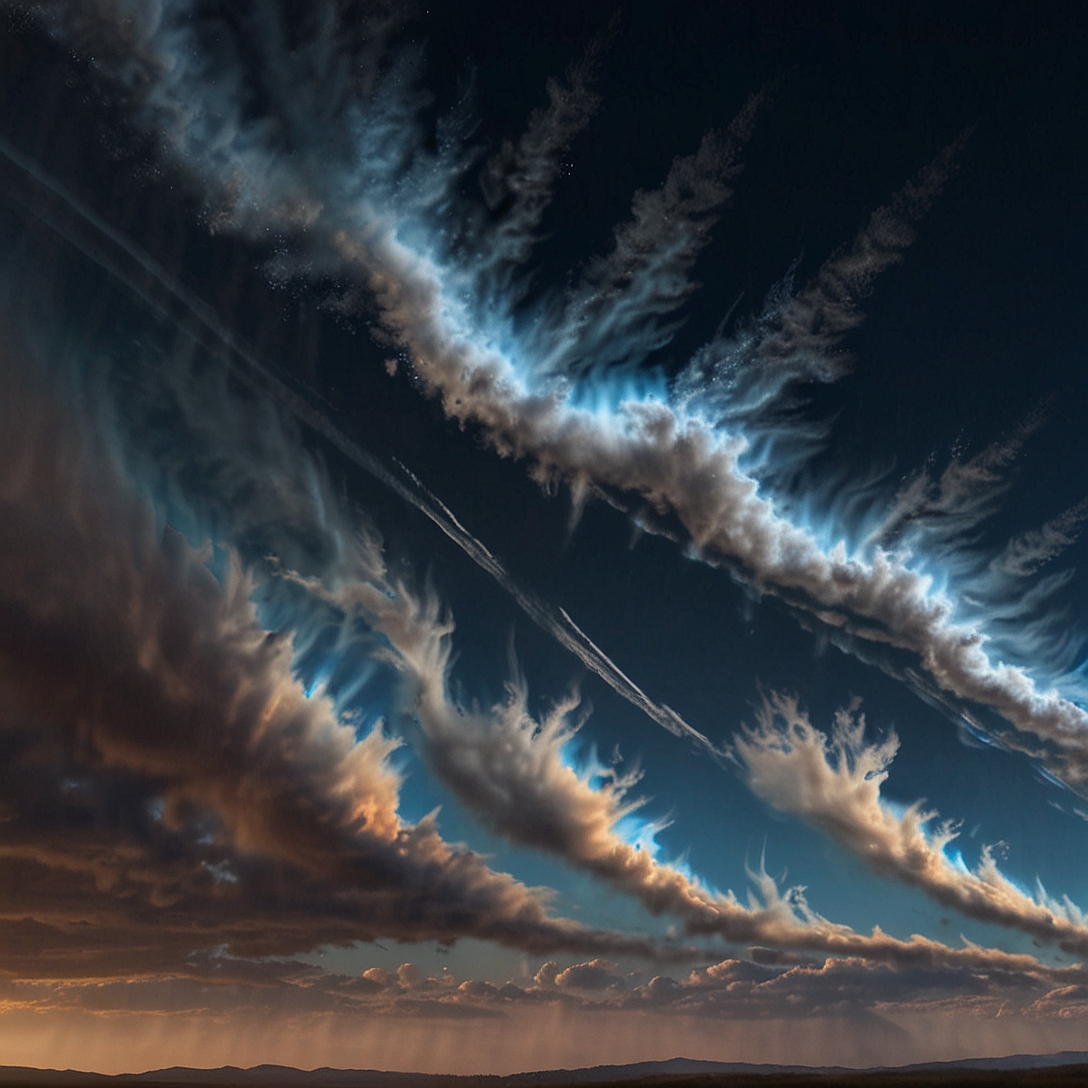}
        \caption{Leonardo Ai}
    \end{subfigure}

    \begin{subfigure}[b]{\textwidth}
        \centering
        \includegraphics[width=0.3\textwidth]{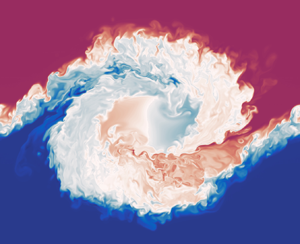}
        \caption{Kelvin–Helmholtz instability; this image, showing a density contour and produced by a numerical simulation, serves as the abstract photo in \cite{lewin2022stratified}, published as open access under a Creative Commons license in the Journal of Fluid Mechanics, allowing readers to freely distribute the content.}
    \end{subfigure}

    \caption{A comparison between the AI-generated images and the numerical simulation result (i.e., ground truth) for the prompt ``Kelvin–Helmholtz instability''}
    \label{Fig8}
\end{figure}


\subsubsection{Kelvin-Helmholtz instability}
\label{Sect274}

Figure \ref{Fig8} displays the collective images generated in response to the prompt ``Kelvin-Helmholtz instability''. Surprisingly, all of the images are relevant to this phenomenon. Images generated by Gemini Advanced, DALL·E 3, and Leonardo Ai attempt to depict the Kelvin-Helmholtz instability in clouds in the sky, which is a common manifestation of this instability in nature (see e.g., \citep{baumgarten2014quantifying}). Images produced by Midjourney, Meta AI, and Runway depict the mixing of different liquids with varying colors, a phenomenon characteristic of Kelvin-Helmholtz instability, as shown in part (g) of Fig. \ref{Fig8}, which serves as a ground truth example. It is worth mentioning that the images generated by Runway differ slightly from the others in terms of context.


\begin{figure}[t]
    \centering
  
    \begin{subfigure}[b]{0.16\textwidth}
        \centering
        \includegraphics[width=\textwidth]{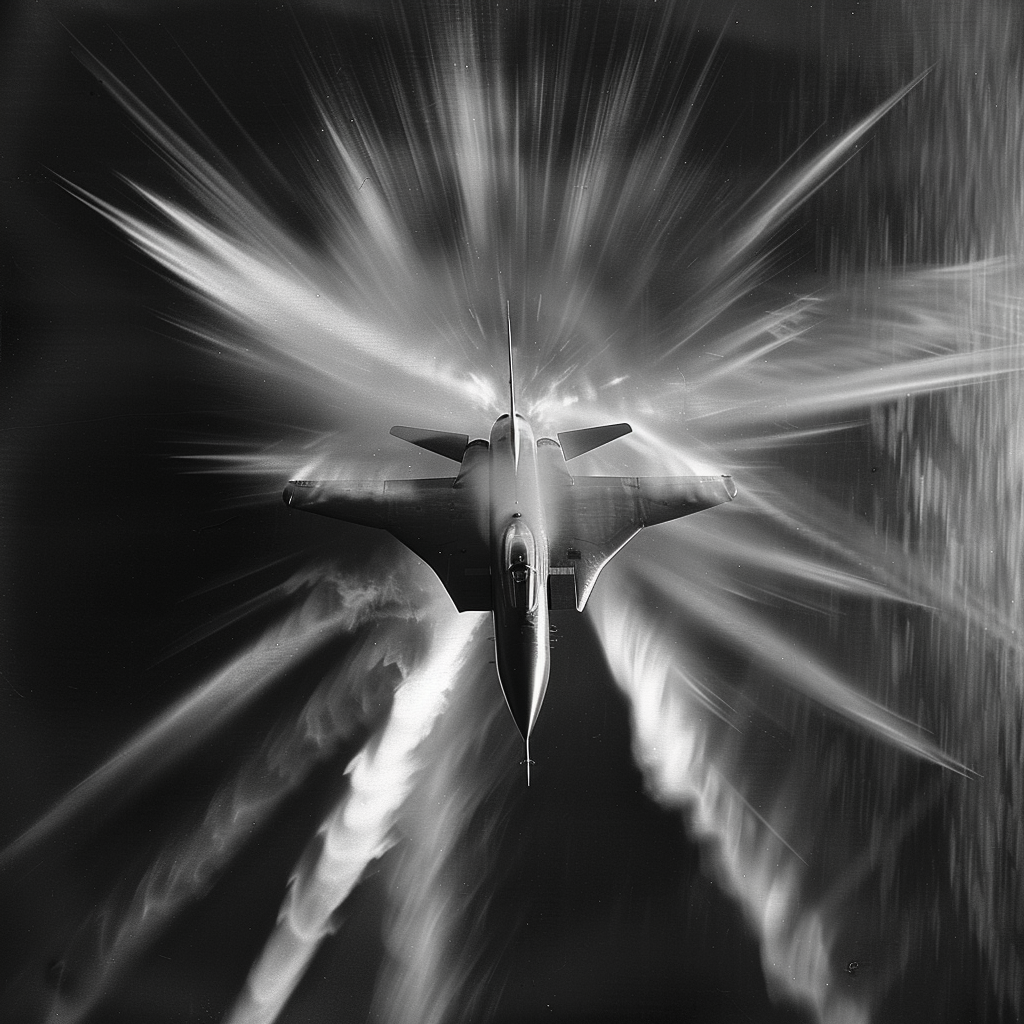}
        \caption{Midjourney}
    \end{subfigure}
    \begin{subfigure}[b]{0.16\textwidth}
        \centering
        \includegraphics[width=\textwidth]{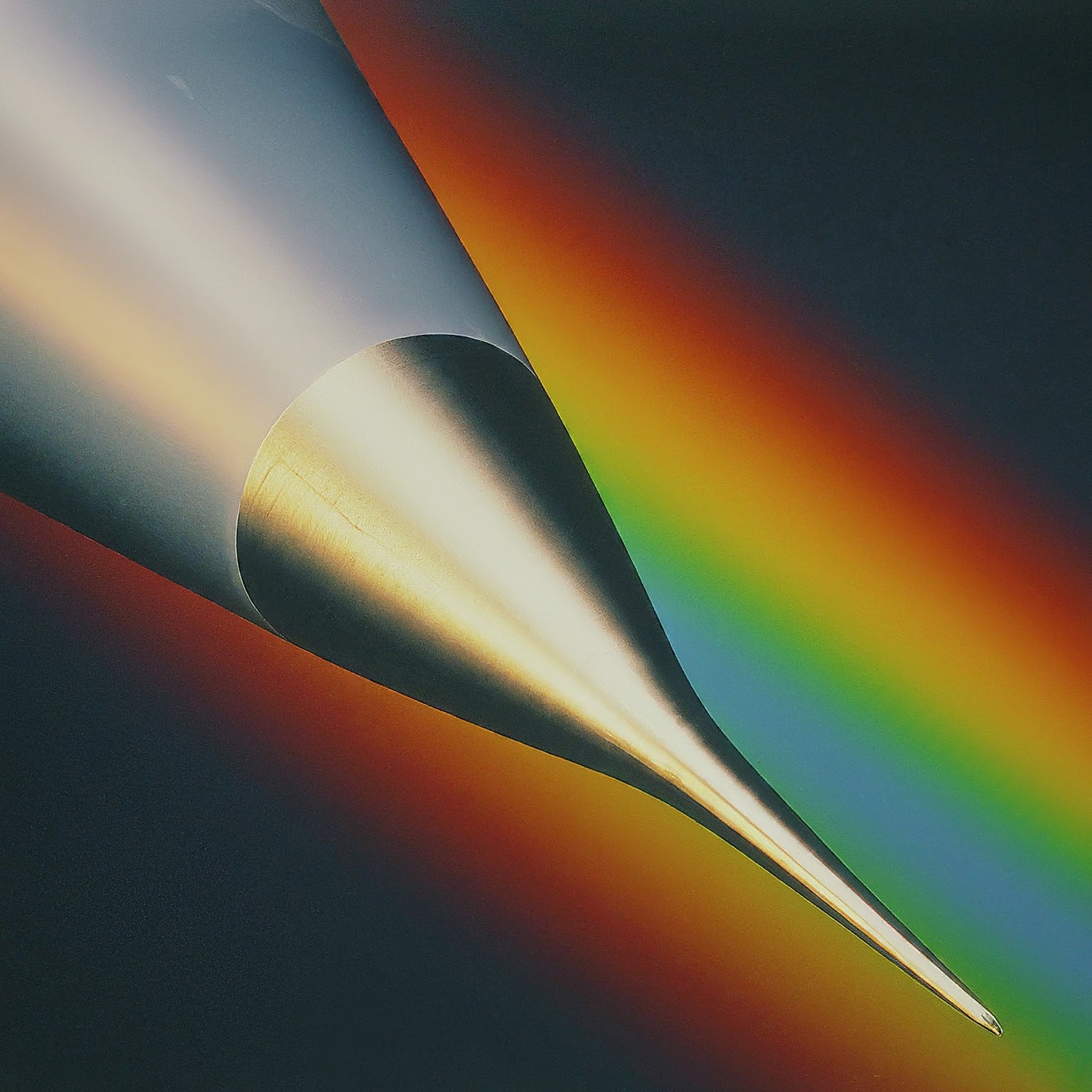}
        \caption{Gemini}
    \end{subfigure}
    \begin{subfigure}[b]{0.16\textwidth}
        \centering
        \includegraphics[width=\textwidth]{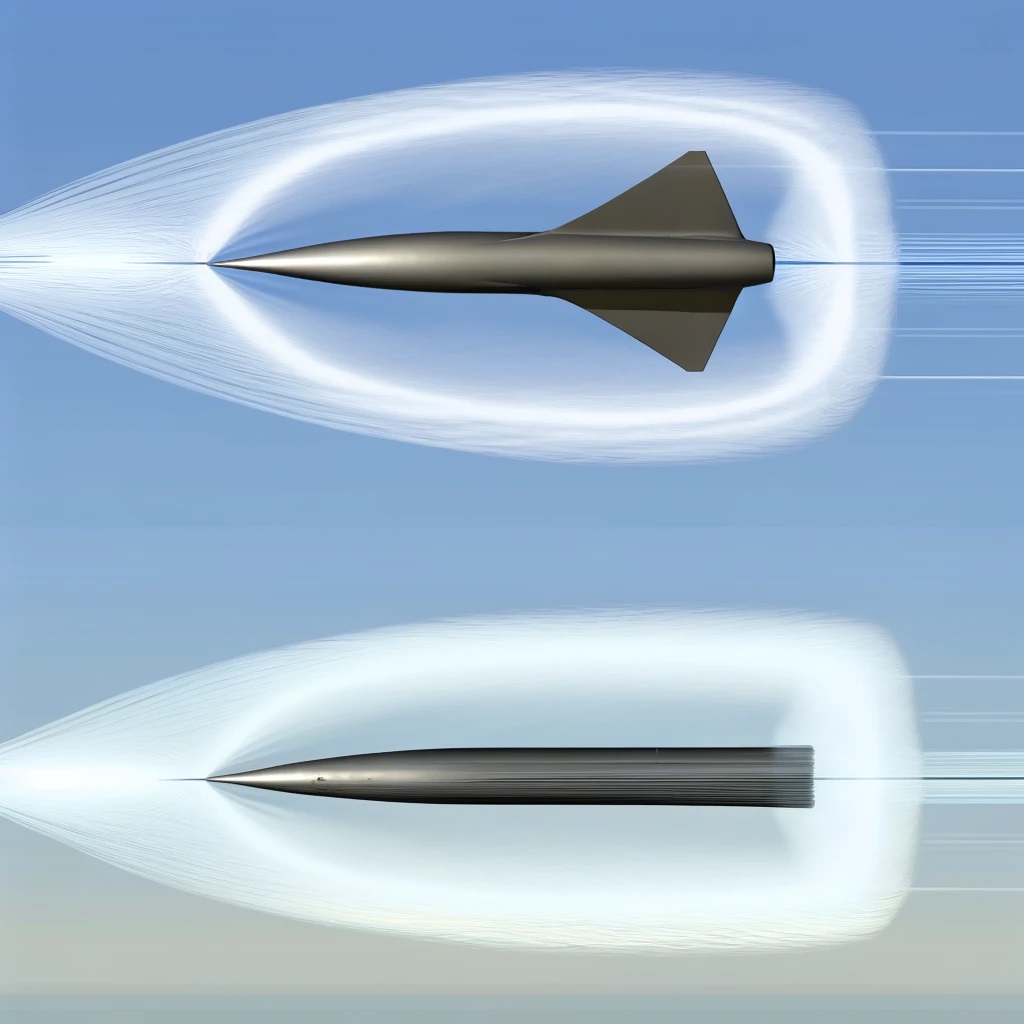}
        \caption{DALL·E 3}
    \end{subfigure}
    \begin{subfigure}[b]{0.16\textwidth}
        \centering
        \includegraphics[width=\textwidth]{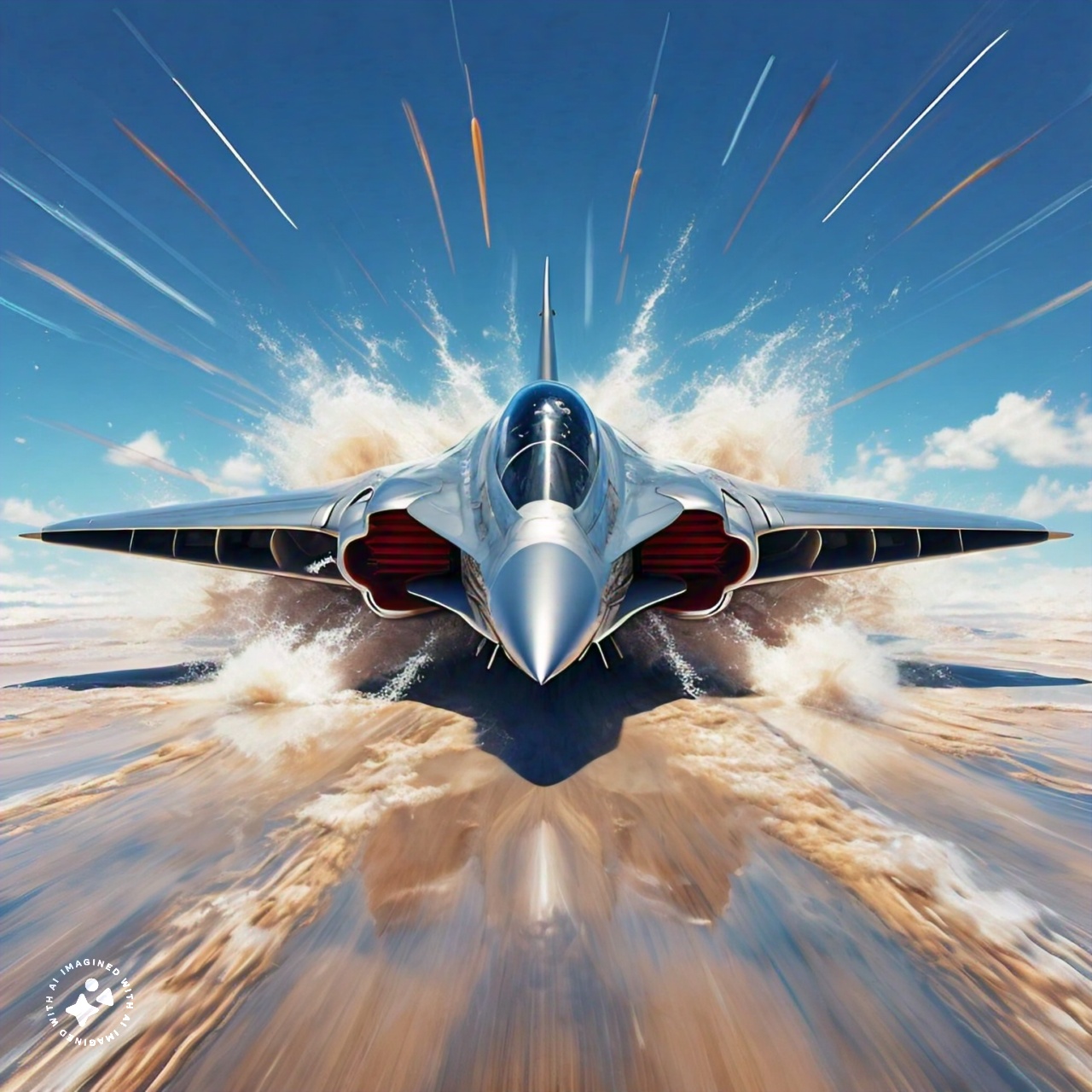}
        \caption{Meta AI}
    \end{subfigure}
    \begin{subfigure}[b]{0.16\textwidth}
        \centering
        \includegraphics[width=\textwidth]{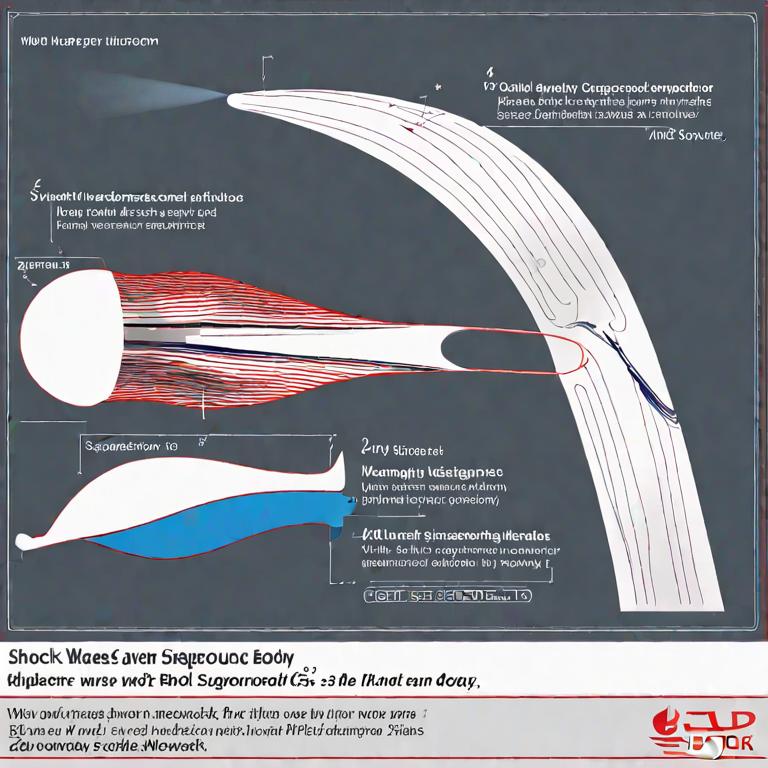}
        \caption{Runway}
    \end{subfigure}
    \begin{subfigure}[b]{0.16\textwidth}
        \centering
        \includegraphics[width=\textwidth]{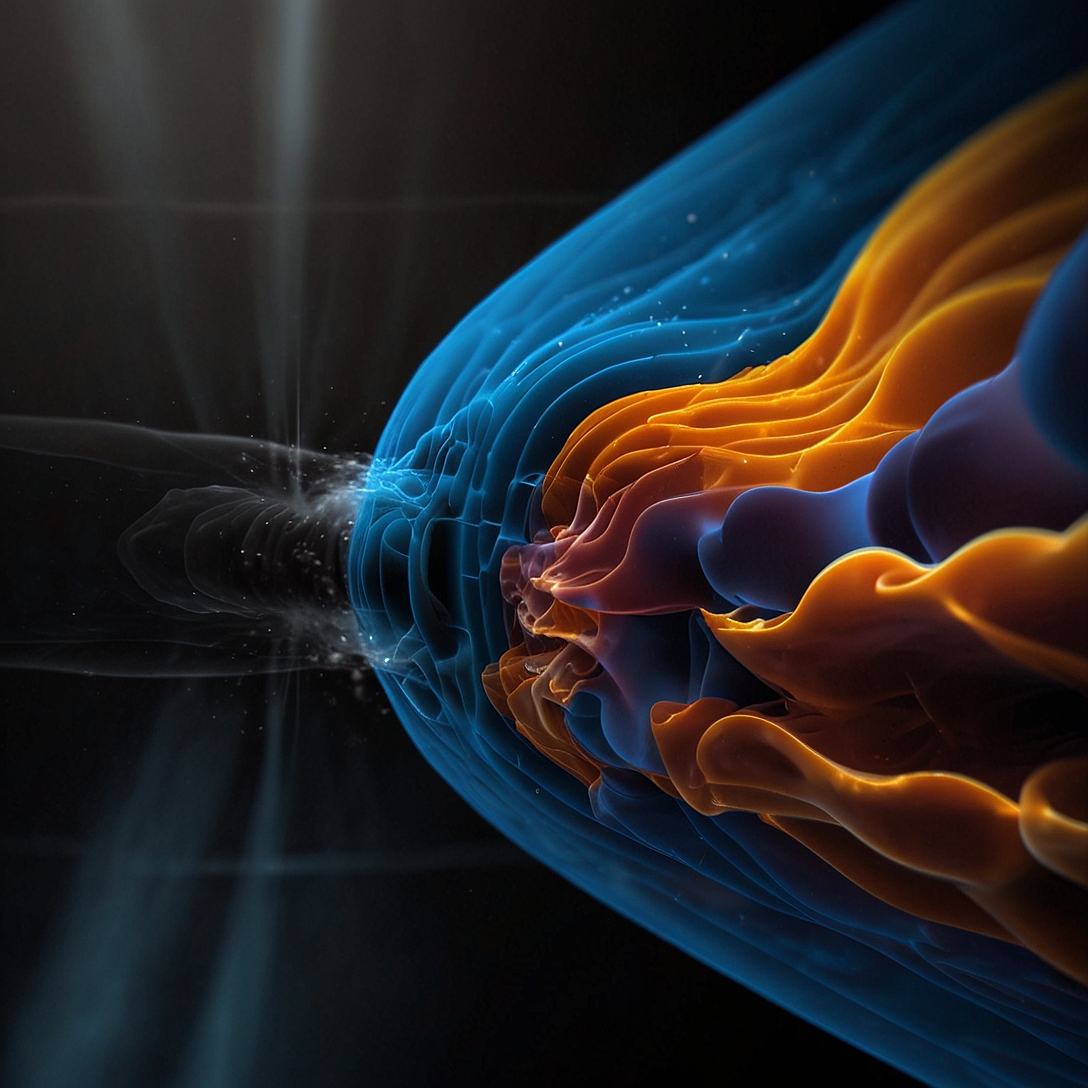}
        \caption{Leonardo Ai}
    \end{subfigure}

    \begin{subfigure}[b]{\textwidth}
        \centering
        \includegraphics[width=0.4\textwidth]{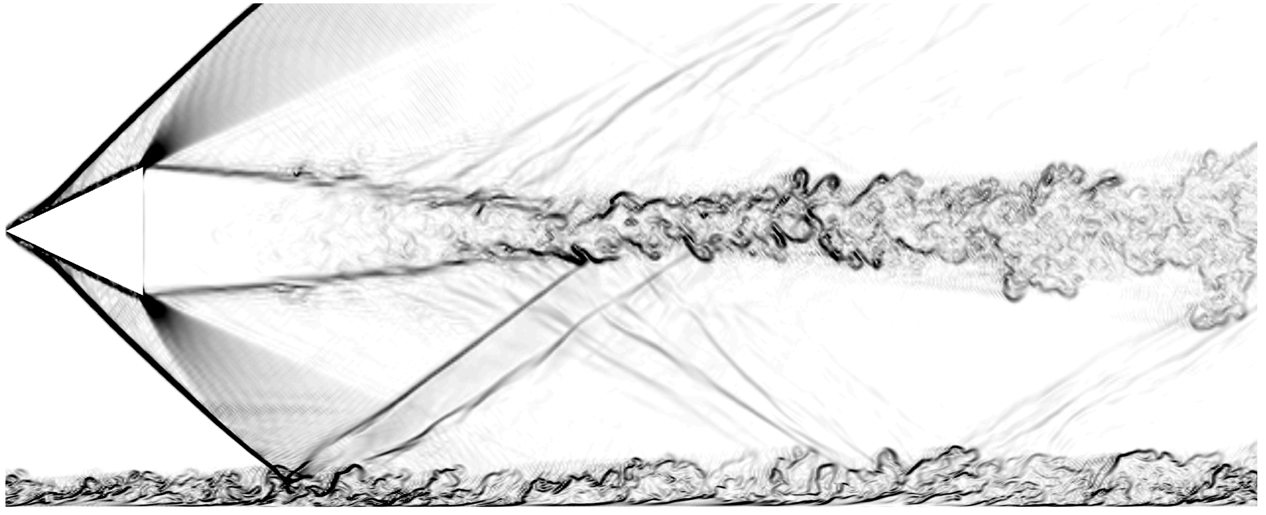}
        \caption{Shock wave around a cone at a Mach number of $M_{\infty} = 2.05$; This image, produced by a numerical simulation, is taken from the top part of Fig. 4 in \cite{zuo2023wall}, published as open access under a Creative Commons license in the Journal of Fluid Mechanics, allowing readers to distribute the content freely.}
    \end{subfigure}

    \caption{A comparison between the AI-generated images and the numerical simulation result (i.e., ground truth) for the prompt ``shock waves on a sharp-nosed supersonic body''}
    \label{Fig5}
\end{figure}


\subsubsection{Shock waves on a sharp-nosed supersonic body}
\label{Sect275}

Figure \ref{Fig5} exhibits a collection of images generated in response to the prompt ``shock waves on a sharp-nosed supersonic body''. A shock can be categorized using the dimensionless number known as the Mach number (see e.g., \citep{davis1966laminar,howe1979attenuation}), formulated as:
\begin{equation}
M_\infty= \frac{V_\infty}{c},
\label{Eq9}
\end{equation}
where $V_\infty$ represents the free stream velocity and $c$ is the sound speed. A Mach number of $M_\infty = 2.05$ indicates a supersonic flow accompanied by shock waves, as illustrated in part (g) of Fig. \ref{Fig5}.

Based on observations from Fig. \ref{Fig5}, the images produced by Gemini Advanced and DALL·E 3 can be considered the most relevant, especially when compared to a ground truth figure shown in part (g) of Fig. \ref{Fig5}. The picture produced by Leonardo Ai is also interesting as it depicts shock boundaries, although it lacks a sharp-nosed body. The result by Runway is difficult to judge. It illustrates an educational poster with text printed below that reads, ``Shock Waes aver Srpeouoc Edoy'', closely mirroring the prompt ``shock waves on a sharp-nosed supersonic body''. Midjourney and Meta AI generate images that feature warplanes as sharp-nosed supersonic bodies, representing shock waves in an artistic rather than a scientific manner.


\begin{figure}[t]
    \centering
    
    \begin{subfigure}[b]{0.16\textwidth}
        \centering
        \includegraphics[width=\textwidth]{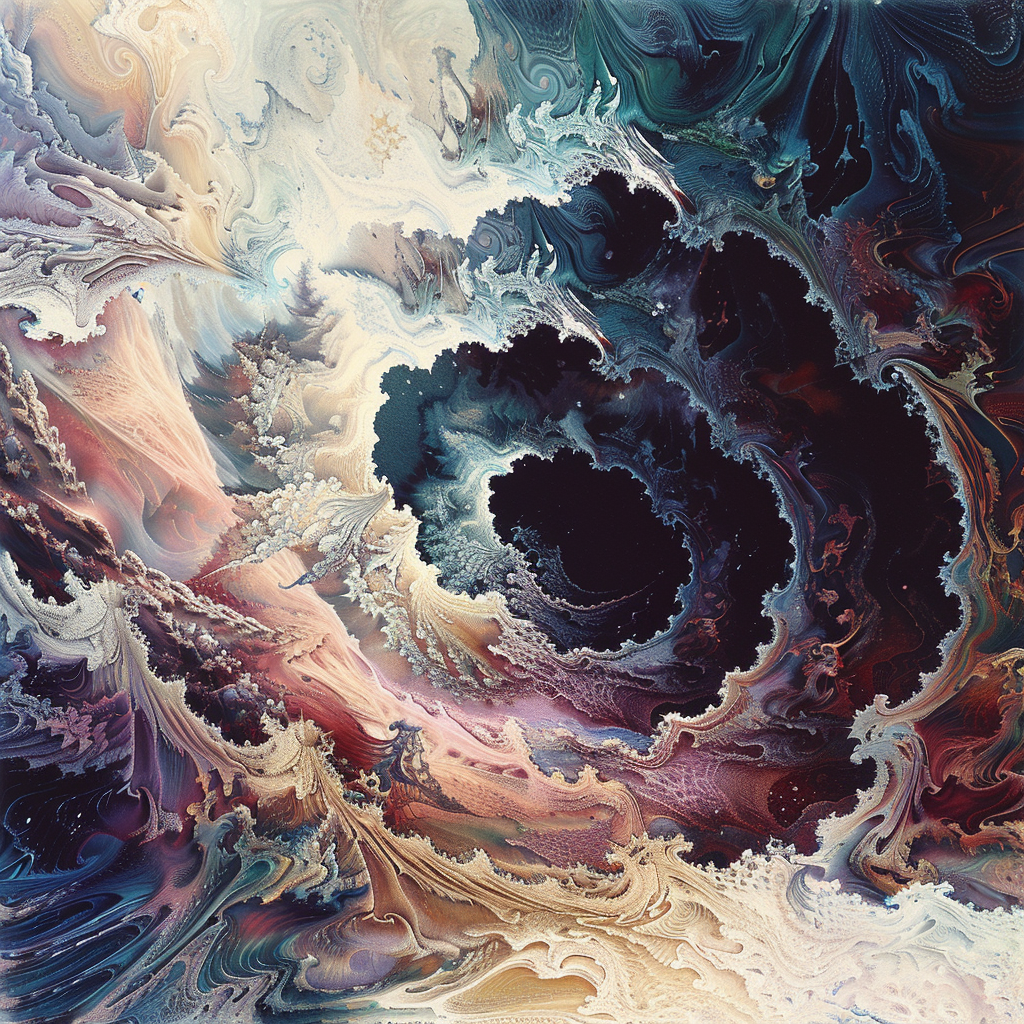}
        \caption{Midjourney}
    \end{subfigure}
    \begin{subfigure}[b]{0.16\textwidth}
        \centering
        \includegraphics[width=\textwidth]{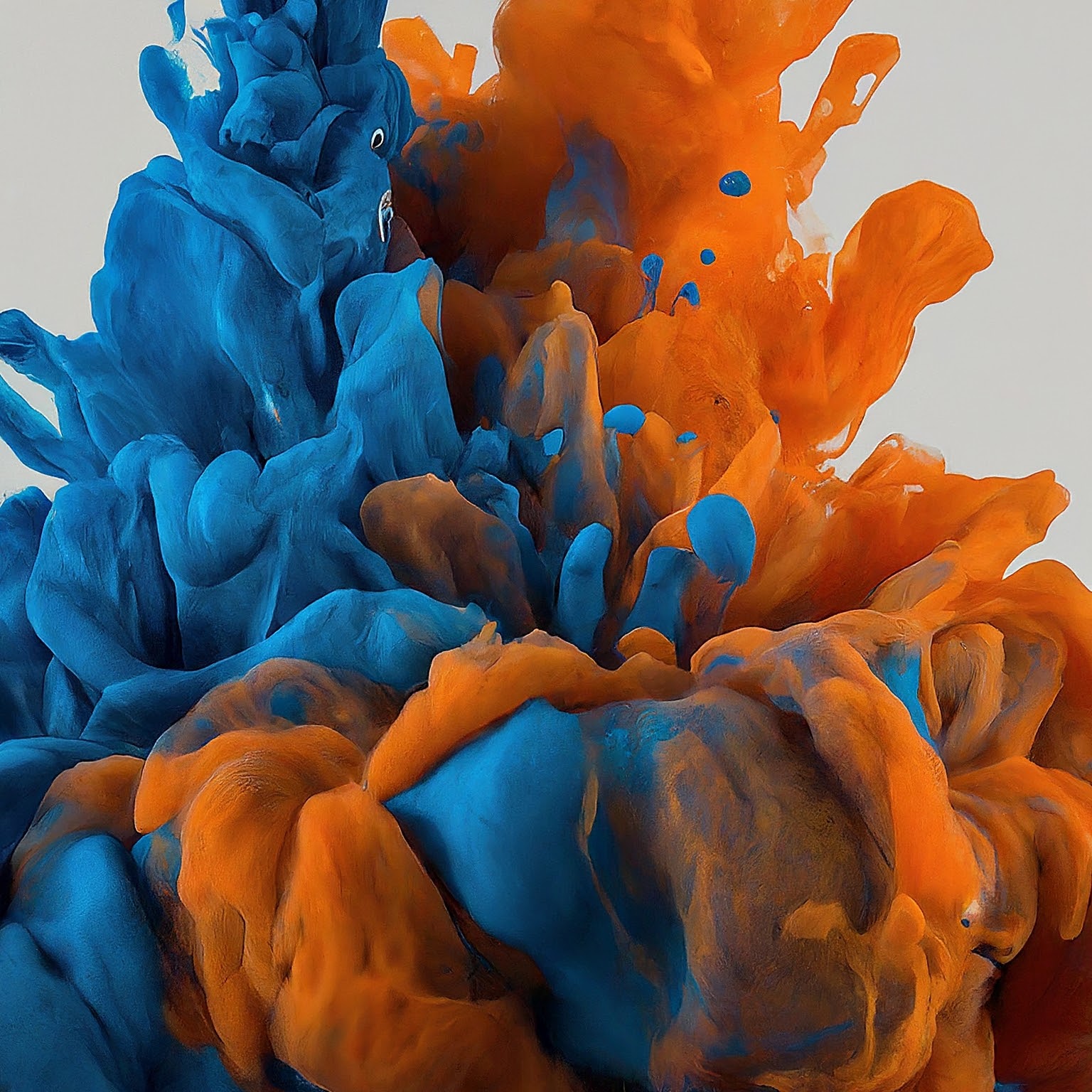}
        \caption{Gemini}
    \end{subfigure}
    \begin{subfigure}[b]{0.16\textwidth}
        \centering
        \includegraphics[width=\textwidth]{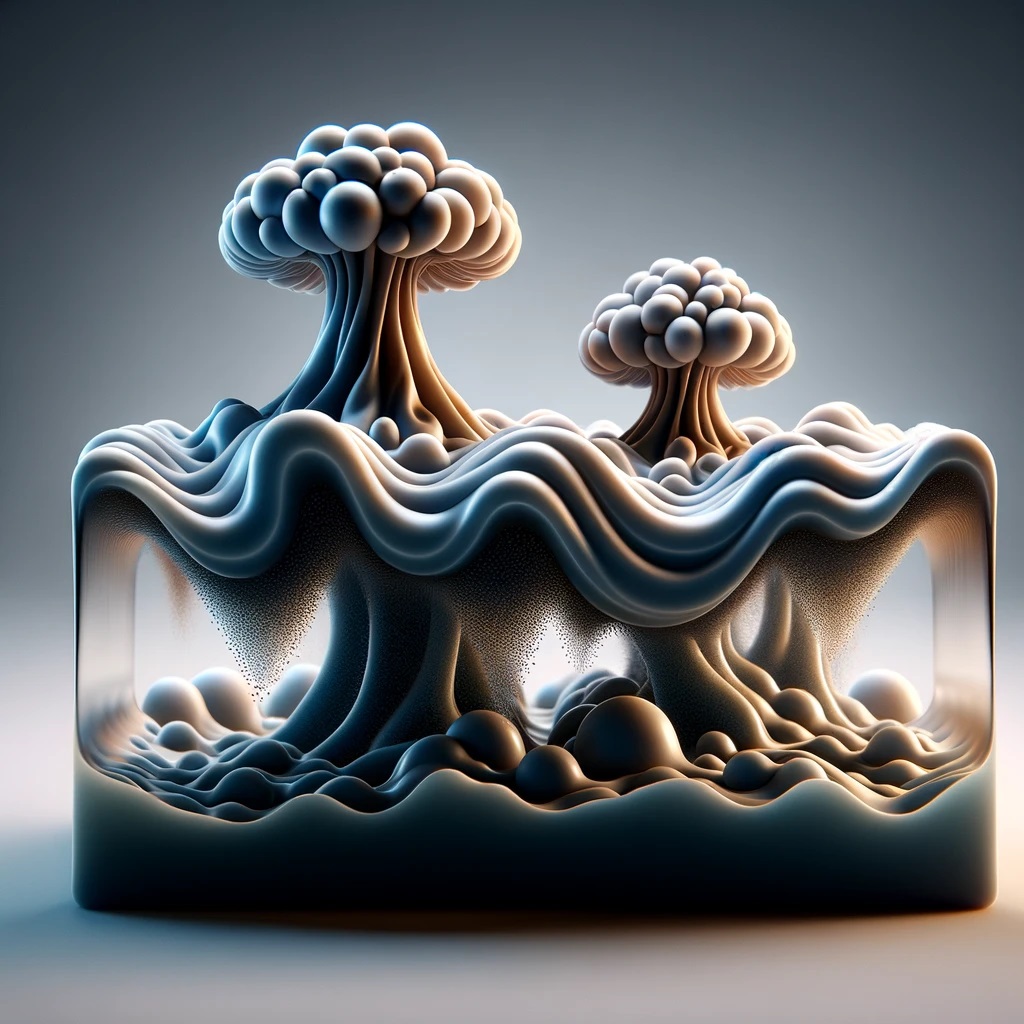}
        \caption{DALL·E 3}
    \end{subfigure}
    \begin{subfigure}[b]{0.16\textwidth}
        \centering
        \includegraphics[width=\textwidth]{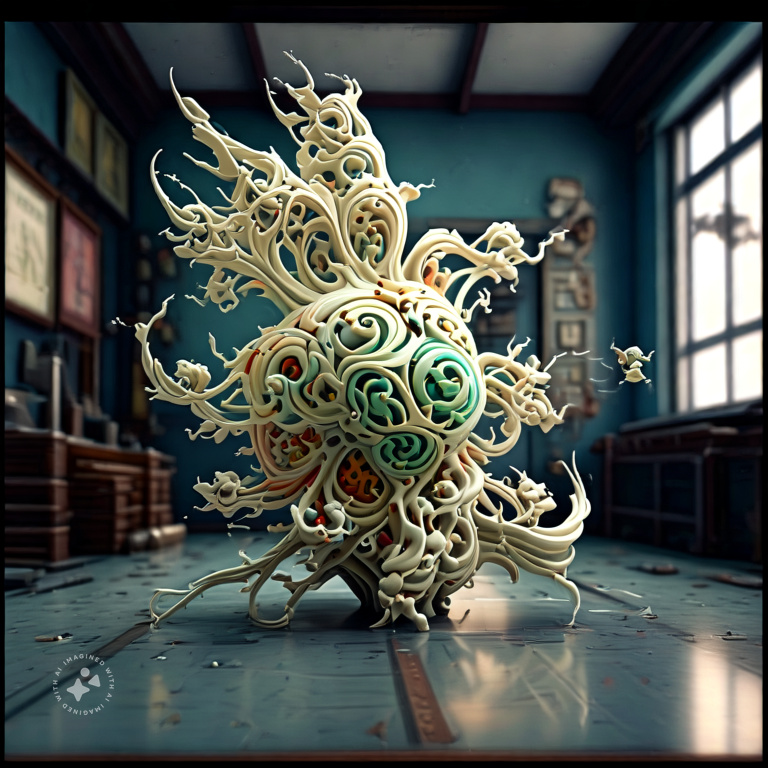}
        \caption{Meta AI}
    \end{subfigure}
    \begin{subfigure}[b]{0.16\textwidth}
        \centering
        \includegraphics[width=\textwidth]{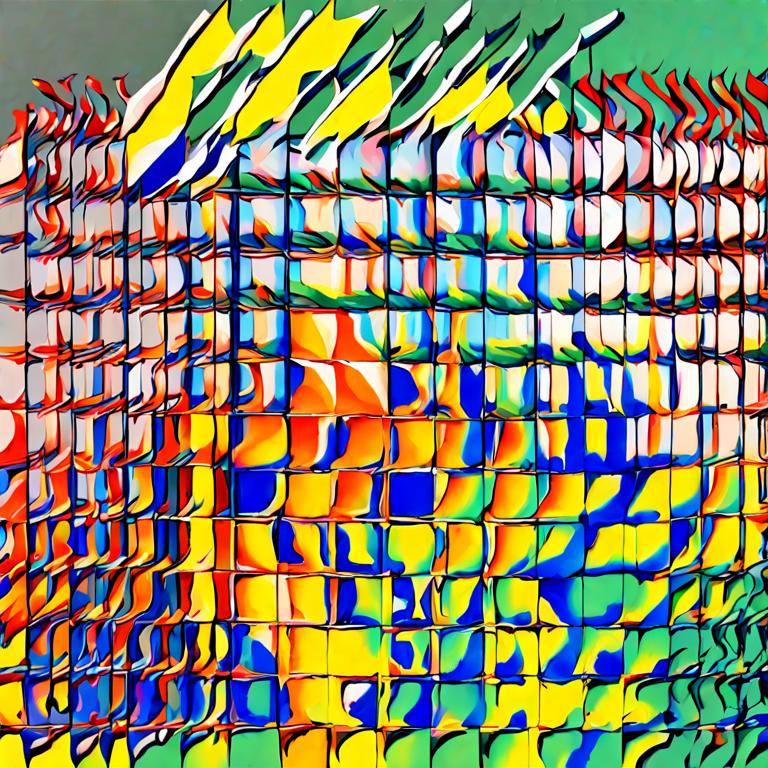}
        \caption{Runway}
    \end{subfigure}
    \begin{subfigure}[b]{0.16\textwidth}
        \centering
        \includegraphics[width=\textwidth]{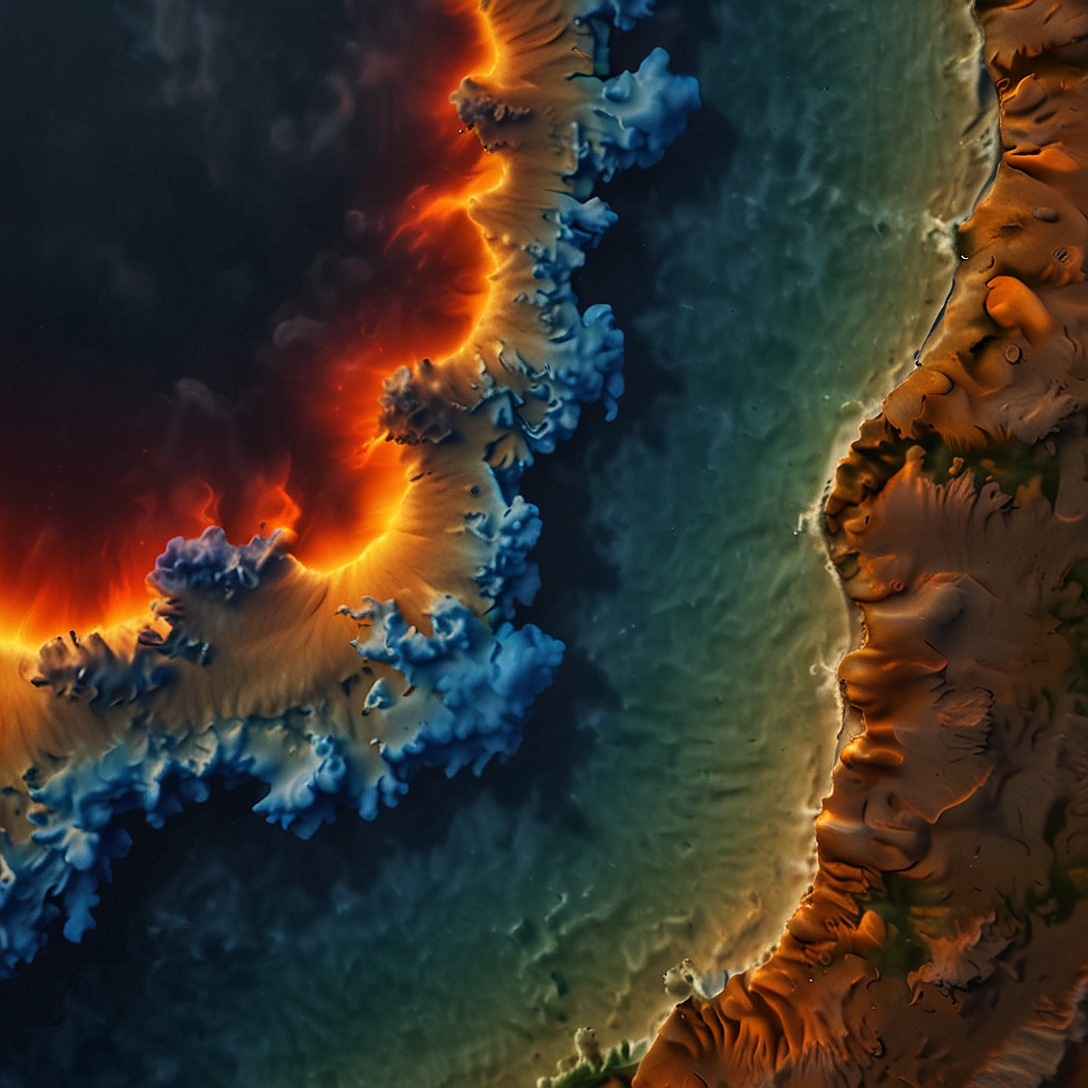}
        \caption{Leonardo Ai}
    \end{subfigure}

    \begin{subfigure}[b]{\textwidth}
        \centering
        \includegraphics[width=0.25\textwidth]{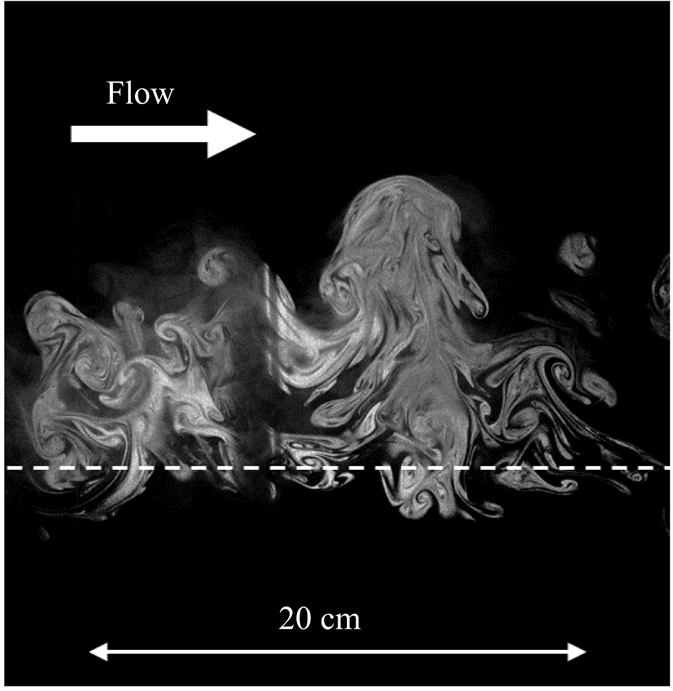}
        \caption{Rayleigh-Taylor instability in a lap experimental setup; the picture is taken from part (a) of Fig. 5 in \cite{suchandra2023dynamics}, published as open access under a Creative Commons license in the Journal of Fluid Mechanics, allowing readers to distribute the content freely.}
    \end{subfigure}

    \caption{A comparison between the AI-generated images and the lap experimental result (i.e., ground truth) for the prompt ``Rayleigh-Taylor instability''}
    \label{Fig1}
\end{figure}


\subsubsection{Rayleigh-Taylor instability}
\label{Sect276}

The next prompt is ``Rayleigh-Taylor instability'', with the generated images displayed in Fig. \ref{Fig1}. An important dimensionless number for investigating Rayleigh-Taylor instability, which can be observed in Fig. \ref{Fig1}, is the Atwood number (see e.g., \citep{luo2020effects,ker1988simulation,andrews2010small}), expressed as:
\begin{equation}
A = \frac{\rho_1 - \rho_2}{\rho_1 + \rho_2},
\label{Eq8}
\end{equation}
where $\rho_1$ and $\rho_2$ are the densities of the heavier and lighter fluids, respectively.

According to observations from Fig. \ref{Fig1}, the image produced by DALL·E 3 is particularly interesting as it suggests the interaction between heavy and light fluids. The result from Gemini Advanced is noteworthy because it displays the interaction of two fluids. The outputs from Midjourney and Leonardo Ai are somewhat similar to the images generated for the ``Kelvin-Helmholtz instability'' prompt, as shown in Fig. \ref{Fig8}. Conversely, the outputs from Meta AI and Runway are not accurate and largely irrelevant.


\begin{figure}[t]
    \centering
   
    \begin{subfigure}[b]{0.16\textwidth}
        \centering
        \includegraphics[width=\textwidth]{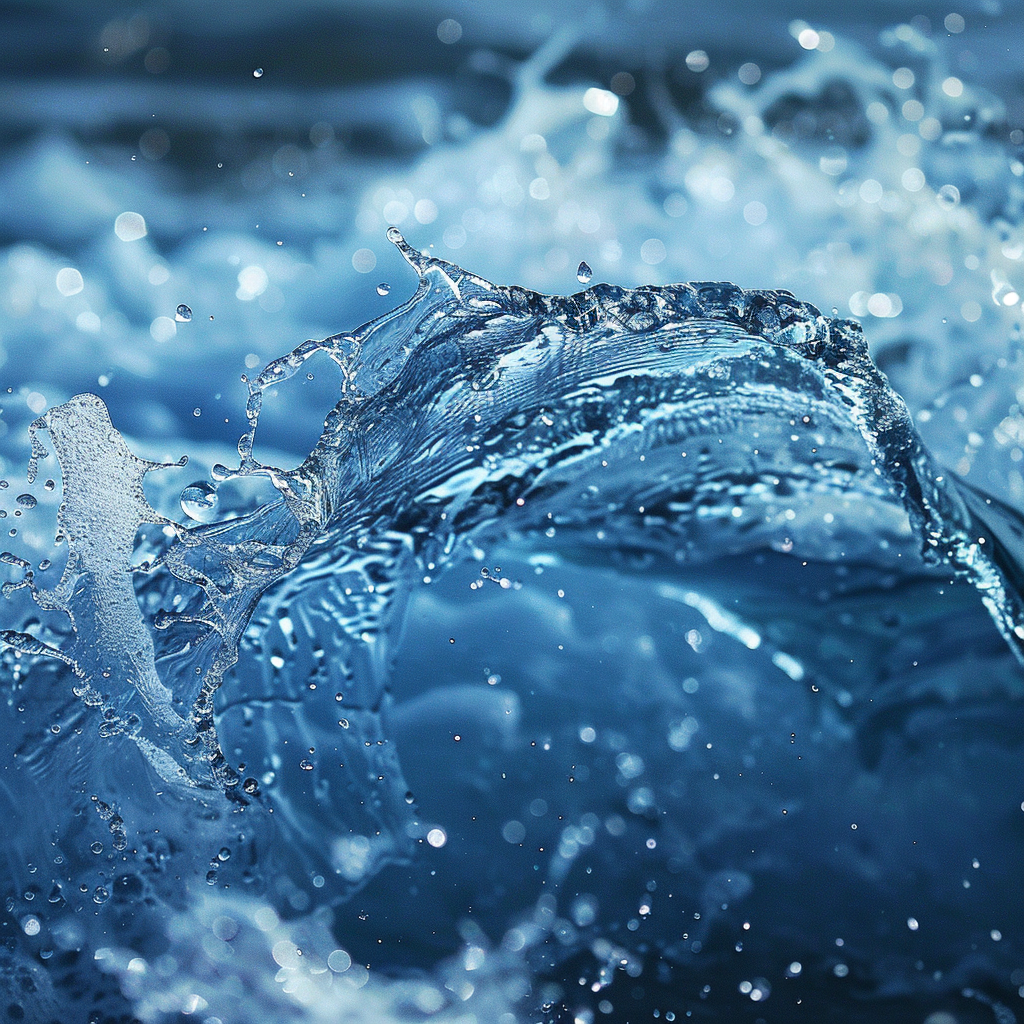}
        \caption{Midjourney}
    \end{subfigure}
    \begin{subfigure}[b]{0.16\textwidth}
        \centering
        \includegraphics[width=\textwidth]{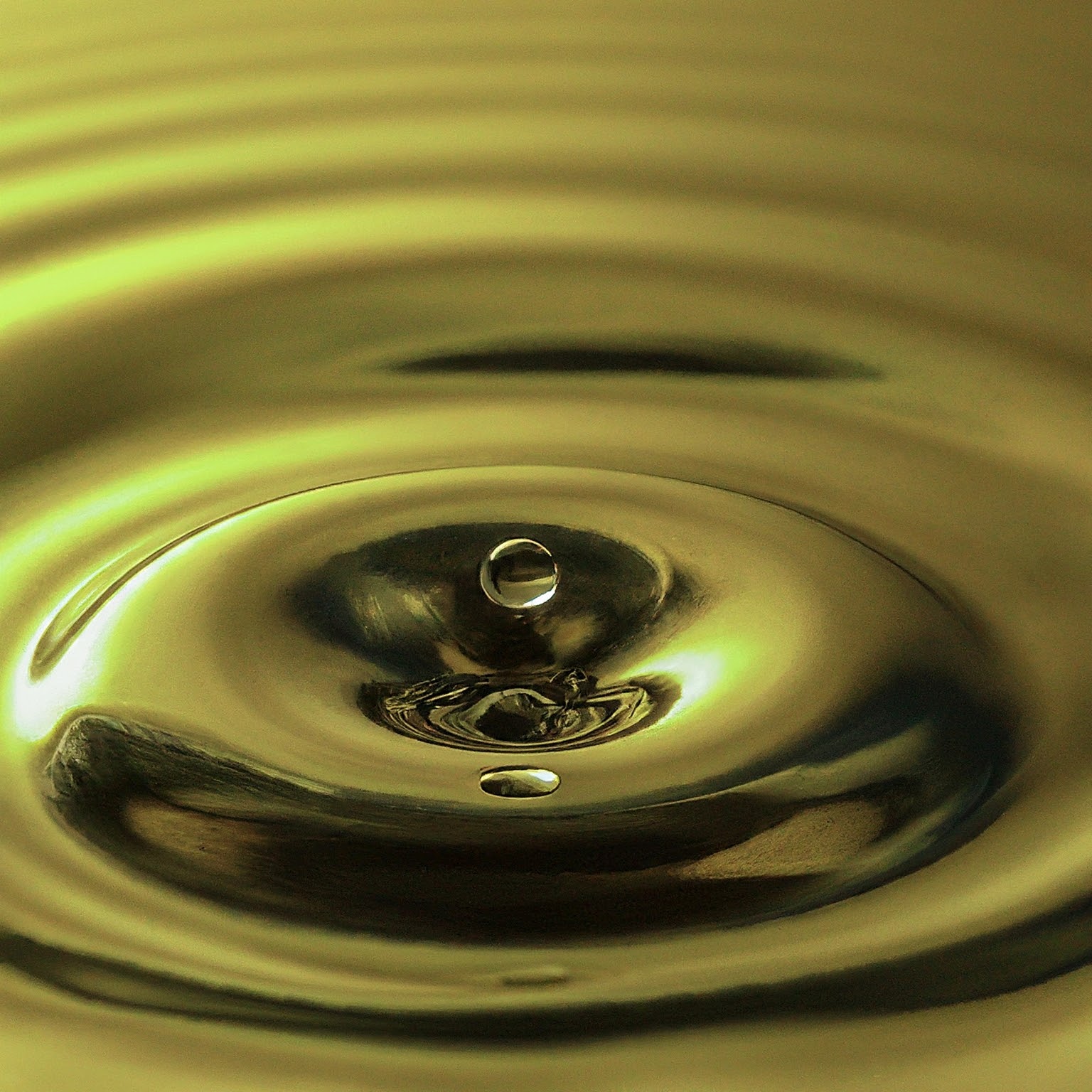}
        \caption{Gemini}
    \end{subfigure}
    \begin{subfigure}[b]{0.16\textwidth}
        \centering
        \includegraphics[width=\textwidth]{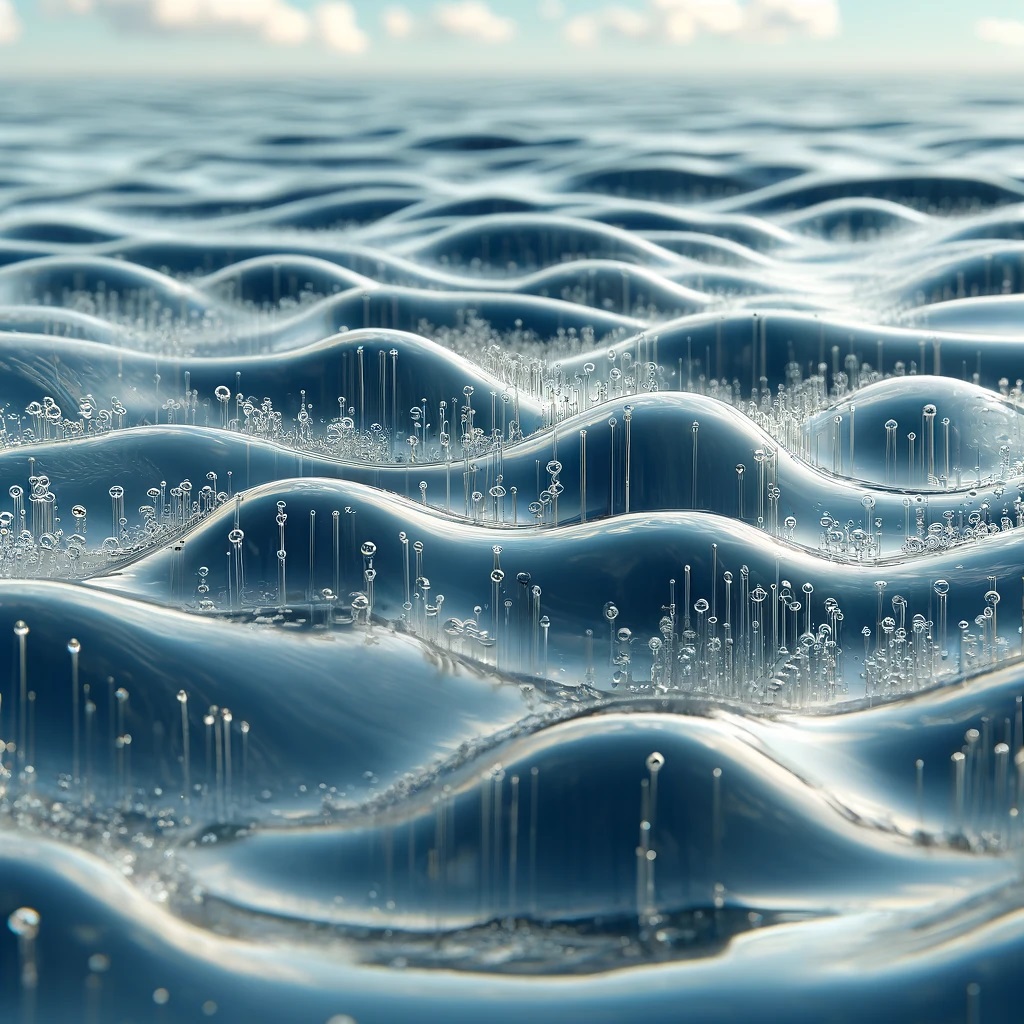}
        \caption{DALL·E 3}
    \end{subfigure}
    \begin{subfigure}[b]{0.16\textwidth}
        \centering
        \includegraphics[width=\textwidth]{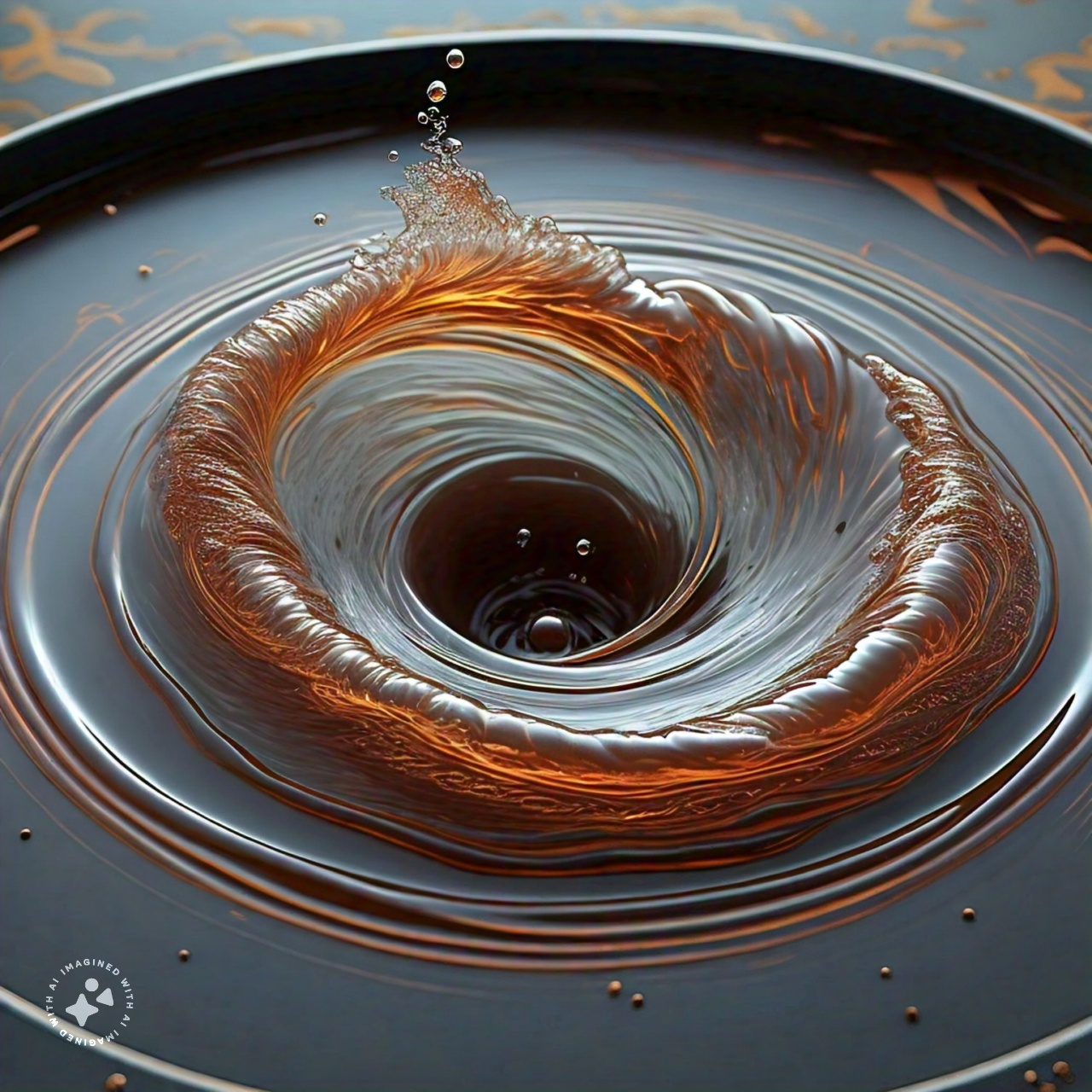}
        \caption{Meta AI}
    \end{subfigure}
    \begin{subfigure}[b]{0.16\textwidth}
        \centering
        \includegraphics[width=\textwidth]{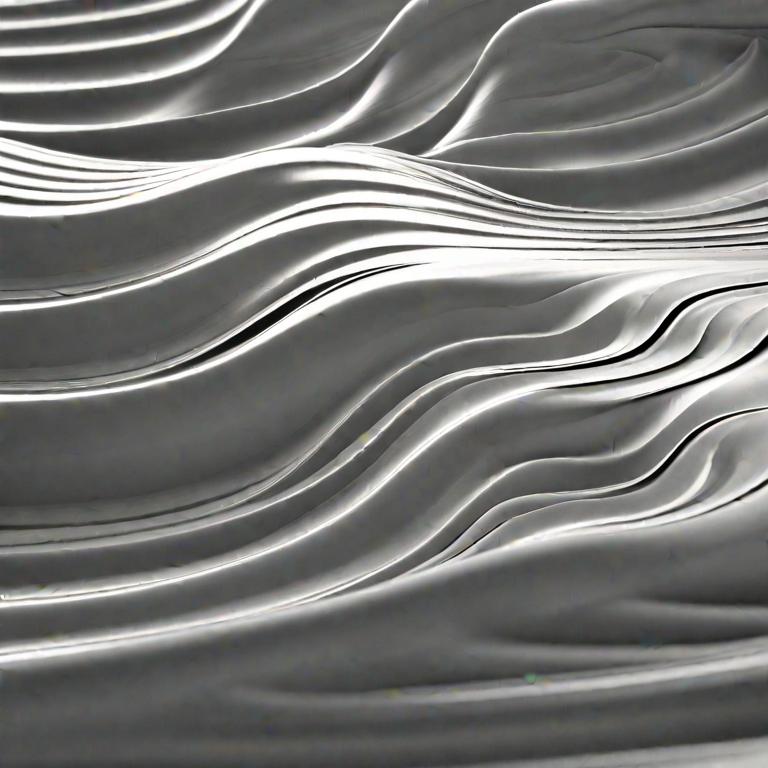}
        \caption{Runway}
    \end{subfigure}
    \begin{subfigure}[b]{0.16\textwidth}
        \centering
        \includegraphics[width=\textwidth]{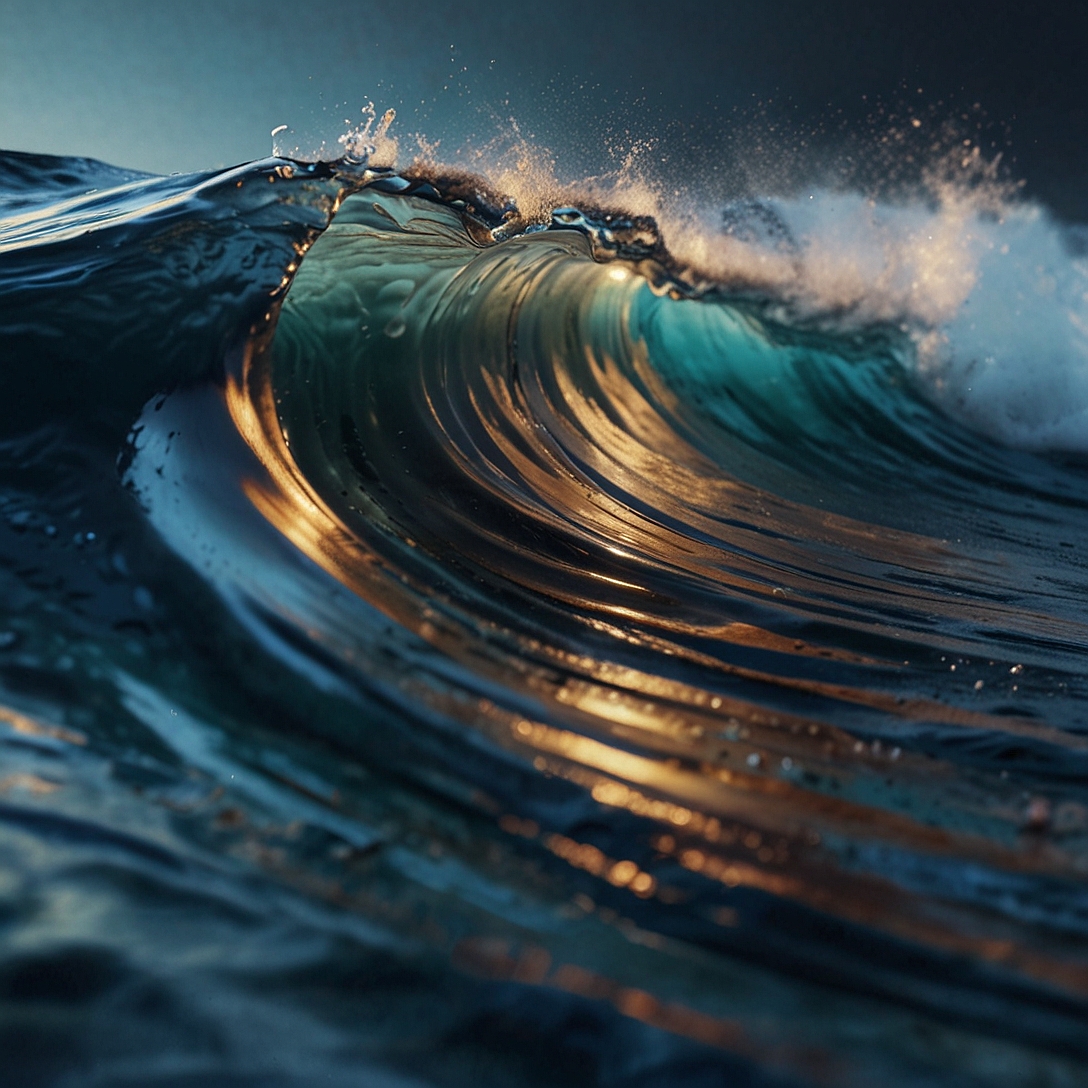}
        \caption{Leonardo Ai}
    \end{subfigure}

    \begin{subfigure}[b]{\textwidth}
        \centering
        \includegraphics[width=0.35\textwidth]{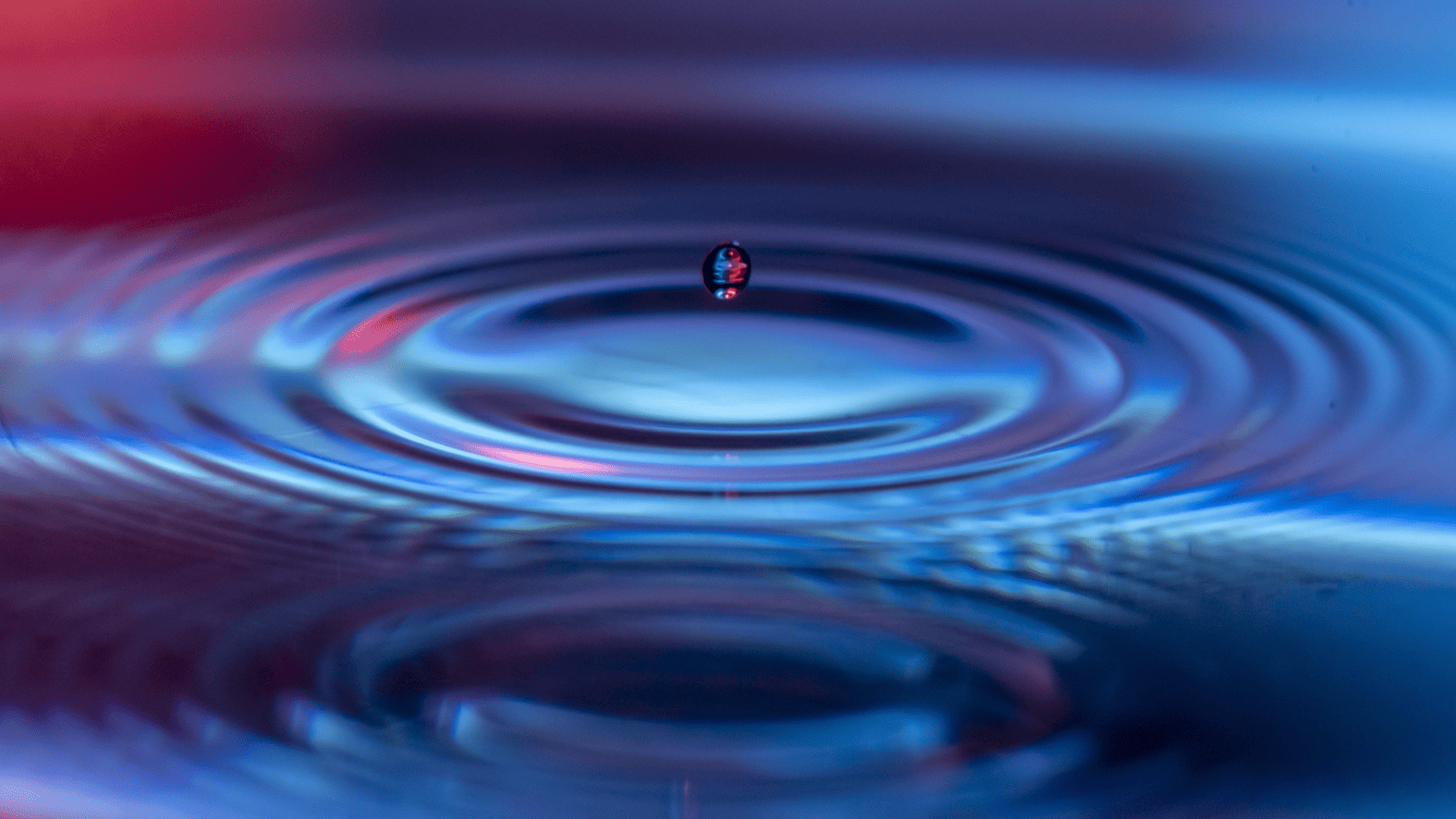}
        \caption{Capillary wave: A small water droplet, approximately 0.4 mm in radius, rebounds from the same fluid bath in a lab experimental setup; this image is Fig. 1 in \cite{alventosa2023inertio}, published as open access under a Creative Commons license in the Journal of Fluid Mechanics, allowing readers to distribute the content freely.}
    \end{subfigure}

    \caption{A comparison between the AI-generated images and the numerical simulation result (i.e., ground truth) for the prompt ``capillary wave''}
    \label{Fig2}
\end{figure}


\subsubsection{Capillary wave}
\label{Sect277}

``Capillary wave'' is the next prompt, and the generated images are collected in Fig. \ref{Fig2}. The most accurate result is generated by Gemini Advanced. One possible question is whether Gemini Advanced uses online information simultaneously for generating images or if it only uses a predefined dataset. Following Gemini Advanced, the output from Meta AI is also relatively accurate and contextually relevant. In contrast, the images generated by Midjourney, Runway, and Leonardo Ai are misleading, as they depict general waves, such as sea waves. Meanwhile, the result from DALL·E 3, while artistic, does not pertain to capillary waves.


\begin{figure}[t]
    \centering
    \begin{subfigure}[b]{0.16\textwidth}
        \centering
        \includegraphics[width=\textwidth]{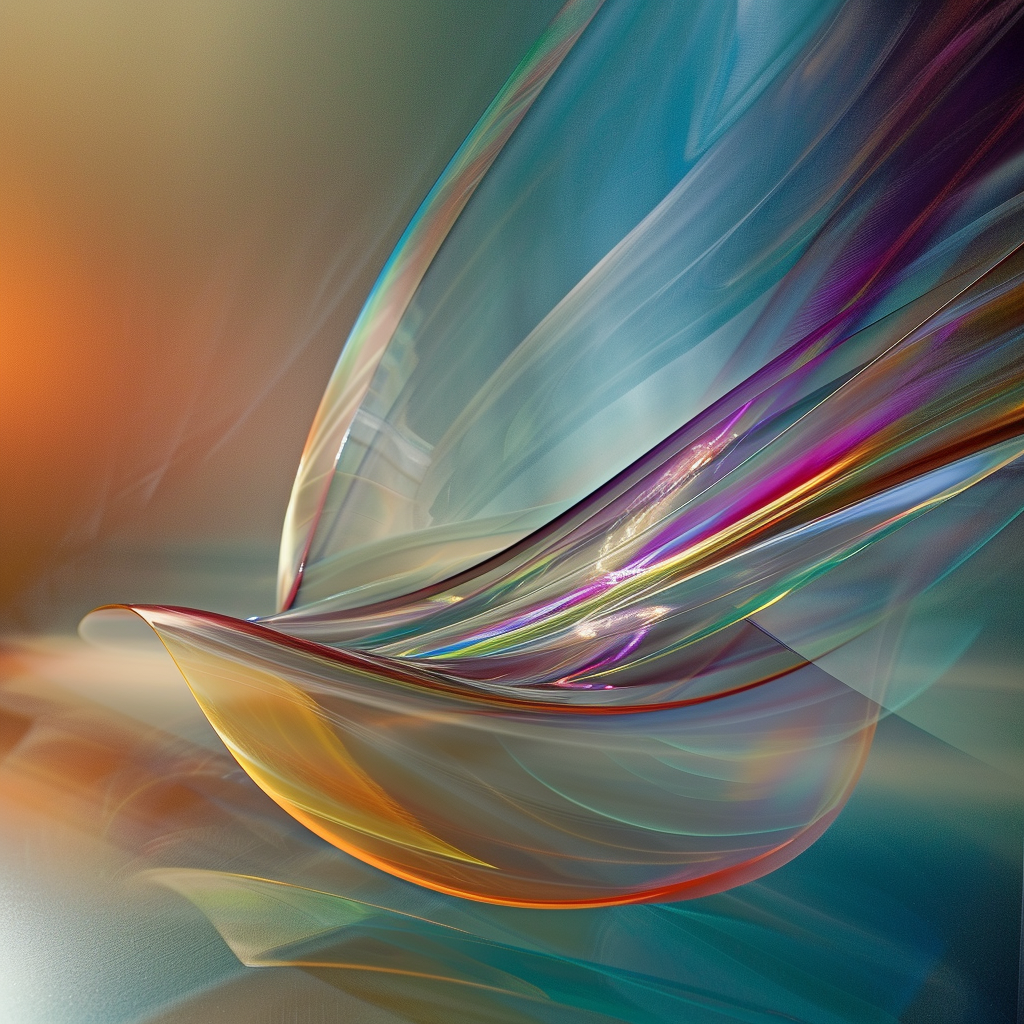}
        \caption{Midjourney}
    \end{subfigure}
    \begin{subfigure}[b]{0.16\textwidth}
        \centering
        \includegraphics[width=\textwidth]{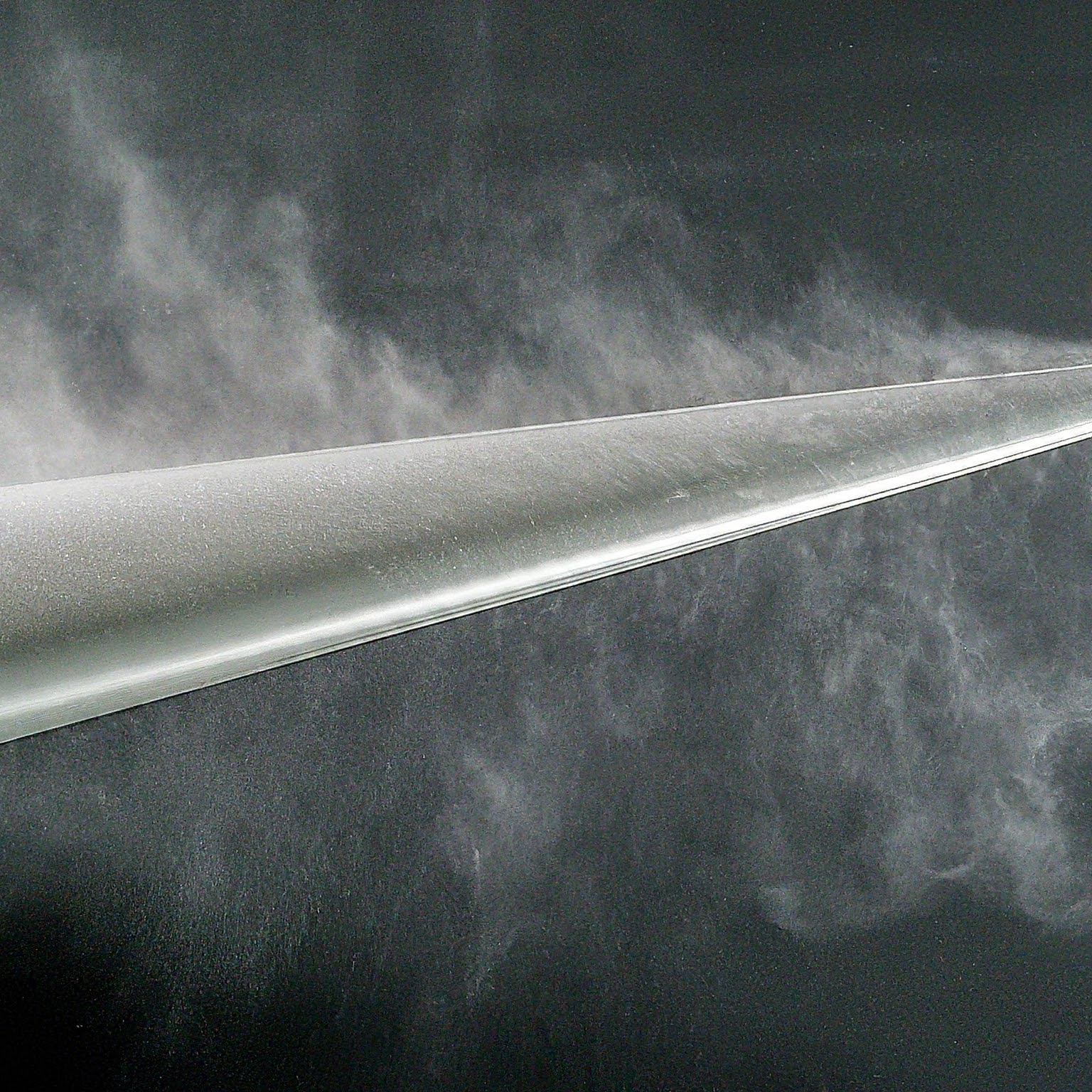}
        \caption{Gemini}
    \end{subfigure}
    \begin{subfigure}[b]{0.16\textwidth}
        \centering
        \includegraphics[width=\textwidth]{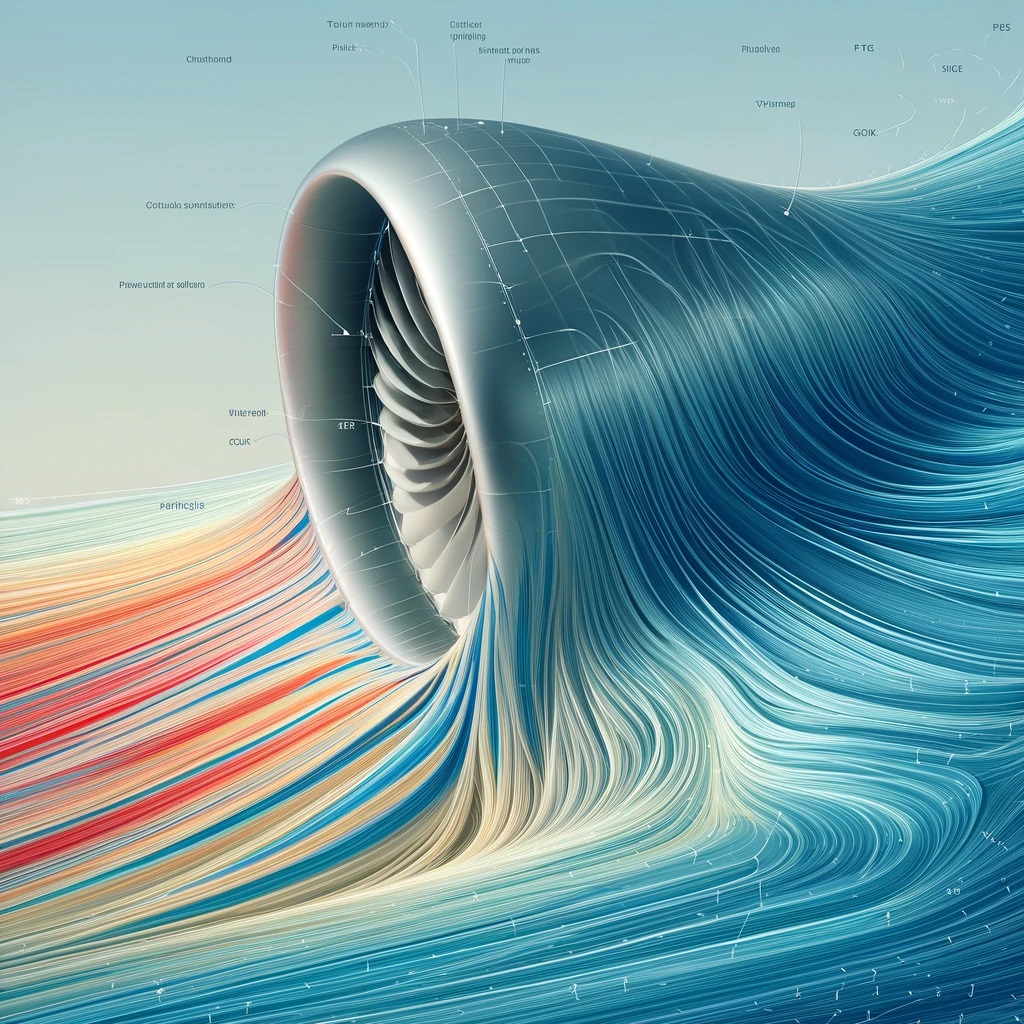}
        \caption{DALL·E 3}
    \end{subfigure}
    \begin{subfigure}[b]{0.16\textwidth}
        \centering
        \includegraphics[width=\textwidth]{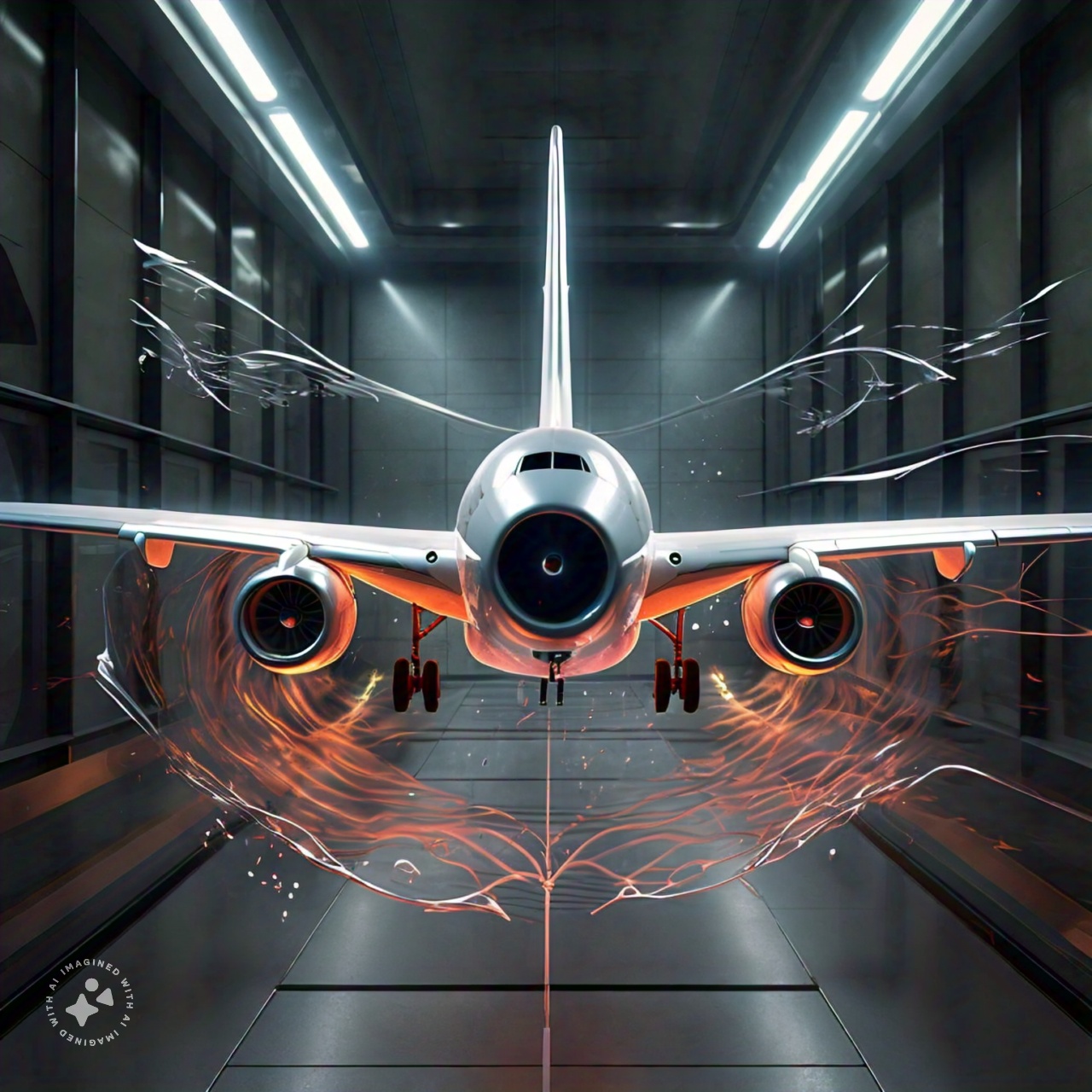}
        \caption{Meta AI}
    \end{subfigure}
    \begin{subfigure}[b]{0.16\textwidth}
        \centering
        \includegraphics[width=\textwidth]{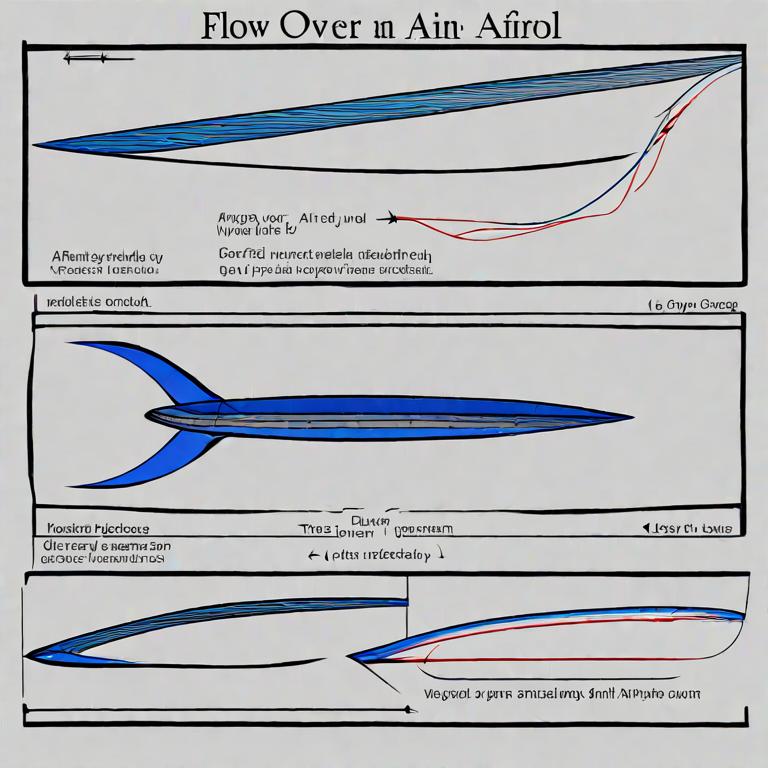}
        \caption{Runway}
    \end{subfigure}
    \begin{subfigure}[b]{0.16\textwidth}
        \centering
        \includegraphics[width=\textwidth]{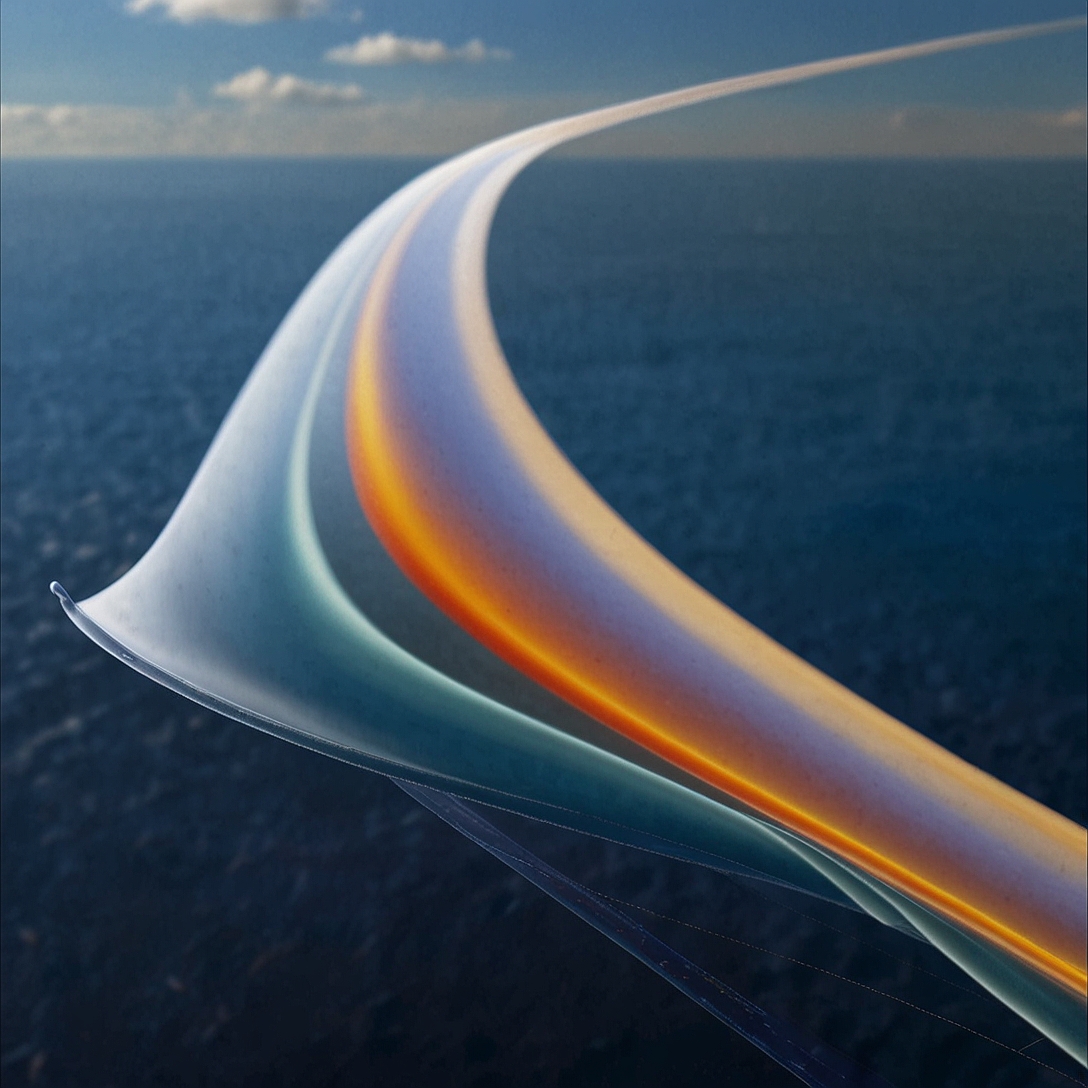}
        \caption{Leonardo Ai}
    \end{subfigure}

    \begin{subfigure}[b]{\textwidth}
        \centering
        \includegraphics[width=0.35\textwidth]{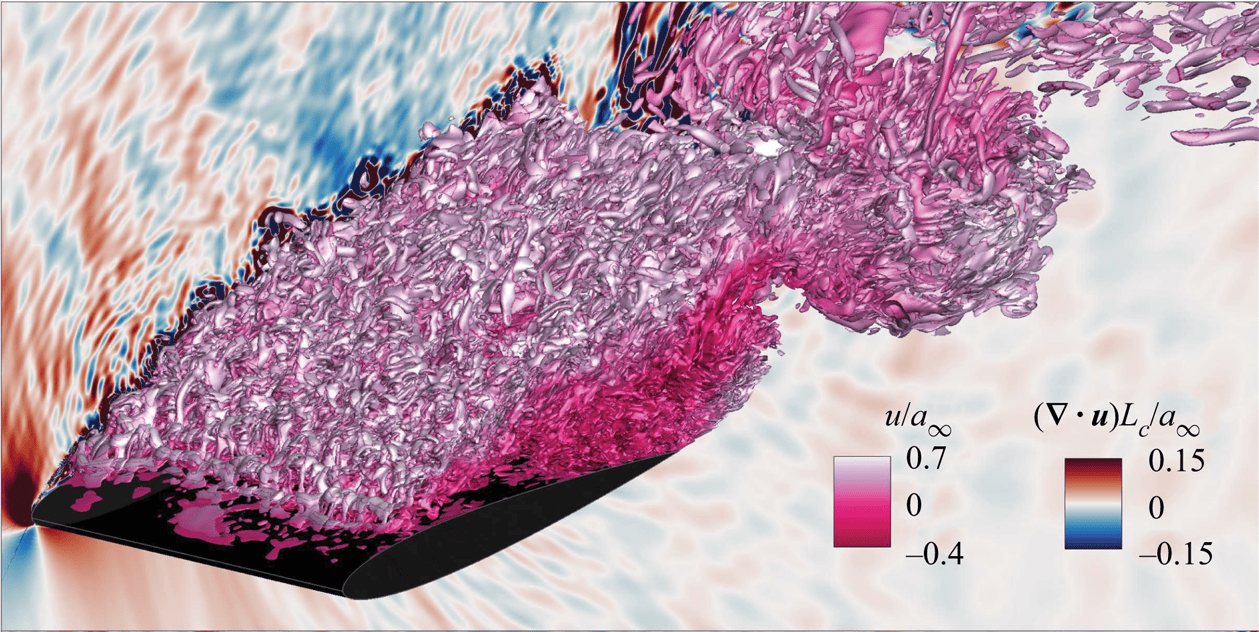}
        \caption{Streamwise vorticity over an airfoil; this image, produced by a numerical simulation, is taken from part (a) of Fig. 1 in \cite{turner2022quadrupole}, published as open access under a Creative Commons license in the Journal of Fluid Mechanics, allowing readers to distribute the content freely.}
    \end{subfigure}

    \caption{A comparison between the AI-generated images and the numerical simulation result (i.e., ground truth) for the prompt ``Flow over an airfoil''}
    \label{Fig10}
\end{figure}


\subsubsection{Flow over an airfoil}
\label{Sect278}

In Fig. \ref{Fig10}, we present a collection of generated images for the prompt ``Flow over an airfoil''. Gemini Advanced produces the most relevant image and it depicts an airplane wing. This wing might resemble NACA airfoils \citep{abbott1945summary}, where the cross-section of the wing can be considered as an airfoil.

The image generated by DALL·E 3 depicts an airplane turbine, while Meta AI produces an image of an airplane in a room. The image generated by Runway is conceptually similar to those produced by Runway in Fig. \ref{Fig9} and Fig. \ref{Fig5}. Additionally, the letters ``Flow over m Ain Afirol'' can be seen in that image. Images generated by Midjourney and Leonardo Ai are artistically similar to each other, yet they do not accurately represent the fluid mechanics concept of flow over an airfoil. Similar to our discussion on the prompt ``Flow past a circular cylinder'' in Sect. \ref{Sect272}, the term ``flow'' may not be understood as ``fluid flow'' by these generative models.


\begin{figure}[t]
    \centering
    \begin{subfigure}[b]{0.16\textwidth}
        \centering
        \includegraphics[width=\textwidth]{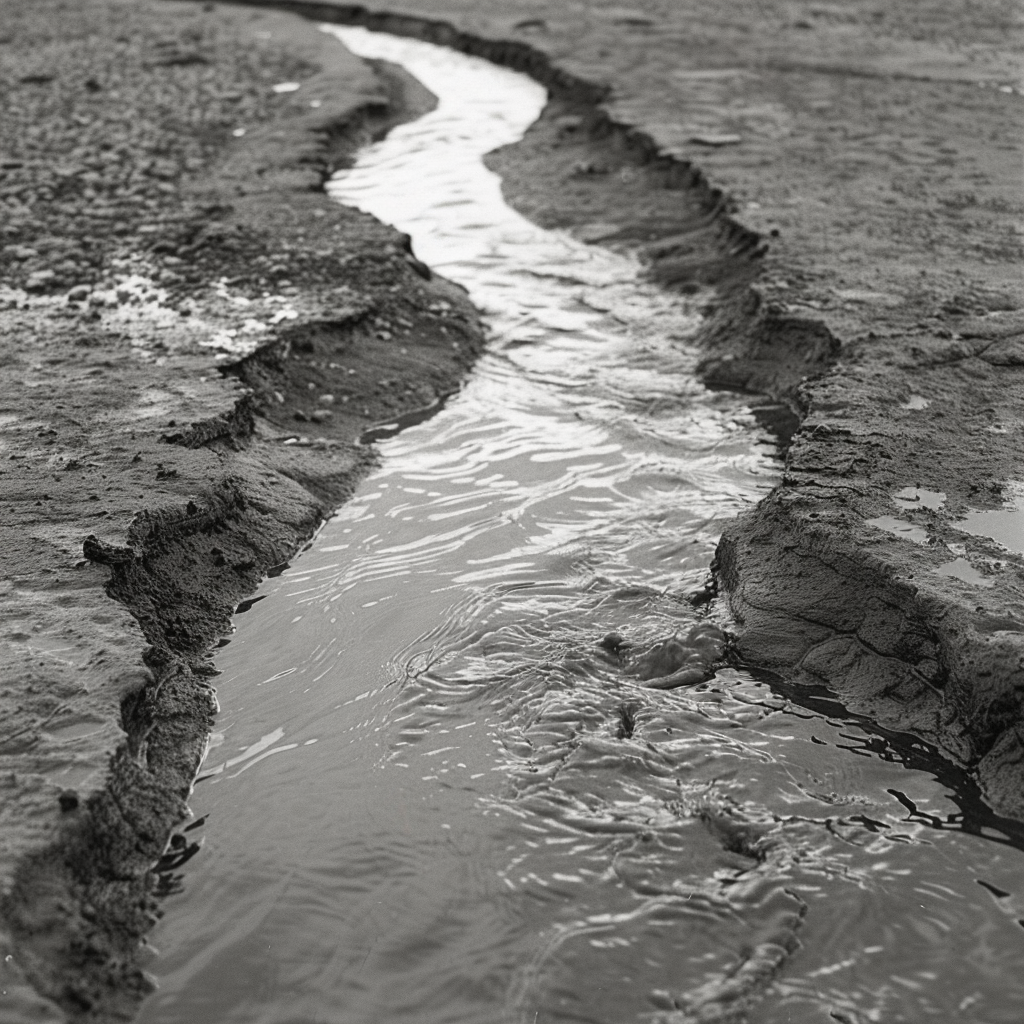}
        \caption{Midjourney}
    \end{subfigure}
    \begin{subfigure}[b]{0.16\textwidth}
        \centering
        \includegraphics[width=\textwidth]{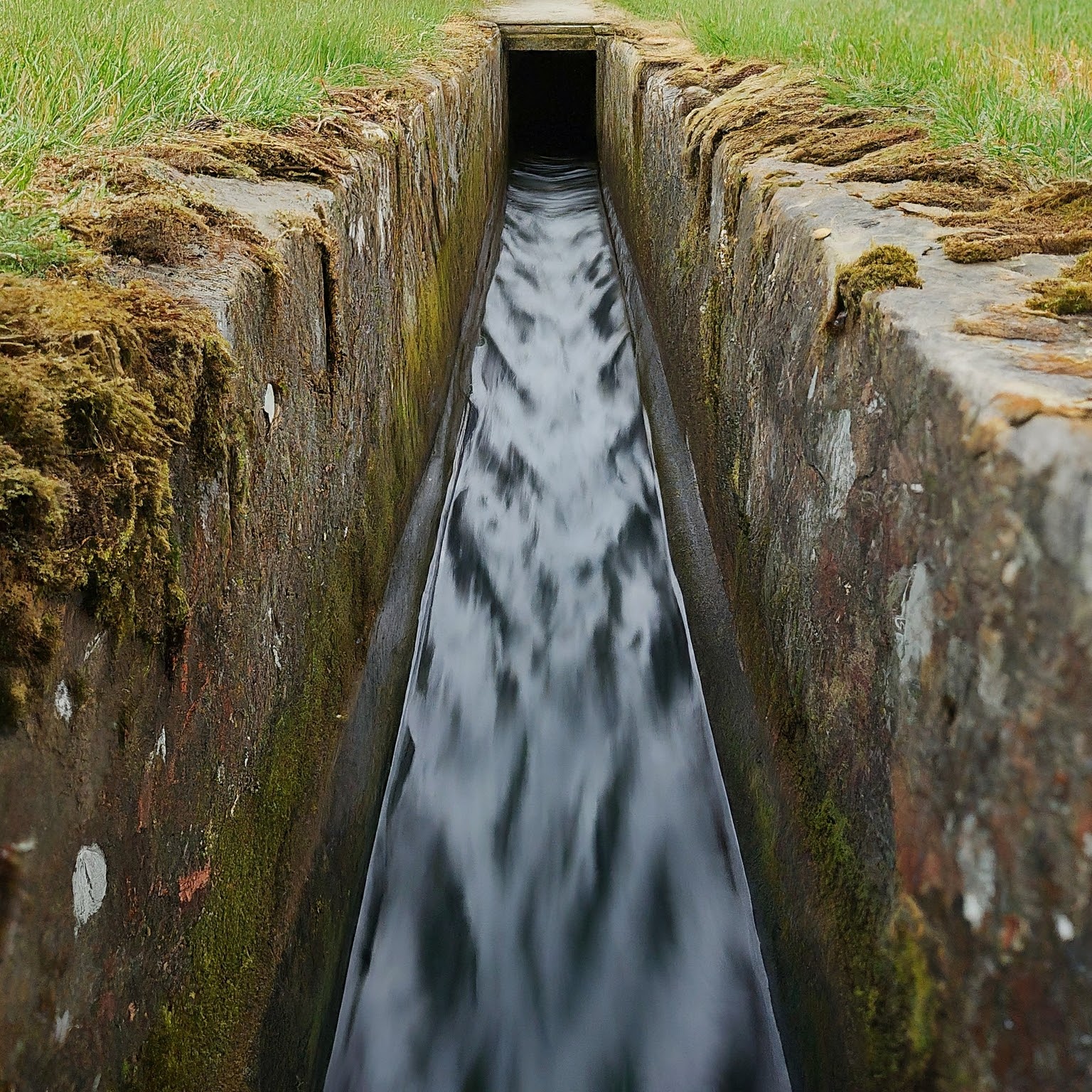}
        \caption{Gemini}
    \end{subfigure}
    \begin{subfigure}[b]{0.16\textwidth}
        \centering
        \includegraphics[width=\textwidth]{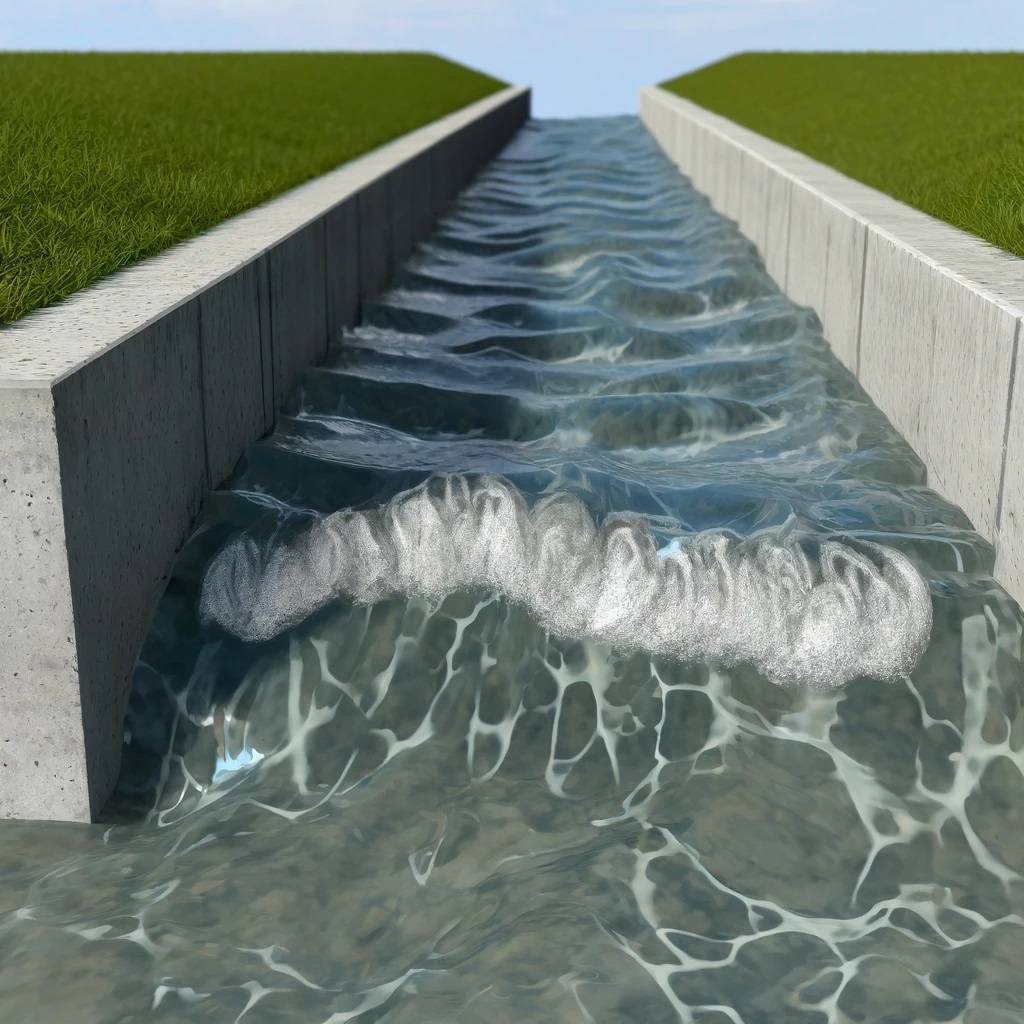}
        \caption{DALL·E 3}
    \end{subfigure}
    \begin{subfigure}[b]{0.16\textwidth}
        \centering
        \includegraphics[width=\textwidth]{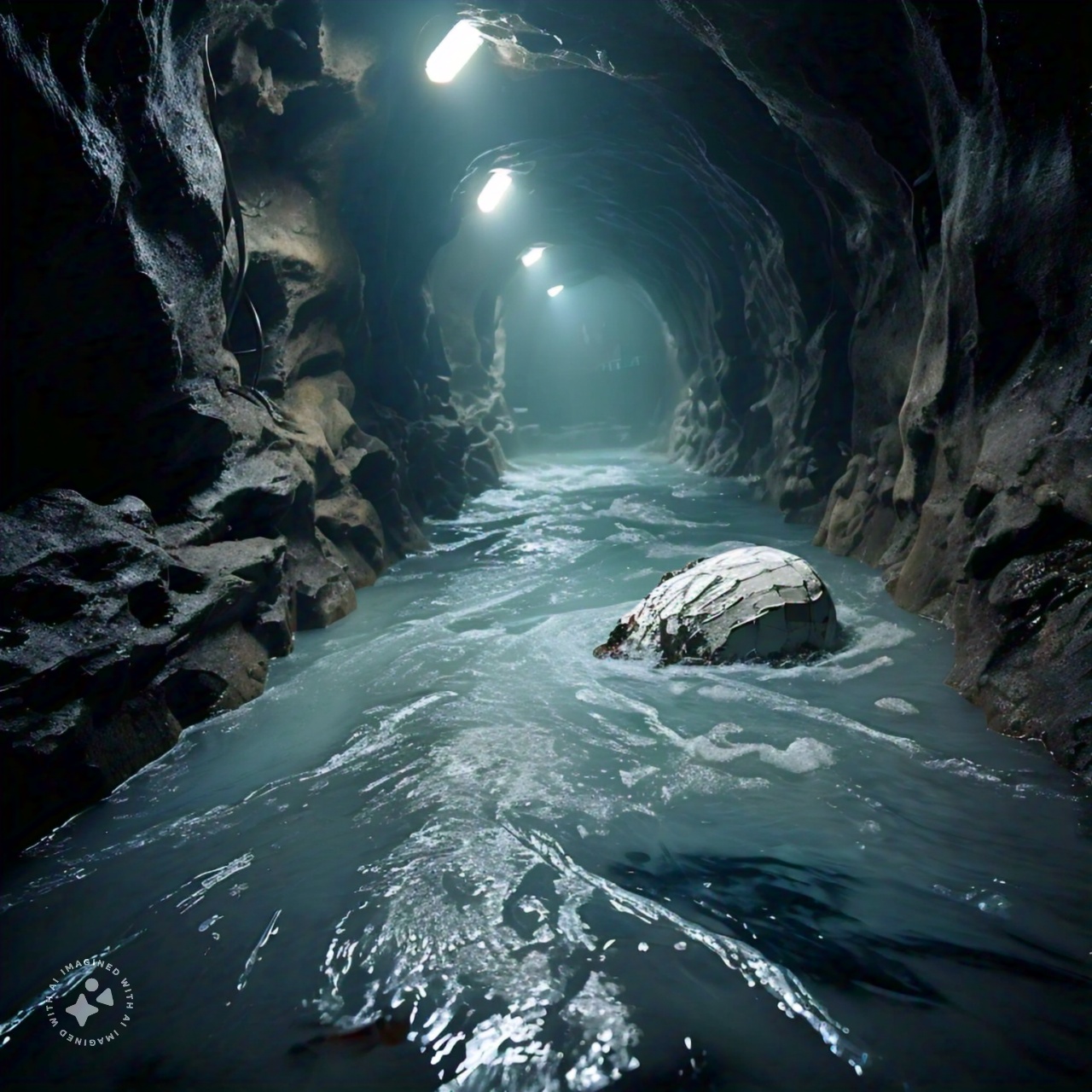}
        \caption{Meta AI}
    \end{subfigure}
    \begin{subfigure}[b]{0.16\textwidth}
        \centering
        \includegraphics[width=\textwidth]{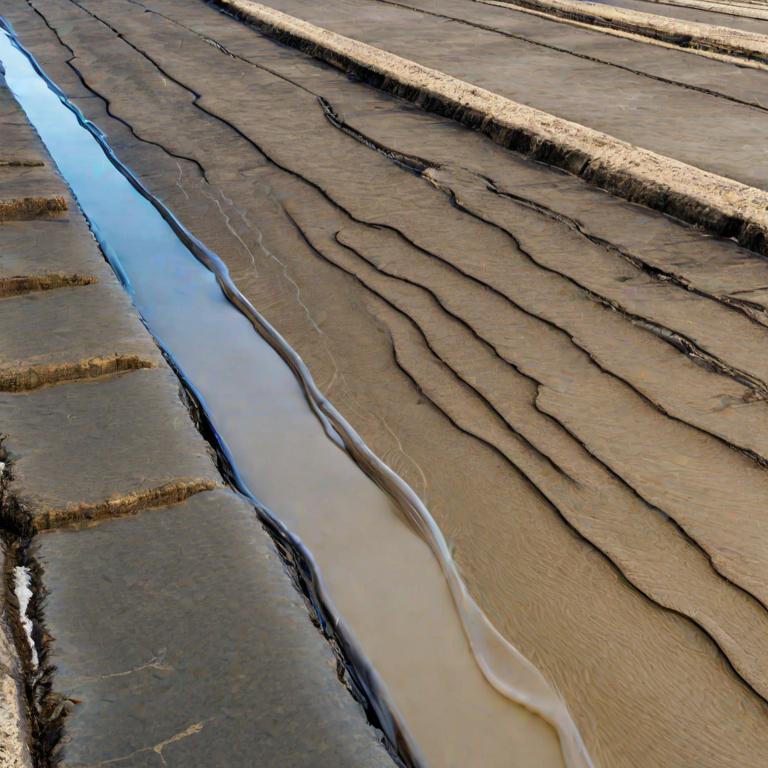}
        \caption{Runway}
    \end{subfigure}
    \begin{subfigure}[b]{0.16\textwidth}
        \centering
        \includegraphics[width=\textwidth]{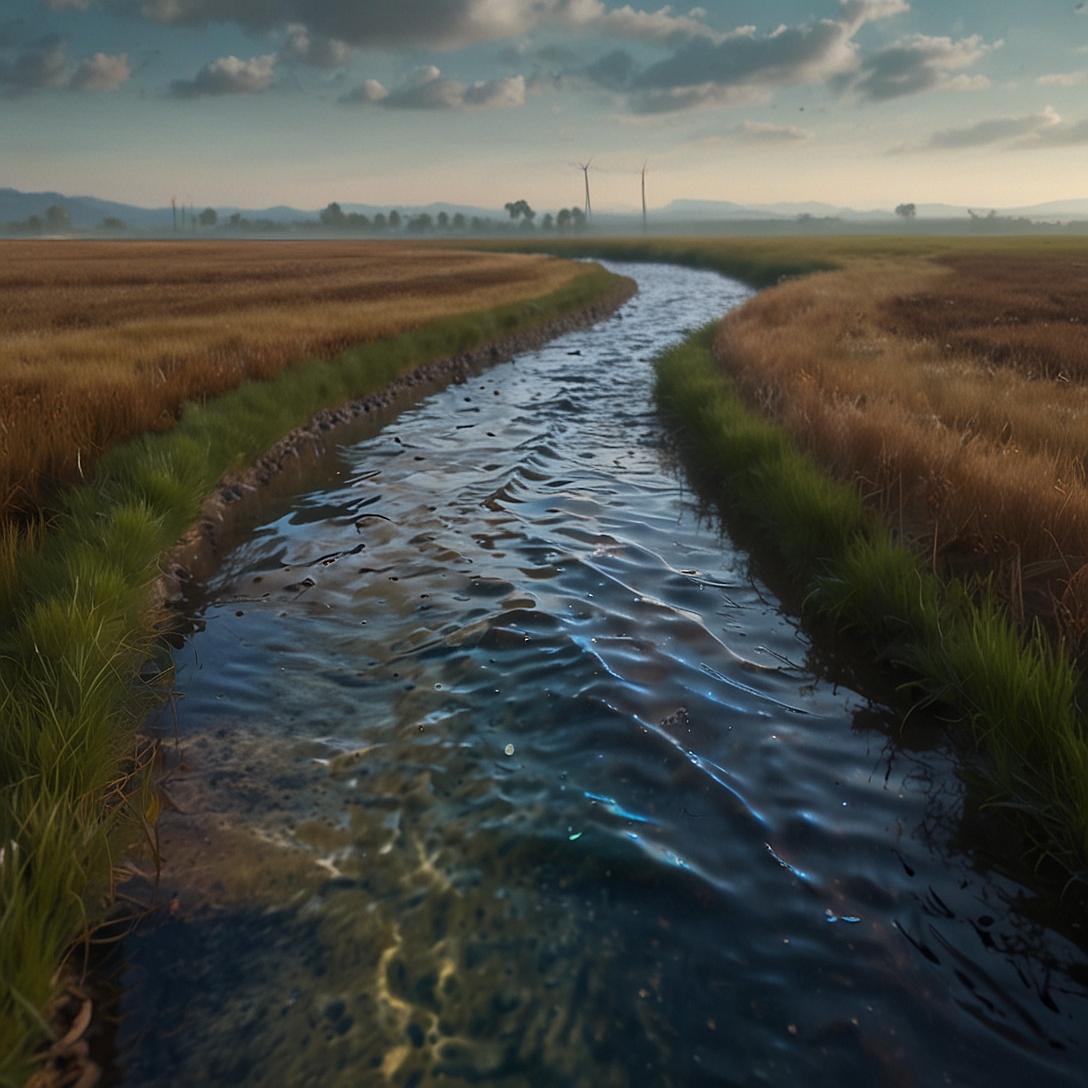}
        \caption{Leonardo Ai}
    \end{subfigure}

    \begin{subfigure}[b]{\textwidth}
        \centering
        \includegraphics[width=0.42\textwidth]{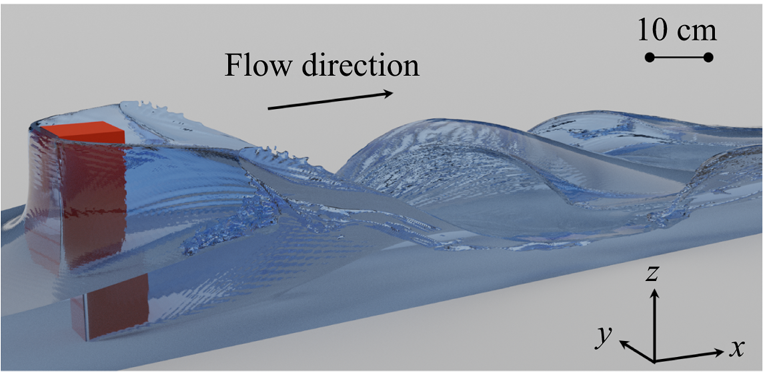}
        \caption{Free-surface flow in a channel around a square cylinder at a Froude number of $Fr=2.5$; a hydraulic jump is observed. The picture, produced by a numerical simulation, is taken from part (c) of Fig. 3 in \cite{eames2024free}, published as open access under a Creative Commons license in the Journal of Fluid Mechanics, allowing readers to distribute the content freely.}
    \end{subfigure}

    \caption{A comparison between the AI-generated images and the numerical simulation result (i.e., ground truth) for the prompt ``Free-surface flow''}
    \label{Fig3}
\end{figure}


\subsubsection{Free-surface channel flow}
\label{Sect279}

The images generated in response to the prompt ``Free-surface channel flow'' are displayed in Fig. \ref{Fig3}. This prompt is intriguing because it is a common term in everyday life and simultaneously a scientific term in fluid dynamics. Interestingly, this phenomenon shares features in both contexts. As illustrated in Fig. \ref{Fig3}, all the generated images depict free surface flow in nature, resembling everyday photos. Conversely, part (g) of Fig. \ref{Fig3} presents a scientific picture associated with this phenomenon.

This simple experiment with prompts suggests the idea of allowing users to choose the context in generative models, particularly those available as applications. This feature would be beneficial when a prompt is common across different contexts or domains of knowledge, enabling users to select the specific context for which they want to generate images. For instance, Leonardo Ai currently allows users to specify a domain of interest for generating images, such as cinematic, environment, general, photography, etc. However, an academic option is not yet available.


\begin{figure}[t]
    \centering
    
    \begin{subfigure}[b]{0.16\textwidth}
        \centering
        \includegraphics[width=\textwidth]{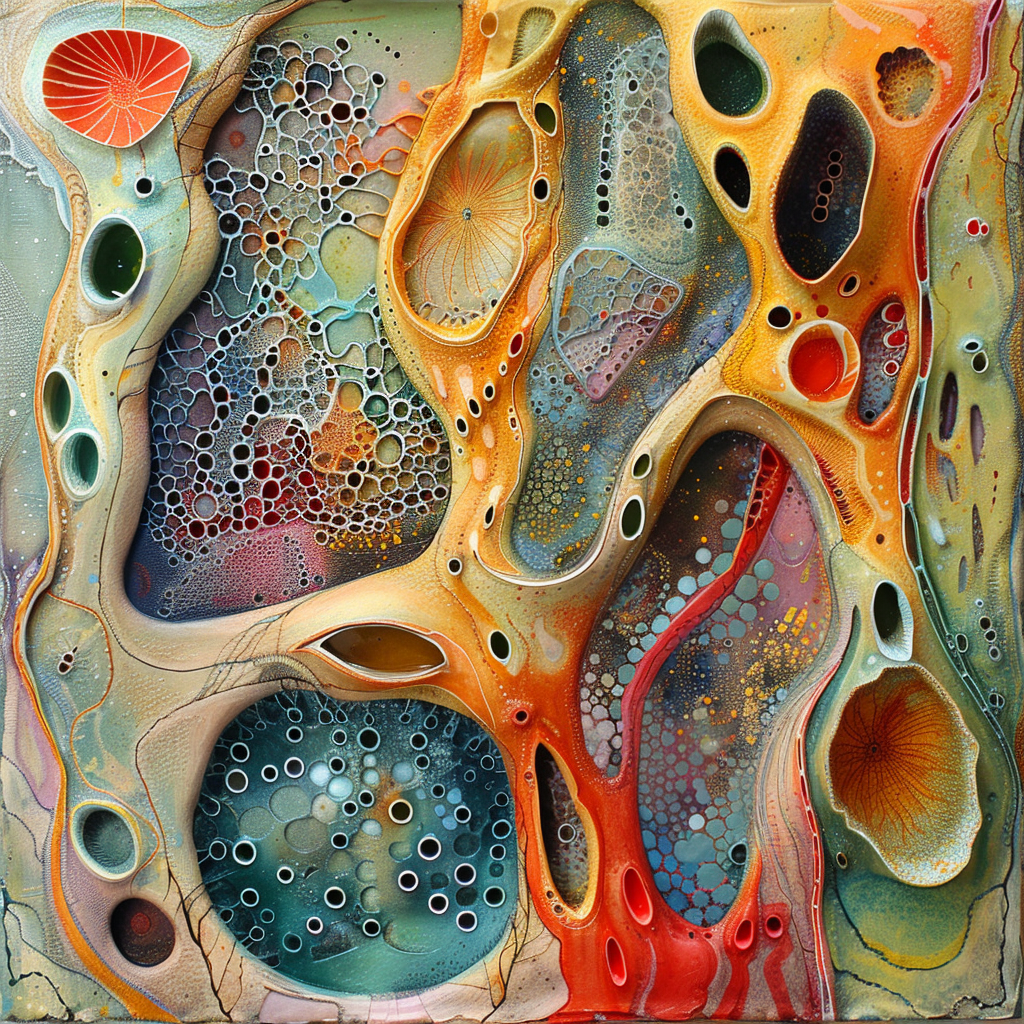}
        \caption{Midjourney}
    \end{subfigure}
    \begin{subfigure}[b]{0.16\textwidth}
        \centering
        \includegraphics[width=\textwidth]{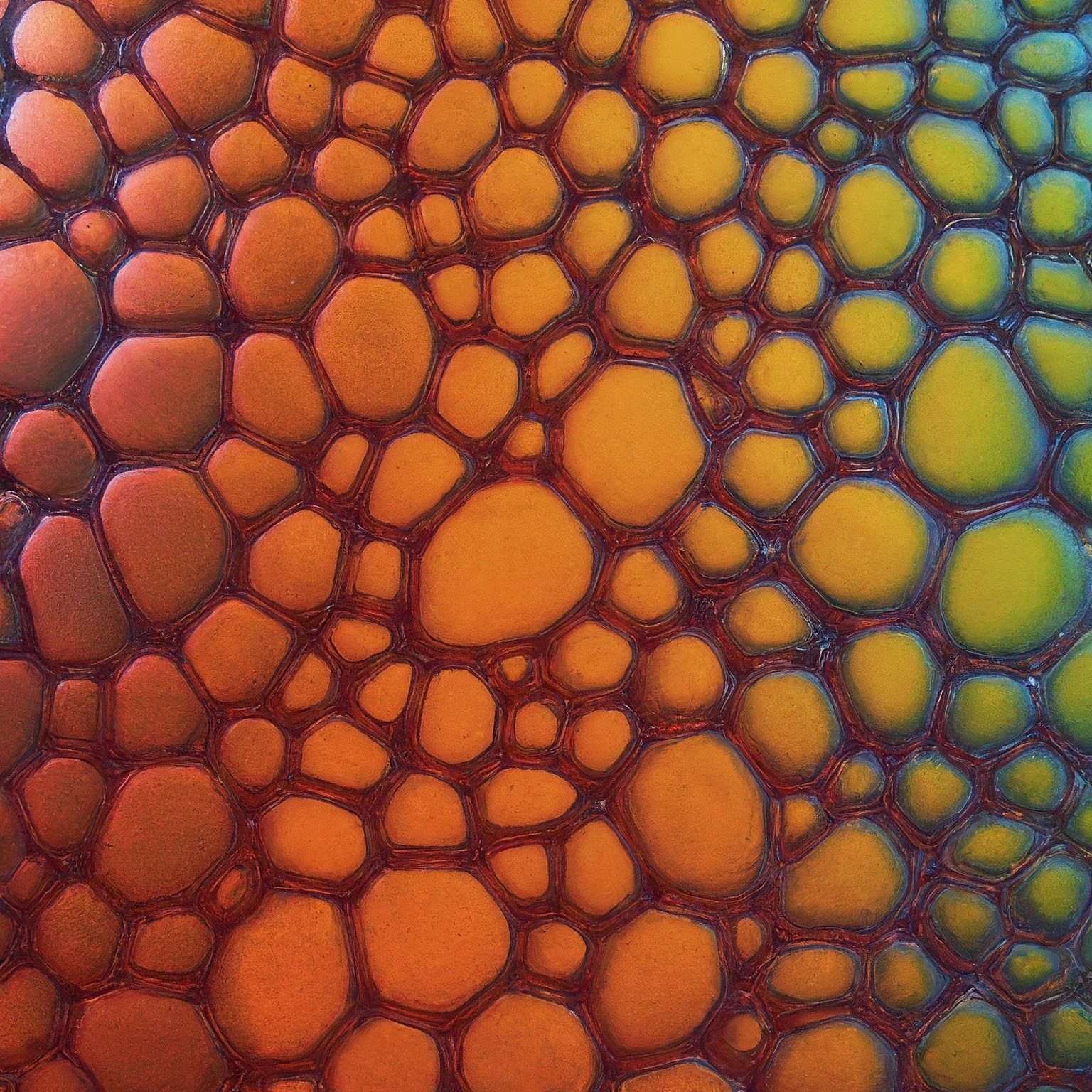}
        \caption{Gemini}
    \end{subfigure}
    \begin{subfigure}[b]{0.16\textwidth}
        \centering
        \includegraphics[width=\textwidth]{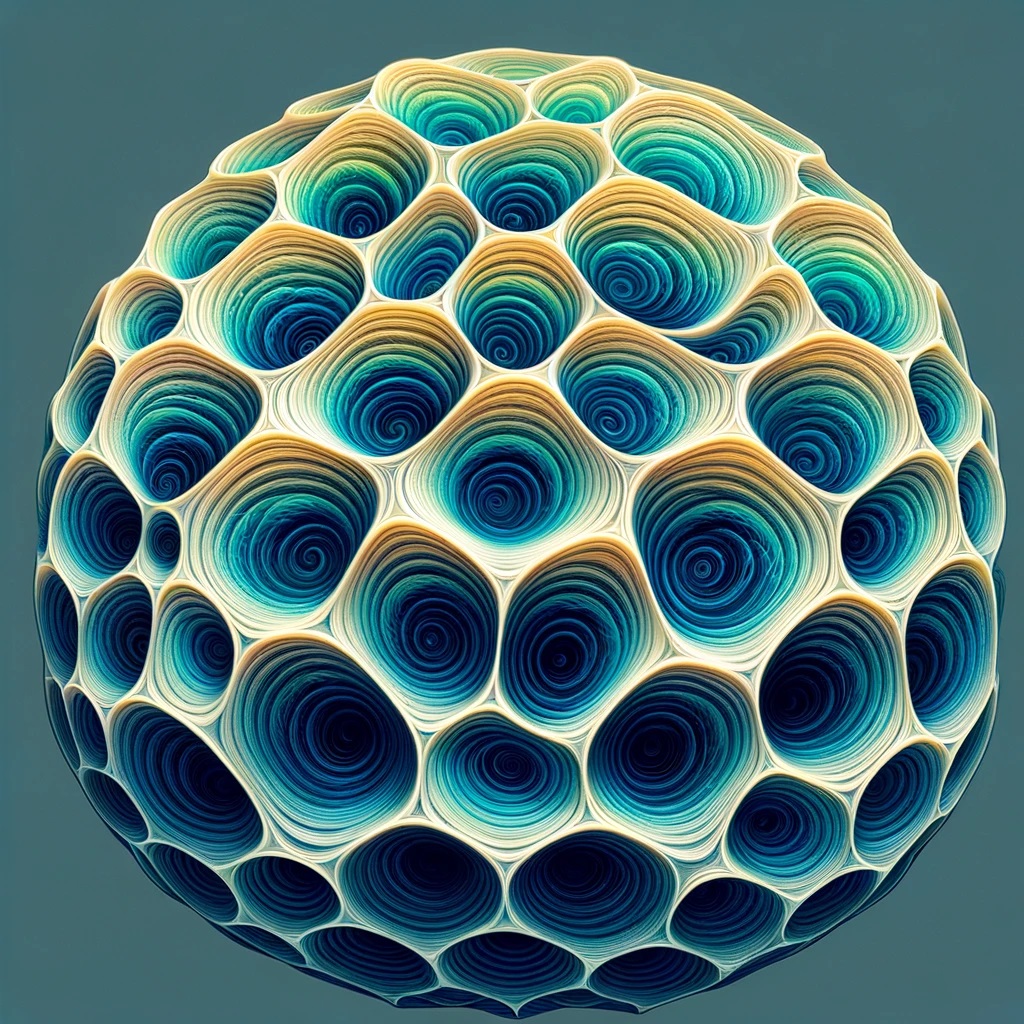}
        \caption{DALL·E 3}
    \end{subfigure}
    \begin{subfigure}[b]{0.16\textwidth}
        \centering
        \includegraphics[width=\textwidth]{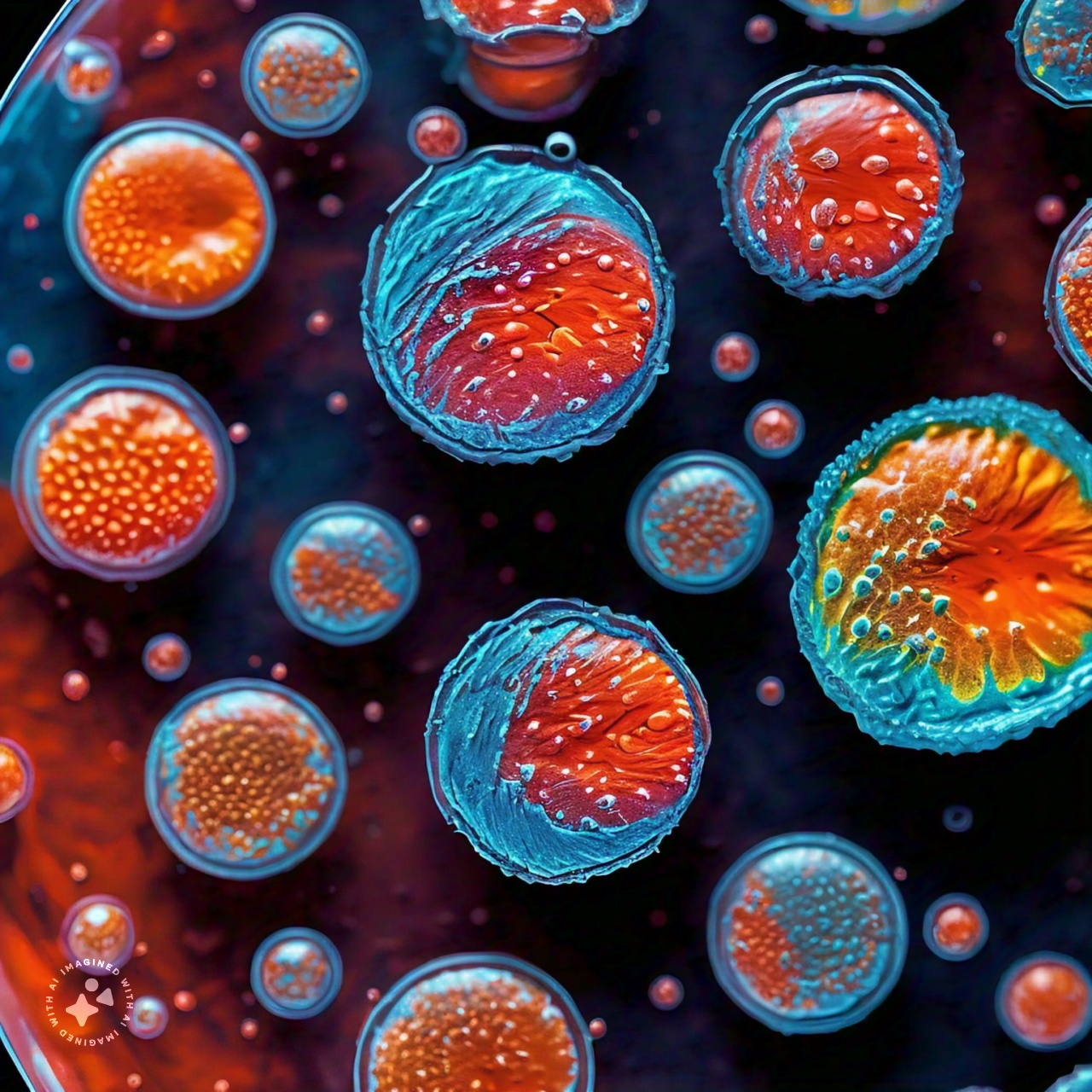}
        \caption{Meta AI}
    \end{subfigure}
    \begin{subfigure}[b]{0.16\textwidth}
        \centering
        \includegraphics[width=\textwidth]{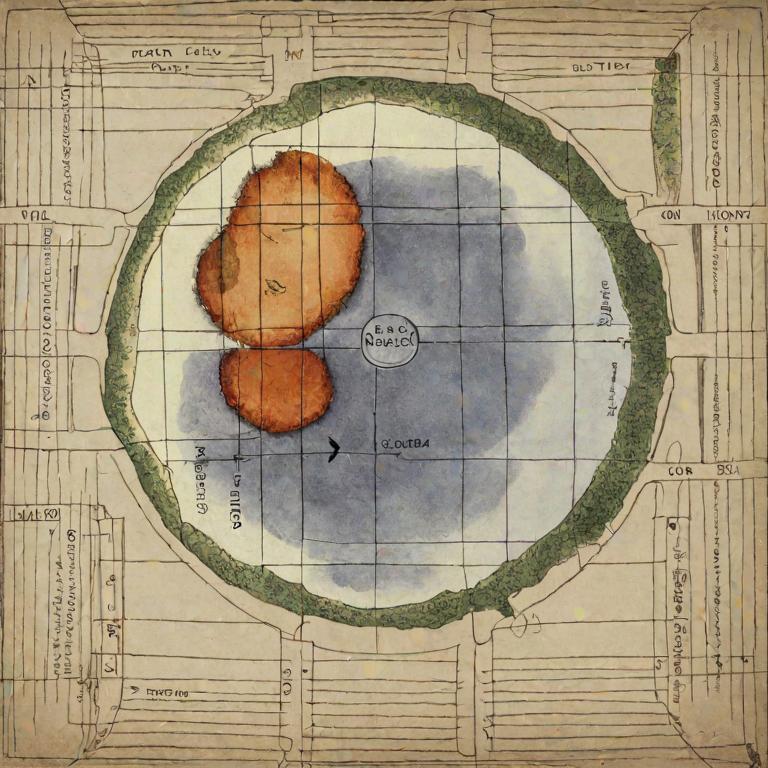}
        \caption{Runway}
    \end{subfigure}
    \begin{subfigure}[b]{0.16\textwidth}
        \centering
        \includegraphics[width=\textwidth]{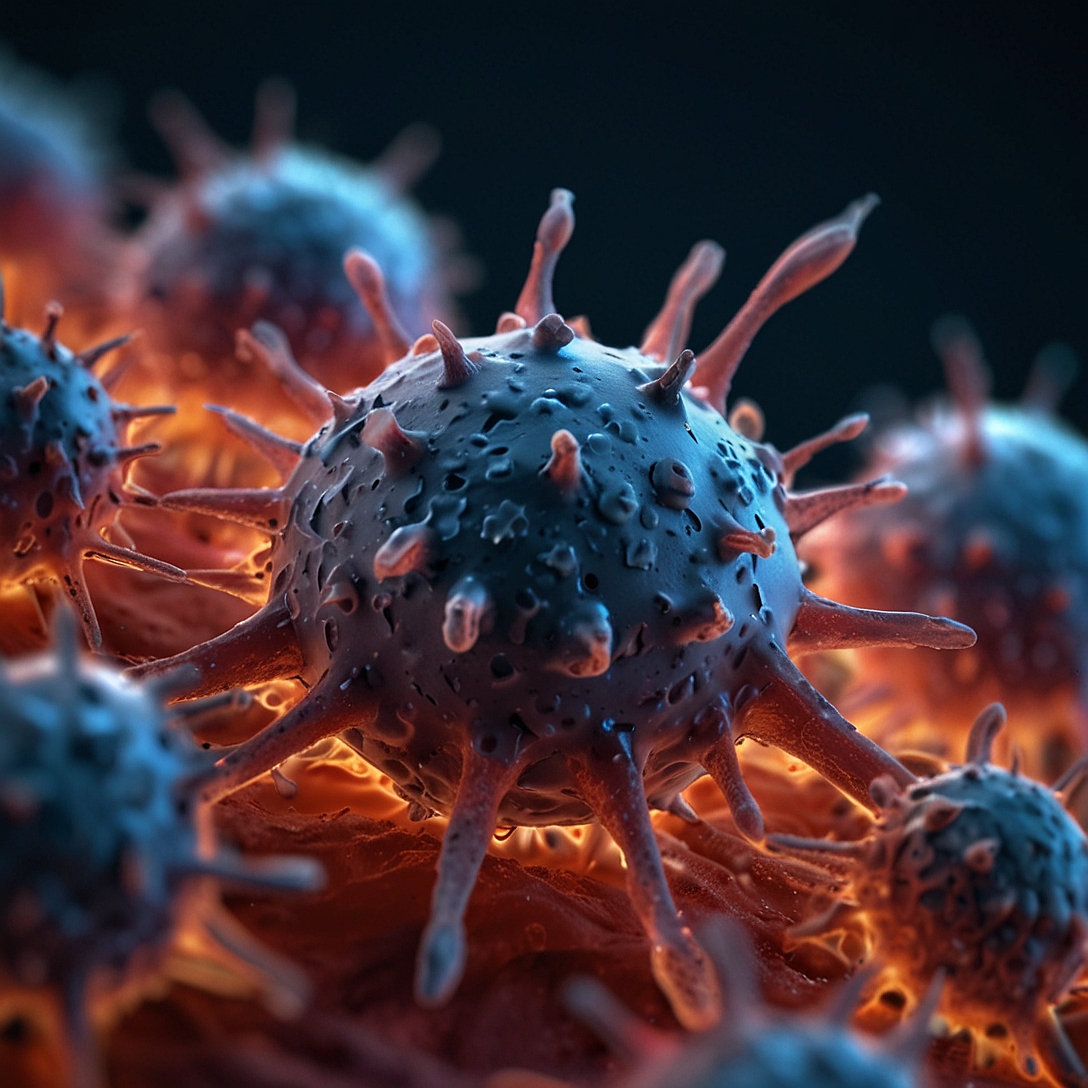}
        \caption{Leonardo Ai}
    \end{subfigure}

    \begin{subfigure}[b]{\textwidth}
        \centering
        \includegraphics[width=0.25\textwidth]{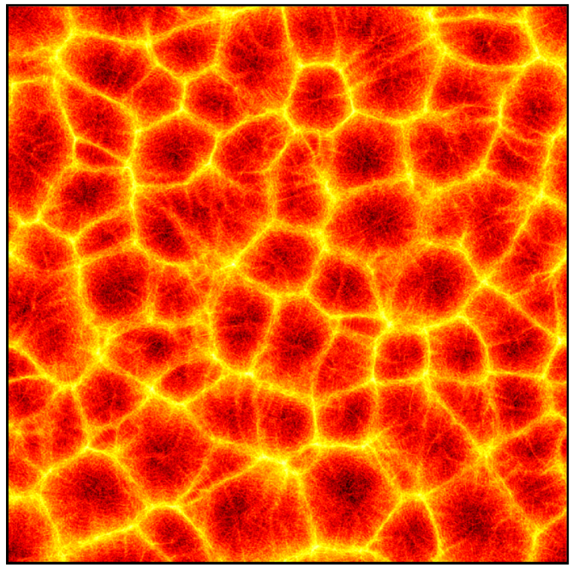}
        \caption{Bénard cells resulting from Rayleigh–Bénard convection in a porous medium. This image, produced by a numerical simulation, is taken from part (b) of Fig. 9 in \cite{de2022strong}, published as open access under a Creative Commons license in the Journal of Fluid Mechanics, allowing readers to distribute the content freely.}
    \end{subfigure}

    \caption{A comparison between the AI-generated images and the numerical simulation result (i.e., ground truth) for the prompt ``Bénard cells''}
    \label{Fig4}
\end{figure}


\subsubsection{Bénard cells}
\label{Sect2710}

In Fig. \ref{Fig4}, we examine the images generated by the models in response to the prompt ``Bénard cells''. Except for the image produced by Gemini Advanced, the rest are irrelevant. Similar to what we discussed in the previous part, this accuracy from Gemini Advanced may suggest it has access to online information. The other models display images of cells, likely reflecting contexts such as biology or medicine, indicating a misunderstanding of ``Bénard cells'' as a technical term in fluid mechanics.



\subsubsection{Flow over an airplane wing}
\label{Sect2711}

\begin{figure}[t]
    \centering
   
    \begin{subfigure}[b]{0.16\textwidth}
        \centering
        \includegraphics[width=\textwidth]{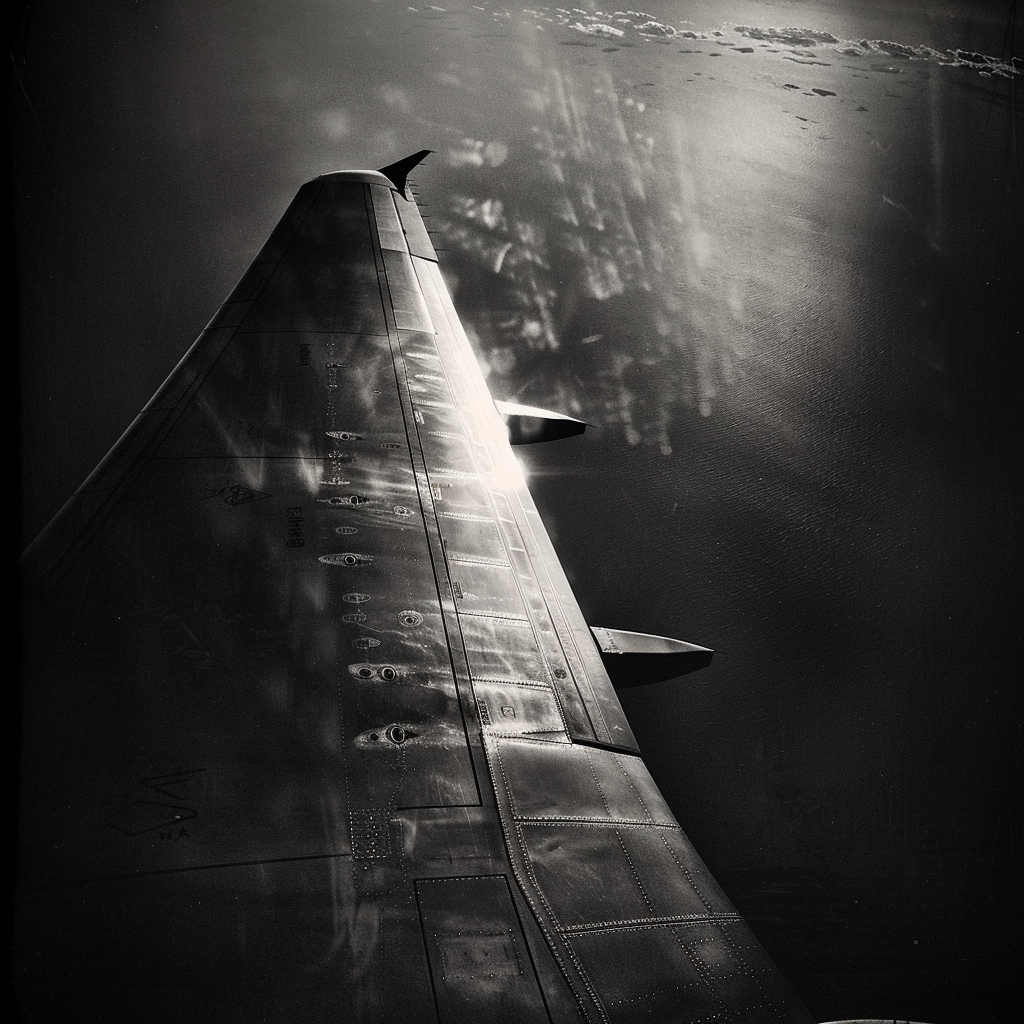}
        \caption{Midjourney}
    \end{subfigure}
    \begin{subfigure}[b]{0.16\textwidth}
        \centering
        \includegraphics[width=\textwidth]{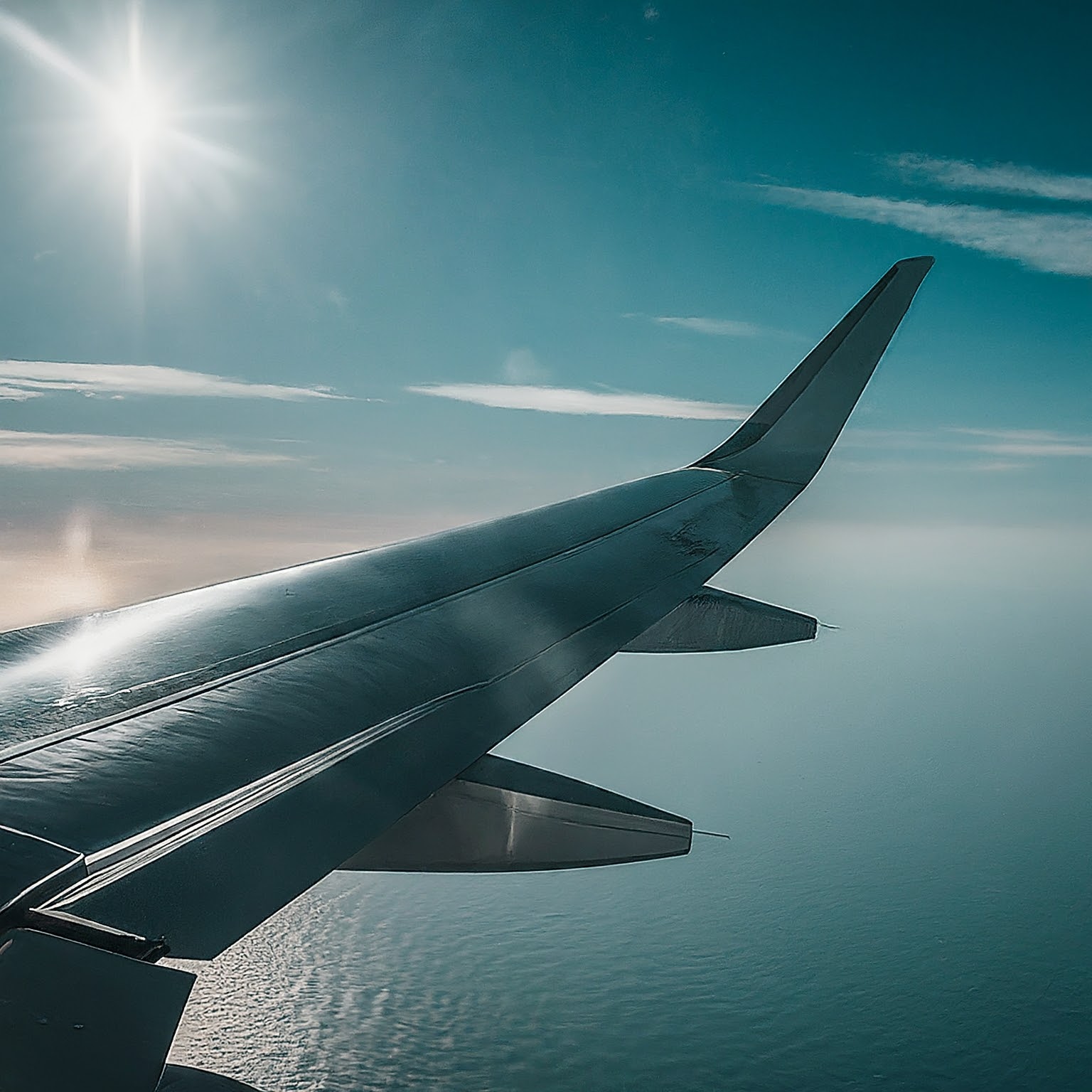}
        \caption{Gemini}
    \end{subfigure}
    \begin{subfigure}[b]{0.16\textwidth}
        \centering
        \includegraphics[width=\textwidth]{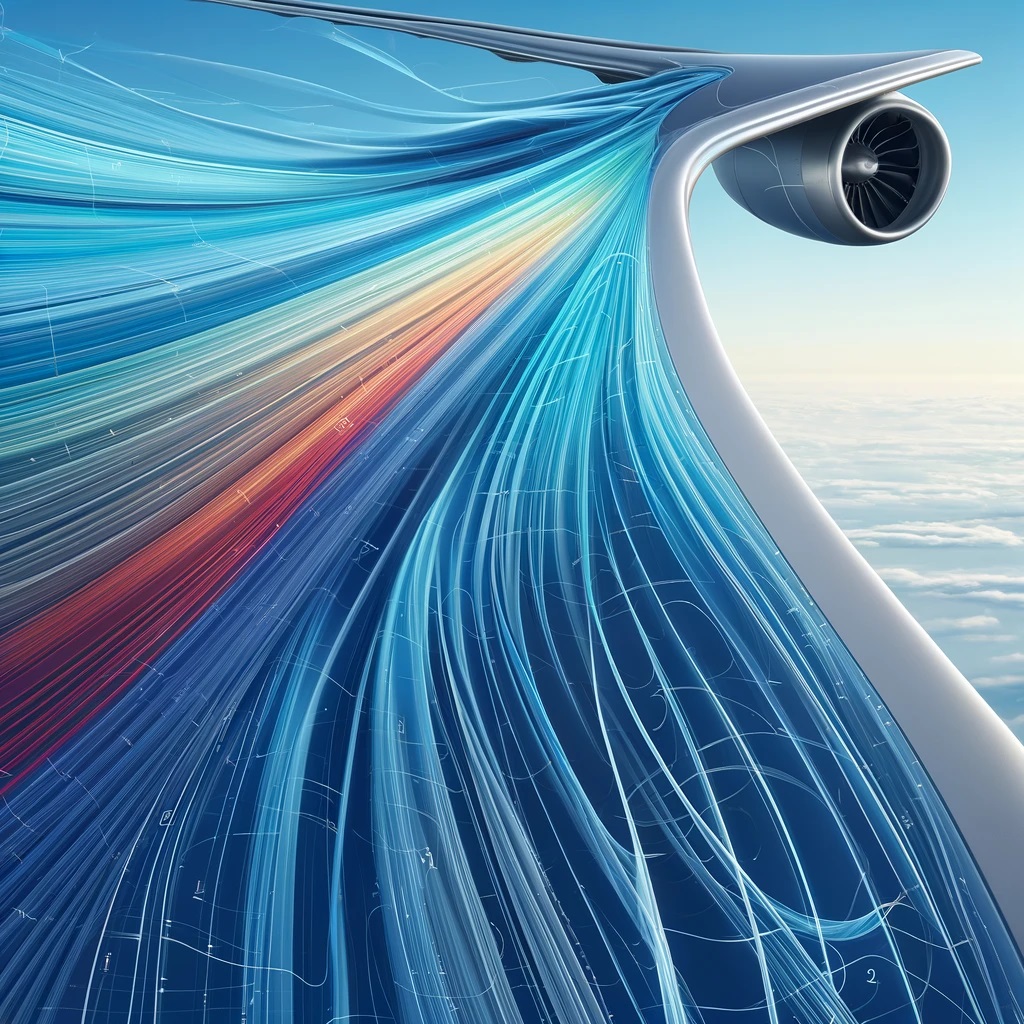}
        \caption{DALL·E 3}
    \end{subfigure}
    \begin{subfigure}[b]{0.16\textwidth}
        \centering
        \includegraphics[width=\textwidth]{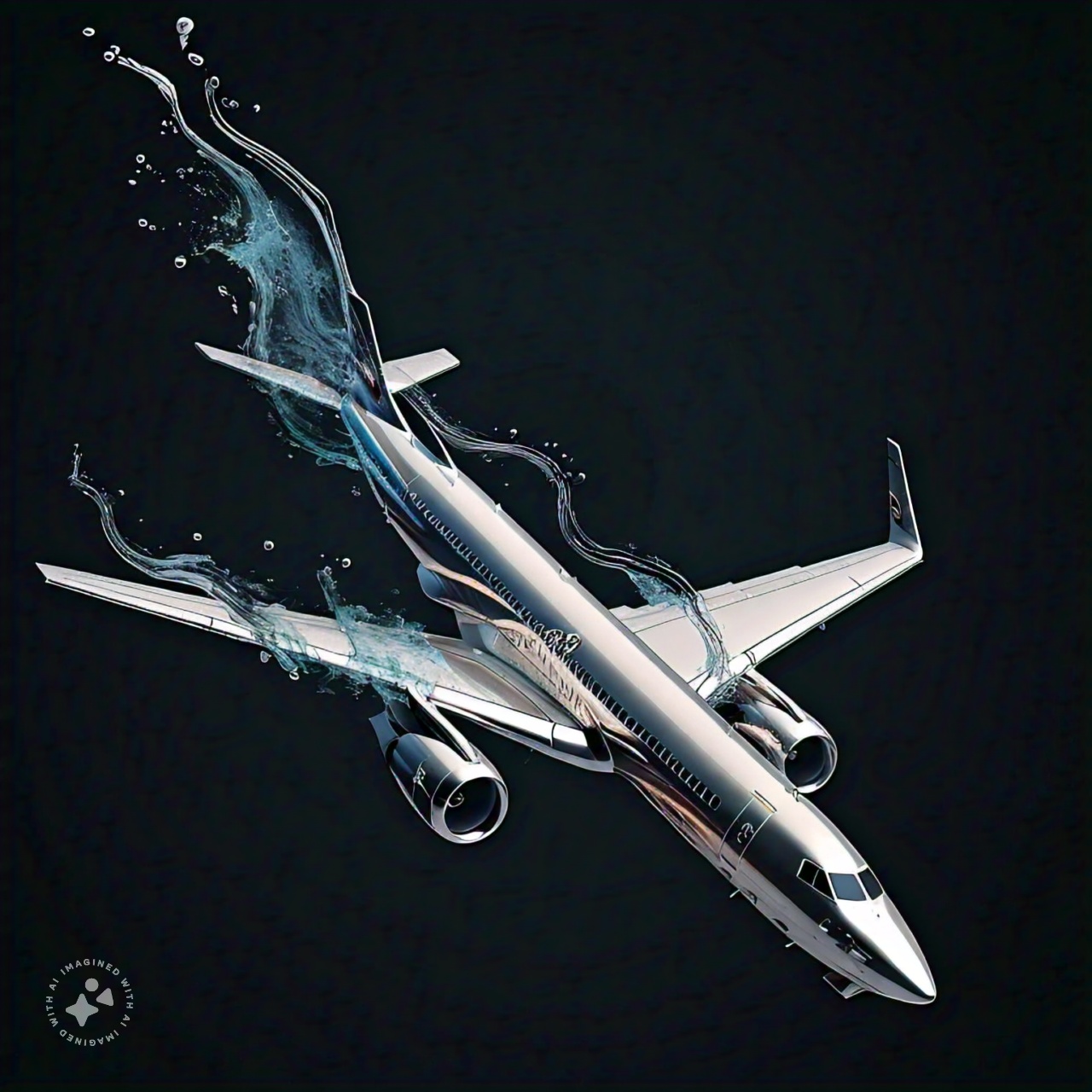}
        \caption{Meta AI}
    \end{subfigure}
    \begin{subfigure}[b]{0.16\textwidth}
        \centering
        \includegraphics[width=\textwidth]{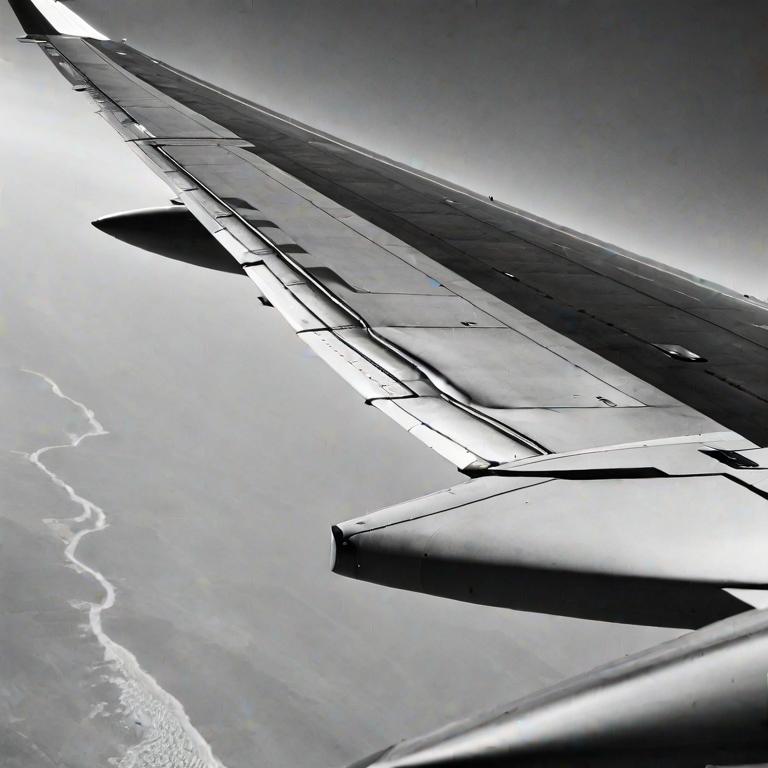}
        \caption{Runway}
    \end{subfigure}
    \begin{subfigure}[b]{0.16\textwidth}
        \centering
        \includegraphics[width=\textwidth]{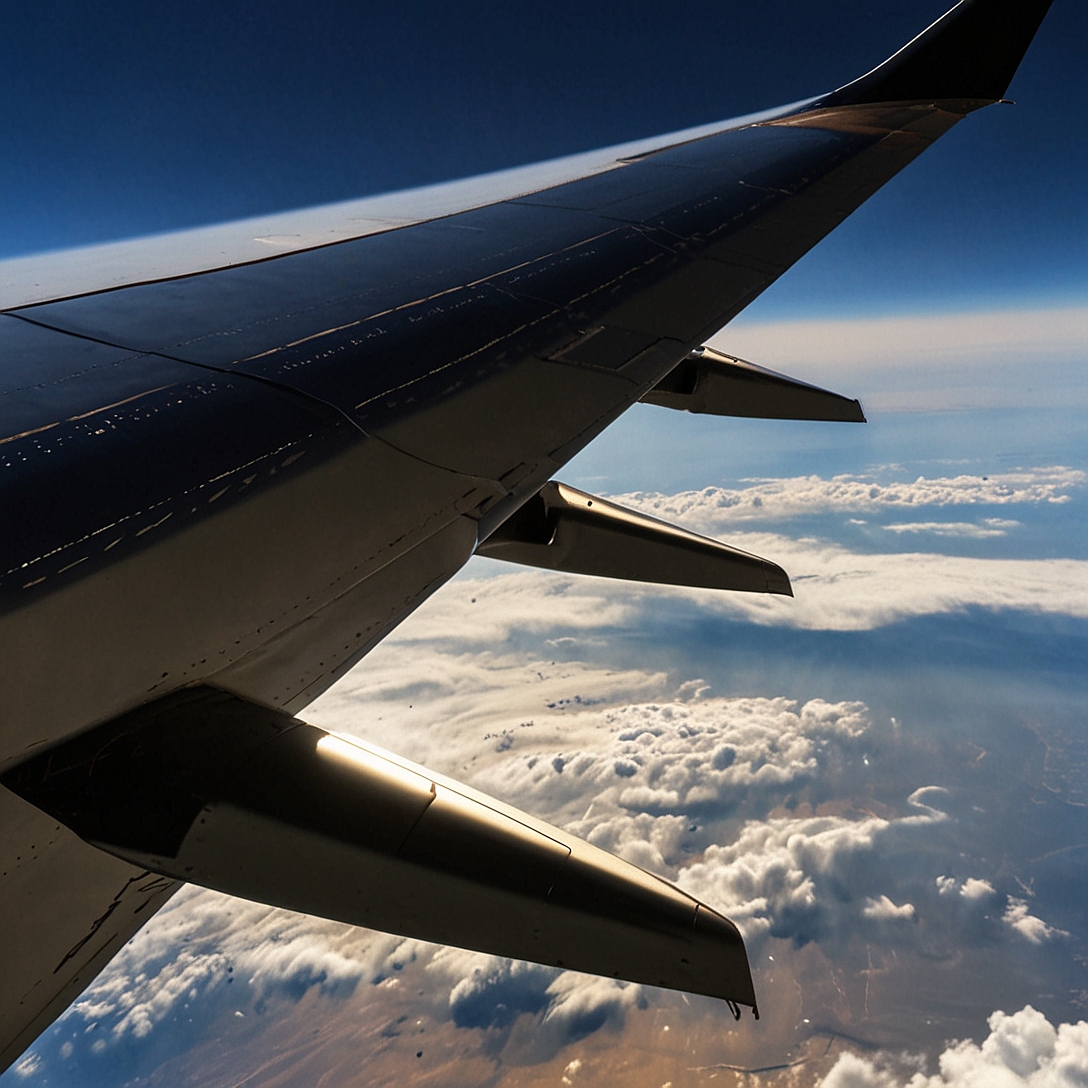}
        \caption{Leonardo Ai}
    \end{subfigure}

    \begin{subfigure}[b]{\textwidth}
        \centering
        \includegraphics[width=0.4\textwidth]{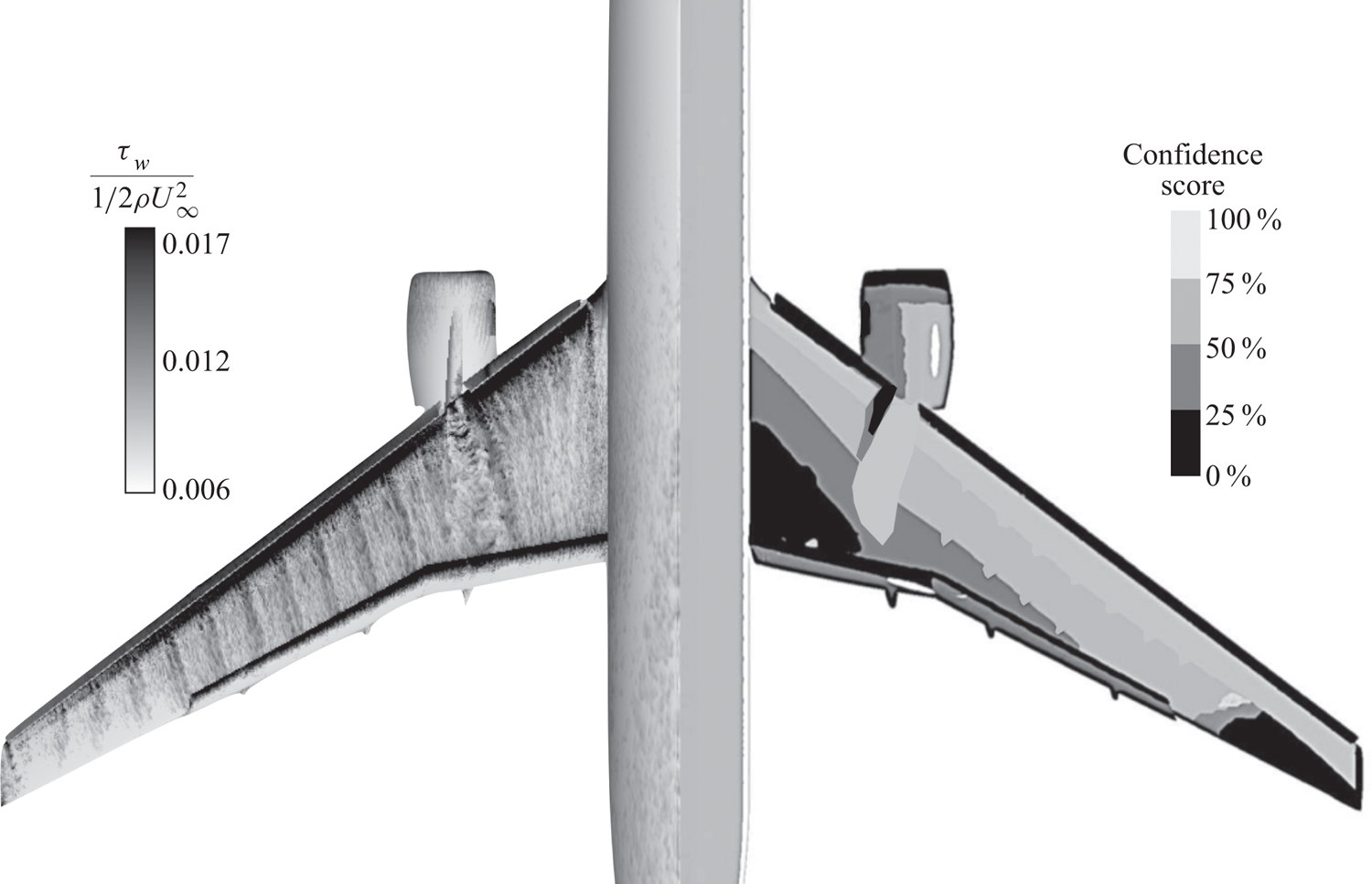}
        \caption{Normalized wall shear stress (left) and confidence map (right) displayed over the NASA Common Research Model High-lift (\url{https://commonresearchmodel.larc.nasa.gov/high-lift-crm/}). This image, produced by a numerical simulation, is Fig. 16 in \cite{lozano2023machine}, published as open access under a Creative Commons license in the Journal of Fluid Mechanics, allowing readers to distribute the content freely.}
    \end{subfigure}

    \caption{A comparison between the AI-generated images and the numerical simulation result (i.e., ground truth) for the prompt ``Flow over an airplane wing''}
    \label{Fig6}
\end{figure}

The last prompt investigated in this technical report is ``Flow over an airplane wing'', with the outputs from the generative models displayed in Fig. \ref{Fig6}. Midjourney, Gemini Advanced, Runway, and Leonardo Ai each generate an image of an airplane wing. These images are very similar to each other, displaying a similar perspective and angle of view. They depict wings of airplanes flying in the sky above clouds. However, it is not clear how the term ``flow'' in our textual prompt is reflected in these images, as can be seen in Fig. \ref{Fig6}.

According to Fig. \ref{Fig6}, the generated image by Meta AI is notable because it shows an airplane moving downward, with water flowing around the wing. The photo appears artistic; however, it can be considered an alternative interpretation of the textual prompt. Indeed, although we did not initially expect to see this image in response to the prompt ``flow over an airplane wing'', the image effectively illustrates the action of an object falling into water, capturing a realistic sense in the context of fluid dynamics.

From what we observe in Fig. \ref{Fig6}, the image generated by DALL·E 3 is moderately similar to what DALL·E 3 produced for the prompt ``flow over an airfoil'', shown in Fig. \ref{Fig10}. Again, we see a turbine attached to an airplane wing. However, this time the wing is depicted in a very cinematic style.


\section{Text to video}
\label{Sect3}

\subsection{Meta AI}
\label{Sect32}

In Sect. \ref{Sect25}, we used Meta AI to generate images of fluid motion from text descriptions. In this subsection, we investigate Meta AI's capability to generate videos from text prompts. It should be noted that Meta AI does not directly create videos from text prompts. Instead, it first generates images, and then users can animate these images by pressing the ``animate'' button. Meta AI automatically animates the images, producing animations approximately three seconds in length.

\subsection{Runway ML}
\label{Sect31}

In Sect. \ref{Sect42}, we used Runway ML for generating images from text, specifically depicting fluid dynamic motion. In this subsection, we shift to exploring Runway ML's text-to-video capabilities, focusing on creating short videos that visually represent fluid dynamic motion based on textual descriptions. Similar to Meta AI, Runway ML first generates a picture and then animates it. However, compared to Meta AI, Runway ML produces images of higher quality.

\subsection{Comparison}
\label{Sect33}

\begin{figure}[htbp]
  \centering 
     
    \begin{subfigure}[b]{0.16\textwidth}
        \centering
        \includegraphics[width=\textwidth]{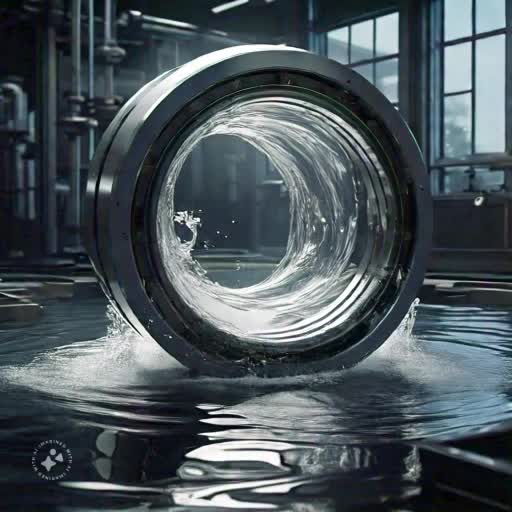}
    \end{subfigure}
    \begin{subfigure}[b]{0.16\textwidth}
        \centering
        \includegraphics[width=\textwidth]{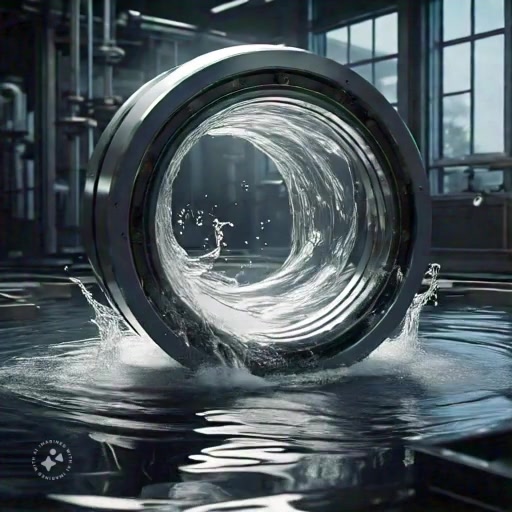}
    \end{subfigure}
    \begin{subfigure}[b]{0.16\textwidth}
        \centering
        \includegraphics[width=\textwidth]{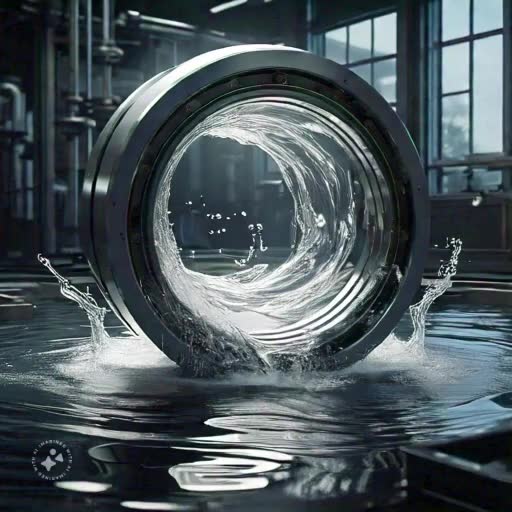}
    \end{subfigure}
    \begin{subfigure}[b]{0.16\textwidth}
        \centering
        \includegraphics[width=\textwidth]{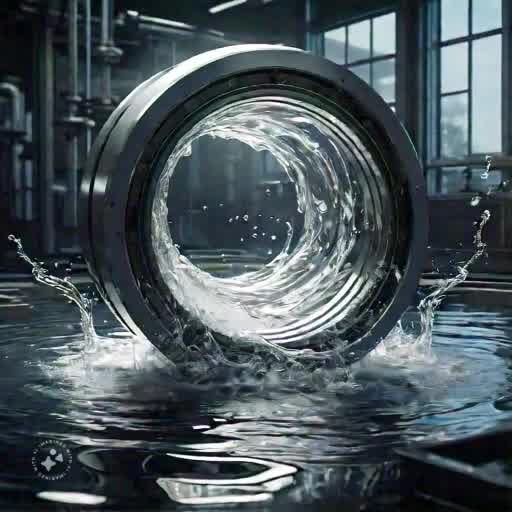}
    \end{subfigure}
    \begin{subfigure}[b]{0.16\textwidth}
        \centering
        \includegraphics[width=\textwidth]{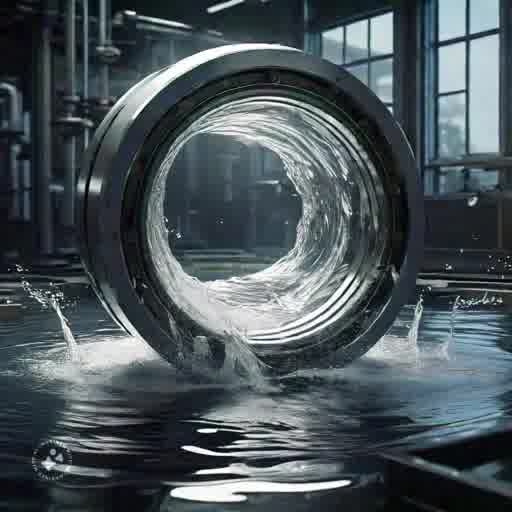}
    \end{subfigure}
    \begin{subfigure}[b]{0.16\textwidth}
        \centering
        \includegraphics[width=\textwidth]{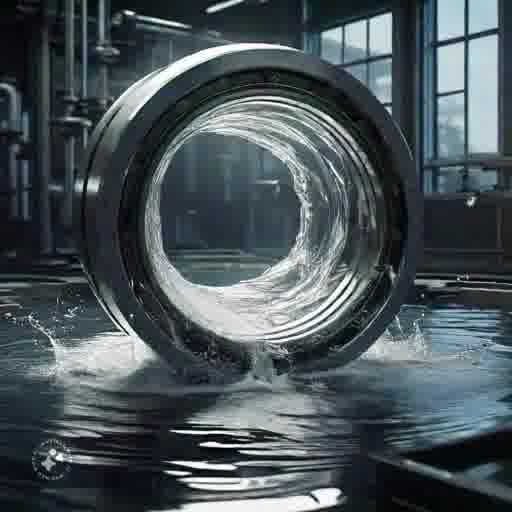}
    \end{subfigure}
    
  \caption{Several sequential frames from the generated video by Meta AI for the prompt ``water past a circular cylinder''. The full generated video is available at this GitHub repository: \url{https://github.com/Ali-Stanford/MisleadingGalleryOfFluidMotionByAI}.}
  \label{Fig120}
\end{figure}

\begin{figure}[htbp]
  \centering 
      \begin{subfigure}[b]{0.16\textwidth}
        \centering
        \includegraphics[width=\textwidth]{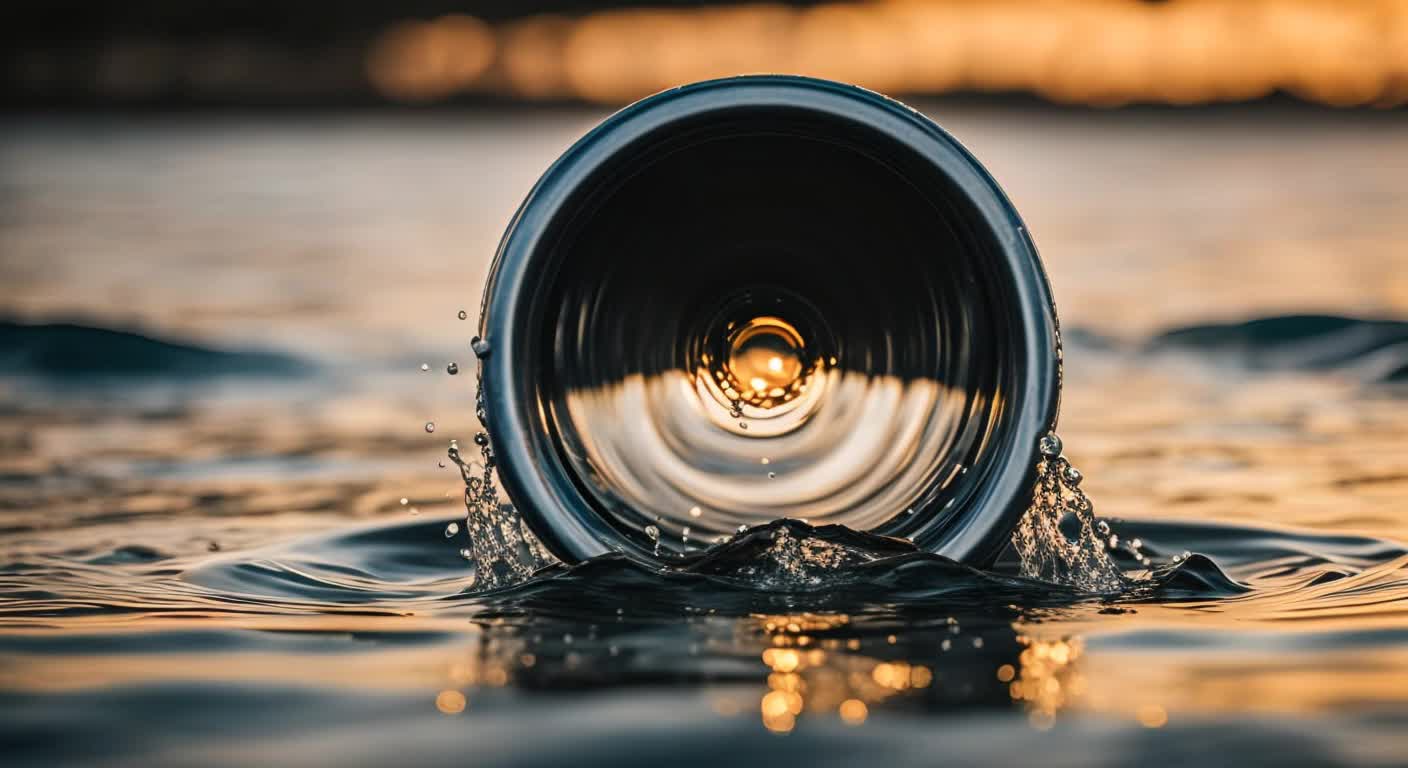}
    \end{subfigure}
    \begin{subfigure}[b]{0.16\textwidth}
        \centering
        \includegraphics[width=\textwidth]{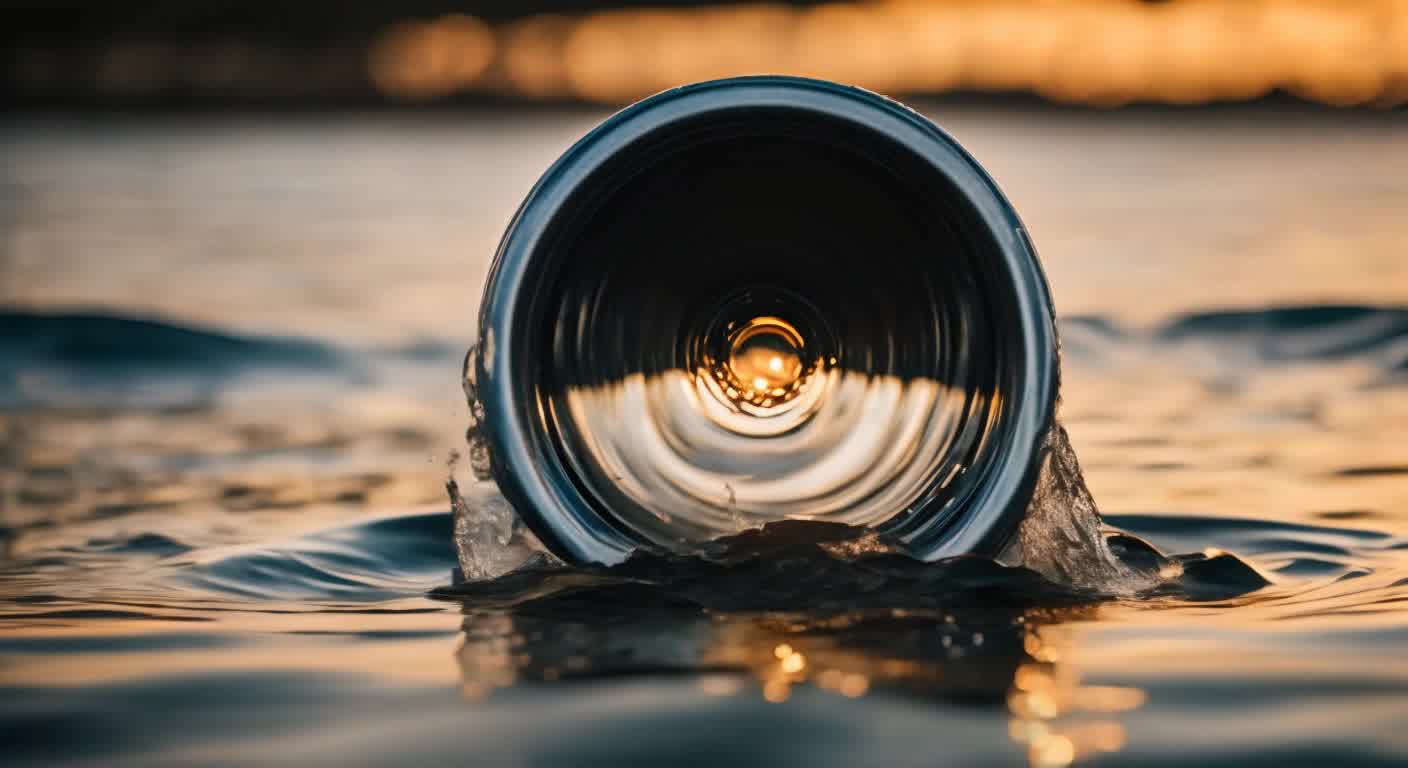}
    \end{subfigure}
    \begin{subfigure}[b]{0.16\textwidth}
        \centering
        \includegraphics[width=\textwidth]{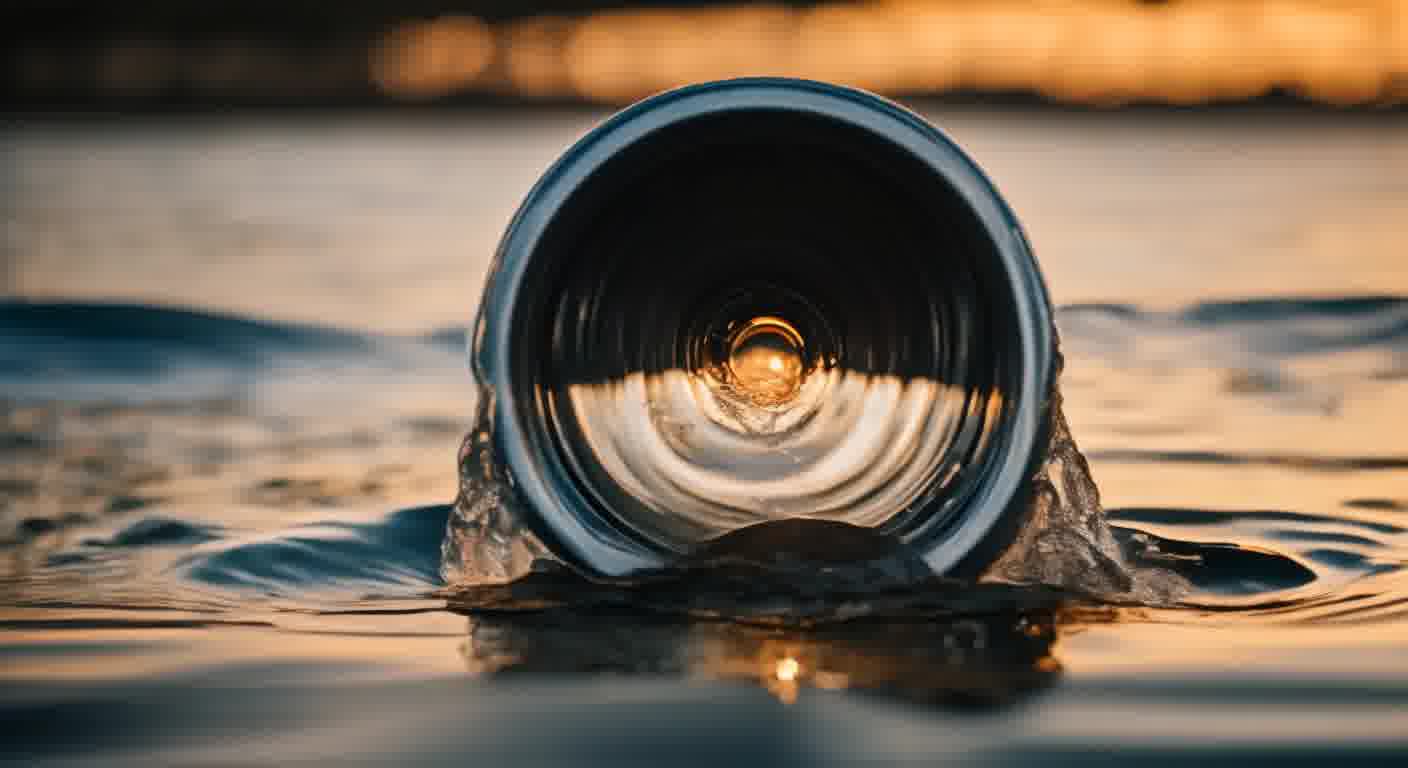}
    \end{subfigure}
    \begin{subfigure}[b]{0.16\textwidth}
        \centering
        \includegraphics[width=\textwidth]{Runway-frame-007.jpg}
    \end{subfigure}
    \begin{subfigure}[b]{0.16\textwidth}
        \centering
        \includegraphics[width=\textwidth]{Runway-frame-014.jpg}
    \end{subfigure}
    \begin{subfigure}[b]{0.16\textwidth}
        \centering
        \includegraphics[width=\textwidth]{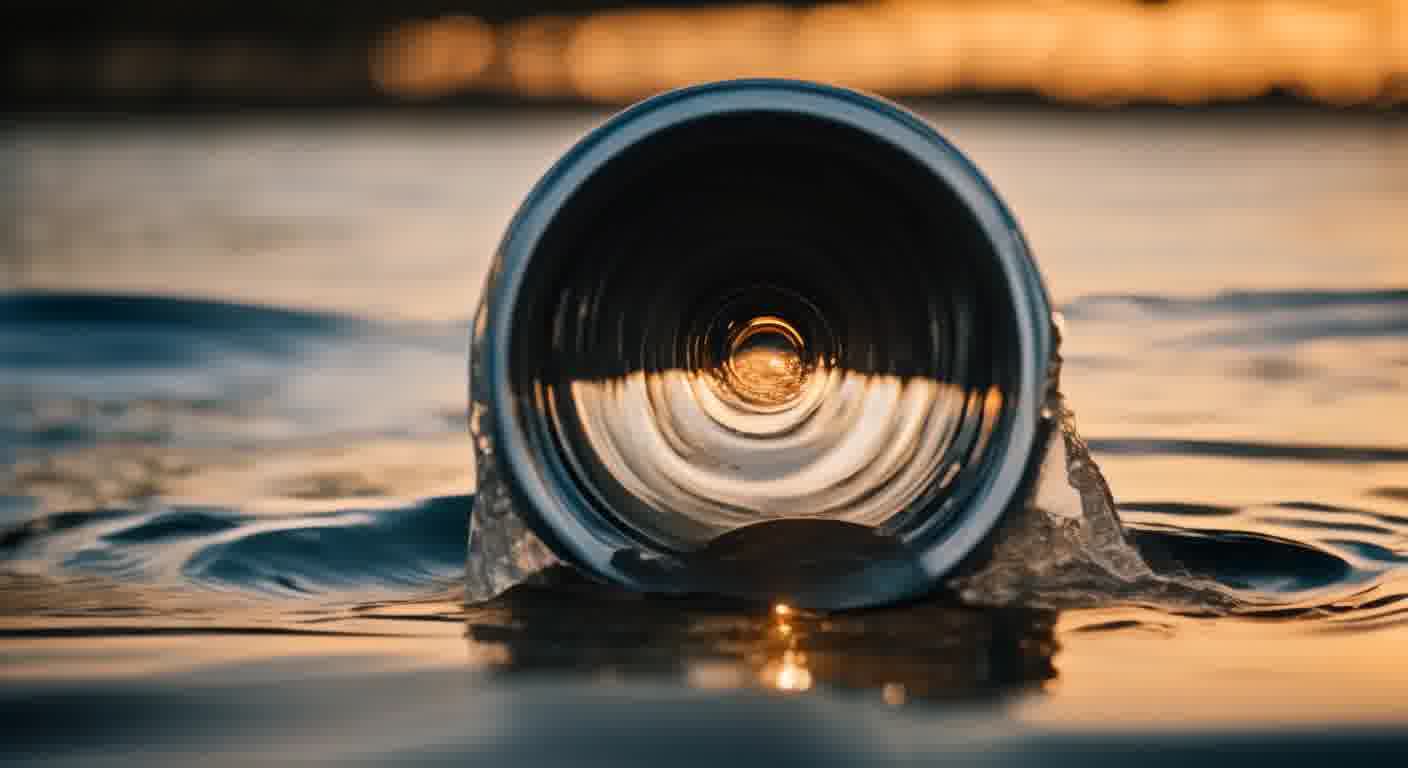}
    \end{subfigure}

  \caption{Several sequential frames from the generated video by Runway ML for the prompt ``water past a circular cylinder''. The full generated video is available at this GitHub repository: \url{https://github.com/Ali-Stanford/MisleadingGalleryOfFluidMotionByAI}.}
  \label{Fig121}
\end{figure}

To compare the performance of Meta AI and Runway ML, we use the textual prompt ``water past a circular cylinder''. The generated videos are available in the repository mentioned in the data availability section. To illustrate the outputs, we display several sequential frames from the generated videos by Meta AI and Runway ML respectively in Fig. \ref{Fig120} and Fig. \ref{Fig121}. The result from Runway ML could be aesthetically valuable. It appears that the cylinder is fixed while water flows around it before entering a river. The interaction between this water and the river water is interesting, as it creates waves. These waves could be considered similar to capillary waves. The image generated by Meta AI is striking, as it appears that the cylinder rotates, causing the water inside to also rotate. Furthermore, the cylinder is partially submerged in pond water, and the interaction between the rotating cylinder and the water surface is displayed. Overall, the generated videos do not deliver the expected and relevant results for the prompt ``water past a circular cylinder''; however, they do generate other interesting fluid dynamics phenomena. Finally, it should be noted that the reason for using the term ``water'' instead of ``flow'' or ``fluid flow'' is intentional, as these generative models have issues interpreting ``flow'' in the context of fluid dynamics. This issue was discussed in detail in Sect. \ref{Sect27}.

\section{Image to text}
\label{Sect4}

\subsection{LLaVA}
\label{Sect41}

LLaVA \footnote{\href{https://llava.hliu.cc}{https://llava.hliu.cc}} was primarily developed by researchers at Microsoft and the University of Wisconsin-Madison \citep{liu2023improved,lin2023video,buscemi2024chatgpt}. It is designed to generate textual descriptions from input images and videos. LLaVA is open-source and free to use, and it has been widely applied in various fields such as biomedicine \citep{li2023llava}, AI agents \citep{liu2023llava}, and solving mathematical geometric problems \citep{gao2023g}. In this subsection, we utilize image-to-text generation, while in Sect. \ref{Sect51}, we employ the video-to-text generation capability of LLaVA.

\subsection{ChatGPT 4}
\label{Sect42}

Beyond the capability of generating images from textual descriptions, ChatGPT 4 is also capable of describing the content of a given image in text \citep{ren2024copyright,lu2024gpt,zhu2023chatgpt,johnson2023image}. In this subsection, we analyze this aspect of ChatGPT 4, focusing on fluid dynamic motion images.

\subsection{Gemini}
\label{Sect43}

In Sect. \ref{Sect25}, we explored Gemini's capability for generating textual descriptions of images depicting fluid motion. In this subsection, we assess Gemini's performance in providing textual descriptions for an image illustrating fluid motion.

\subsection{Comparison}
\label{Sect28}

In this section, we explore the capabilities of three different generative models: ChatGPT-4, Gemini Advanced, and LLaVA, in generating textual descriptions of images depicting fluid motion. We upload a photo of fluid motion to these applications and then ask each to describe the image. To compare these models, we consider two different images. The first image is the Von Karman vortex street shown in part (g) of Fig. \ref{Fig7}. The generated textual descriptions are provided in the following box.

\begin{tcolorbox}
{\fontfamily{lmss}\selectfont
\textbf{ChatGPT-4}\\
\includegraphics[width=5mm]{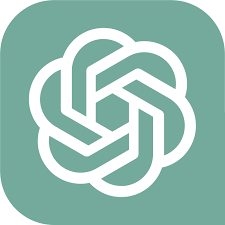}
The image appears to capture a fluid dynamic phenomenon known as a Karman vortex street. This occurs when a fluid flows around a cylindrical object, causing the flow to separate and form vortices on the downstream side. The photograph shows several swirling patterns with bright yellow lines on a dark background, indicating the path of the fluid flow and the vortices created. The yellow lines could be a dye or some form of visual aid to show the flow pattern in the fluid, often used in fluid dynamics studies to visualize the motion. There's a sequence of alternating vortices that create a somewhat symmetric pattern across the image. The overall aesthetic is quite artistic and mesmerizing, capturing both the beauty and complexity of fluid motion.\\

\textbf{Gemini Advanced}\\
\includegraphics[width=5mm]{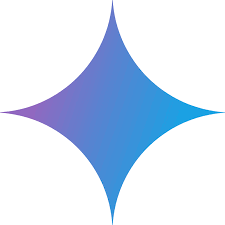}
The image you sent me is a close-up of a lemon peel. The peel is a bright yellow color, and it has a rough, textured surface. The black background makes the yellow of the peel stand out. There is a small white letter “C” in the bottom right corner of the image.\\

\textbf{LLaVA}\\
The image appears to be a close-up photograph of a substance with a flowing, swirling pattern. The substance has a yellowish hue and is captured in motion, creating a dynamic, abstract shape that resembles a snake or a ribbon. The background is dark, which contrasts with the bright yellow of the substance, making the swirling pattern stand out. The image has a shallow depth of field, with the focus on the swirling substance, causing the background to be blurred. There is a letter "C" in the top right corner, which might be a watermark or a part of a logo. The overall effect is quite mesmerizing and could be interpreted as a representation of movement or fluid dynamics.}
\end{tcolorbox}

As indicated in the above box, ChatGPT-4 provides the best description compared to Gemini Advanced and LLaVA. ChatGPT-4 clearly mentions the ``Karman vortex street'' and describes the phenomenon by addressing keywords such as ``vortex'' and ``flow past a cylinder''. It also recognizes the colors yellow and black in the image. More importantly, ChatGPT-4 notes the yellow color as indicative of an experimental technique used to visualize the vortex.

While the LLaVA description is not entirely accurate, it contains significantly more accurate information compared to Gemini Advanced. For instance, LLaVA addresses the swirling pattern and includes terms such as ``movement of fluid dynamics'' in its description. On the other hand, the Gemini Advanced response is almost unrelated. Interestingly, both Gemini Advanced and LLaVA refer to the letter ``C'' in the picture. Note that as mentioned earlier, LLaVA is free, whereas Gemini Advanced costs \$12 per month.

Another interesting and important point is that when we ask DALL·E 3, embedded in ChatGPT-4, to generate a picture of the Von Kármán vortex street in Sect. \ref{Sect2}, the result is inaccurate, as exhibited in part (c) of Fig. \ref{Fig7}. However, it provides an accurate textual description when it analyzes an actual picture of the Von Kármán vortex street.

\begin{figure}[t]
  \centering 
  \includegraphics[width=0.4\textwidth]{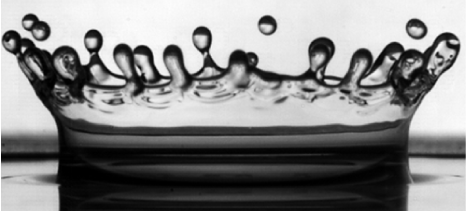}
  \caption{Corona splash in a lab experimental setup; this image is taken from part (c) of Fig. 2 in \cite{burzynski2020splashing}, published as open access under a Creative Commons license in the Journal of Fluid Mechanics, allowing readers to distribute the content freely.}
  \label{Fig12}
\end{figure}

The second image depicts another relatively famous phenomenon in fluid dynamics: the corona splash, which occurs at the moment a droplet impacts a liquid surface, as shown in Fig. \ref{Fig12}. The textual descriptions by ChatGPT-4, Gemini Advanced, and LLaVA are provided in the following box.

\begin{tcolorbox}
{\fontfamily{lmss}\selectfont
\textbf{ChatGPT-4}\\
\includegraphics[width=5mm]{gpt.png}
This image captures a moment of fluid dynamics beautifully. It's a black and white photograph showing a crown-like splash of water. The base of the splash is broad and relatively calm, which suggests that it is probably the surface of water in a container. From this base, several columns of water rise and arc outwards, with droplets separating from the tips of these arcs, forming a pattern that resembles a tiara or a crown. The splash is symmetric, and the physics behind this phenomenon involve the surface tension of the water and the forces exerted upon the initial impact, likely from a droplet falling into the water. The clarity and sharpness of the image suggest a high-speed photograph to freeze the motion.\\

\textbf{Gemini Advanced}\\
\includegraphics[width=5mm]{gemini.png}
The image you sent is a black and white photo of a crown made of water droplets. The water droplets are arranged in a circular pattern, with some droplets larger than others. The background is dark and blurry. The overall effect is one of fragility and impermanence, as the water droplets appear to be on the verge of falling apart.\\

\textbf{LLaVA}\\
The image shows a glass of water with water droplets being poured into it. The droplets are captured in mid-air, creating a dynamic and splashy effect. The water appears to be clear, and the glass is transparent, allowing us to see the water inside. The background is a plain, light color, which contrasts with the dark tones of the water and the droplets, emphasizing the action of the pouring. The image is in black and white, which adds a timeless and artistic quality to the scene.}
\end{tcolorbox}

Similar to the previous investigation, and now focusing on the image shown in Fig. \ref{Fig12}, the output from ChatGPT-4 is significantly more accurate and informative than those generated by Gemini Advanced and LLaVA. ChatGPT-4 clearly mentions the term ``splash''. More interestingly, it addresses the physics of surface tension and accurately recognizes that the photo was taken with a high-speed camera. Both Gemini Advanced and LLaVA describe the water splashing and droplets, and their results are more relevant compared to the image of the Von Karman vortex street.

\begin{figure}[htbp]
  \centering 
  \includegraphics[width=0.9\textwidth]{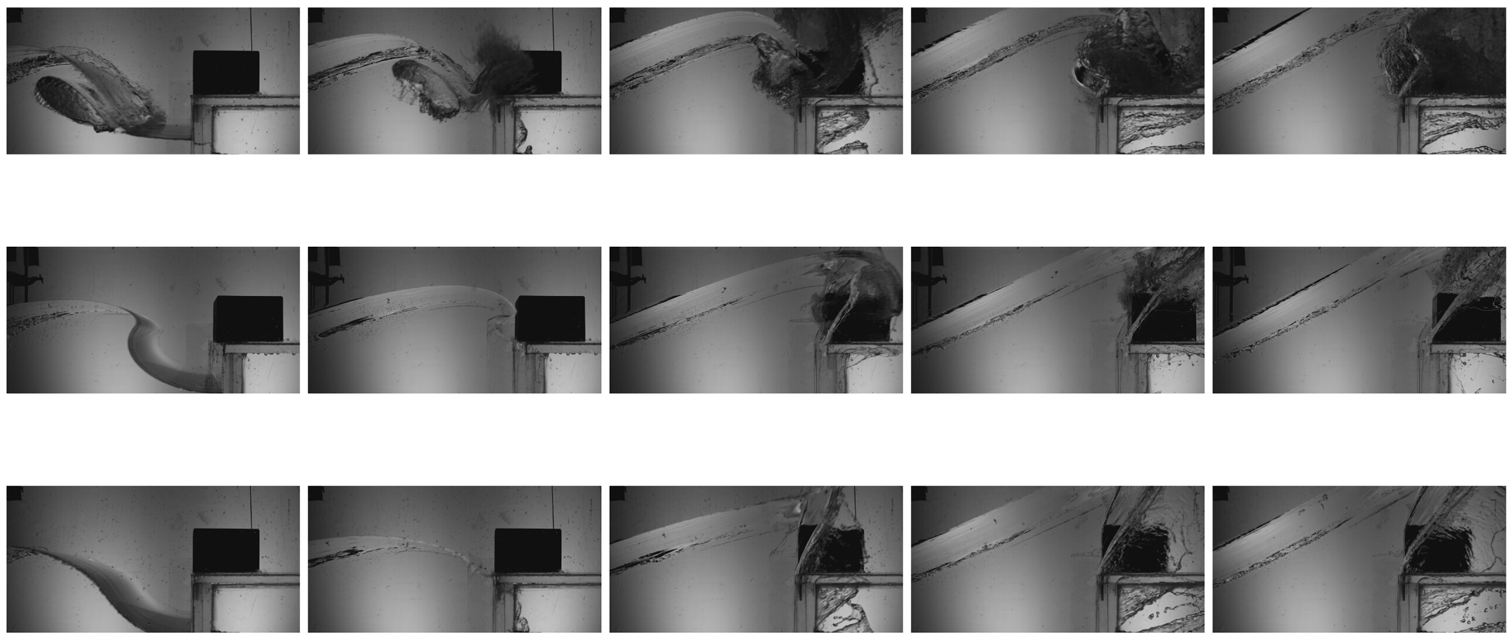}
  \caption{Breaking wave in a lab experimental setup: The corresponding movie is available as a ``supplementary movie'' at \url{https://doi.org/10.1017/jfm.2021.841}. This image is Fig. 3 in \cite{steer2021breaking}, published as open access under a Creative Commons license in the Journal of Fluid Mechanics, allowing readers to distribute the content freely.}
  \label{Fig13}
\end{figure}


\begin{figure}[htbp]
  \centering 
  \includegraphics[width=0.53\textwidth]{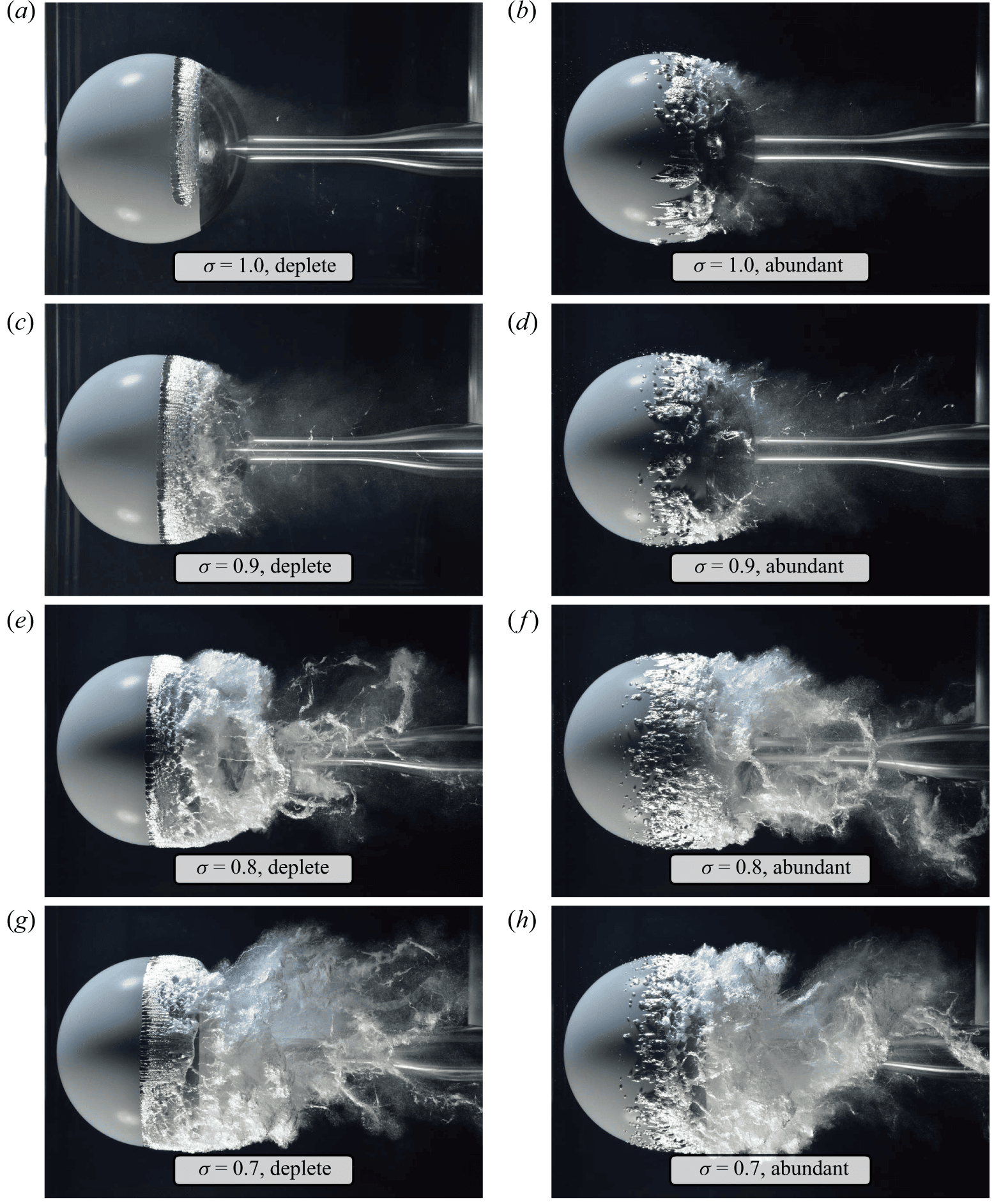}
  \caption{Drops and bubbles around a sphere showing nucleation effects on cavitation in a lab experimental setup: The corresponding movie is available as ``supplementary movie 2'' at \url{https://doi.org/10.1017/jfm.2022.511}. This image is taken from parts (a)--(h) of Fig. 6 in \cite{brandner2022nucleation}, published as open access under a Creative Commons license in the Journal of Fluid Mechanics, allowing readers to distribute the content freely.}
  \label{Fig14}
\end{figure}


\section{Video to text}
\label{Sect5}

\subsection{Video-LLaMA}
\label{Sect51}

In this subsection, we evaluate the performance of Video-LLaMA \footnote{\href{https://huggingface.co/spaces/DAMO-NLP-SG/Video-LLaMA}{https://huggingface.co/spaces/DAMO-NLP-SG/Video-LLaMA}} \citep{zhang2023video} in providing textual descriptions for a short movie that shows fluid motion. We consider two short movies, both of which are taken from laboratory experimental studies published in the Journal of Fluid Mechanics. These movies are available as supplementary materials on the website of each journal paper. Therefore, potential audiences can refer to these websites to watch the movies. Additionally, we have included frames from the movies in Figures 13 and 14, as presented in the associated journal papers. The first movie depicts breaking waves, as illustrated by the frames shown in Fig. \ref{Fig13}. The accompanying textual description from Video-LLaMA is provided in the box below.

 \begin{tcolorbox}
{\fontfamily{lmss}\selectfont
\textbf{Video-LLaMA}\\
\includegraphics[width=5mm]{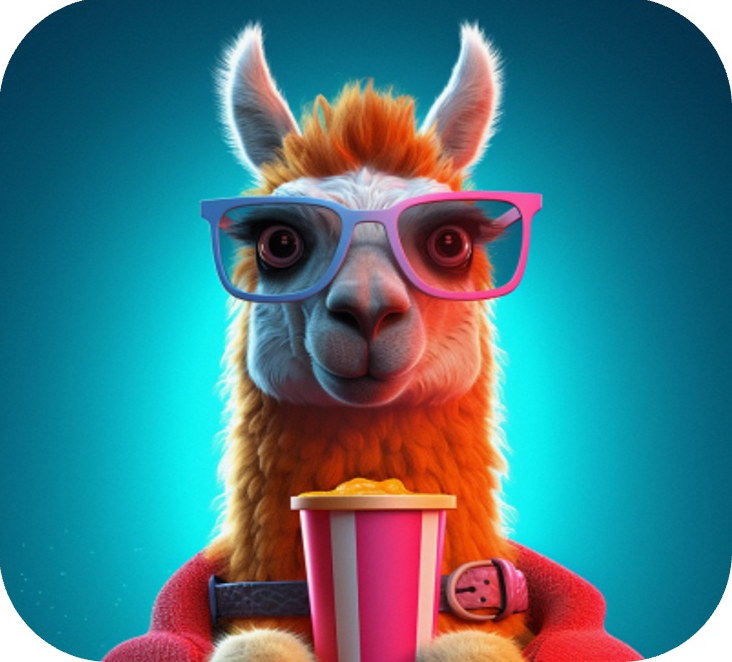}
In this video, we see a black and white photograph of a room with water pouring on a table or floor. We see a variety of objects in the room, including a table, glasses, and a toilet. There is also a skateboard in the room and a clock in the background. We see different angles and views of the pouring water as it covers the table or floor.}
\end{tcolorbox}

The only accurate information in the textual description by Video-LLaMA is relevant to water being poured on a table or the floor. Video-LLaMA also attempts to identify objects, and the only accurately identified object is a table. However, it fails to address the main phenomenon depicted in the movie, which is wave breaking.

The second movie depicts drops and bubbles around a sphere, and some of its frames can be observed in Fig. \ref{Fig14}. The text describing the movie, generated by Video-LLaMA, can be read as follows.

\begin{tcolorbox}
{\fontfamily{lmss}\selectfont
\textbf{Video-LLaMA}\\
\includegraphics[width=5mm]{videoLLaMA.png}
The video consists of a black background with various objects and actions happening around it. The objects in the video include a metal ball on a table, a metal plate, a black liquid, a bottle of water, a cup, and a black mirror. The actions in the video include the presence of smoke, water on a mirror, and various objects moving.}
\end{tcolorbox}

Similar to the previous examination, Video-LLaMA cannot correctly predict and describe the phenomenon occurring in the movie. The only correct information it contains pertains to the presence of a ball and water. It seems that Video-LLaMA views the movies as a sequence of pictures; however, it is unable to make acceptable reasoning among the frames.

%
%

\section{Summary and conclusion}
\label{Sect6}

This technical report comprehensively evaluated the capabilities of several generative AI models to generate text-to-image, text-to-video, image-to-text, and video-to-text in the context of fluid dynamics. Specifically, we investigated prominent models such as Midjourney, Dall·E3, Gemini Advanced, Meta AI, Runway ML, Leonardo Ai, Video-LLaMA, and LLaVA. Our findings demonstrated that the outputs generated by these AI models generally exhibited discrepancies when compared to realistic images obtained from numerical simulations or laboratory experiments, where fluid dynamics events are scientifically and accurately represented. The limitations of these generative models could be concerning in educational contexts, where they might potentially be used with students. Therefore, it is crucial to be aware of these models' shortcomings in handling domain-specific knowledge.

Our conjecture is that these limitations arise because these generative models have not been trained or refined with empirical data on fluid dynamics phenomena. This issue likely stems from the fact that such data, including images and videos, are often copyrighted by scientific journals and publishers, making them not publicly available. As a potential solution, educational institutions could request that the corresponding companies customize and refine these generative models specifically for fluid dynamics by providing sufficient data. We hope that this technical report serves as a starting point for further collaboration between AI companies developing these models and domain-specific experts in the fluid mechanics community to bridge the existing gap.

Finally, it should be noted that similar assessments have been conducted in other scientific disciplines for generative models (see e.g., \citep{mesko2023imperative,megahed2024generative,theis2015note,yang2020evaluation}). As a future direction, such investigations could be extended to additional areas of mechanical engineering and engineering mechanics \citep{xue2020machine,regenwetter2022deep,kashefi2022physicsPIPN,kashefi2023physicsElasticity,pal2024domain,ni2024mechagents,polverini2024understanding,luu2024bioinspiredllm,li2024integrating}. This expansion would provide further insights into the capabilities and limitations of these generative AI models across various engineering disciplines.

\subsection*{Data availability}

The generated videos and high-quality images are available at the following GitHub repository: \url{https://github.com/Ali-Stanford/MisleadingGalleryOfFluidMotionByAI}.









\bibliographystyle{unsrtnat}

\bibliography{references}  







\end{document}